\advance\voffset by 1mm
\advance\hoffset by -7mm
\documentstyle[12pt]{article}
\makeatletter
\@addtoreset{equation}{section}
\makeatother

\newcommand{\artref}[4]{{\sc #1} {\it #2} {\bf #3} #4}
\newcommand{\bookref}[2]{{\sc #1}, #2}

\newcommand{\dl}{\delta^{\Lambda}(0)}
\newcommand{\pif}{\underline \pi}
\newcommand{\gle}{G^{\Lambda}_{1}}
\newcommand{\glz}{G^{\Lambda}_{2}}
\newcommand{\gl}{G^{\Lambda}}
\newcommand{\en}{{\rm n}}

\newcommand{\bfg}{{\bf g}}
\newcommand{\dlf}{\delta^{\Lambda}}
\newcommand{\Dl}{D^{\Lambda}(0)}

\newcommand{\DL}{\Delta^{\Lambda}(0)}
\newcommand{\DLF}{\Delta^{\Lambda}}
\newcommand{\ganz}{{\kern+.25em\sf{Z}\kern-.78em\sf{Z}\kern+.78em\kern-.65em}}
\newcommand{\reell}{{\kern+.25em\sf{R}\kern-.78em\sf{I}\kern+.78em\kern-.25em}}
\newcommand{\posganz}{{\kern+.25em\sf{N}\kern-.86em\sf{I}
\kern+.86em\kern-.25em}}

\begin{document}
\null\vspace{25mm}
\noindent
\begin{center}
{\Large {\bf Goldstone Bosons in a Finite Volume: \\ [2mm]
the Partition Function to Three Loops }}
\footnote{Work supported by Schweizerischer Nationalfonds}
\vspace{10mm}

\noindent {\large W. Bietenholz}
\vspace{3mm}

\noindent
Department for Theoretical Physics \\
University of Bern \\
Sidlerstrasse 5, CH-3012 Bern, Switzerland
\footnote{Present address: CBPF, Rua Dr. Xavier Sigaud 150,
22290 Rio de Janeiro, Brazil}
\vspace{24mm}
\end{center}

\noindent {\small {\em Abstract.}
A system of Goldstone bosons -- stemming from a symmetry
breaking $O(N) \to O(N-1)$ -- in a finite volume at finite temperature
is considered.
In the framework of dimensional regularization, the partition function is
calculated to the 3-loop level for 3 and 4 dimensions, where the
Polyakov method for the measure of the path integral is applied.

Although the underlying theory is the non-linear $\sigma $-model, it will
be shown that the 3-loop result is renormalizable in the sense that
all the singularities can be absorbed by the coupling constants
occurring so far. In finite volume this property is highly non-trivial.
Thus the method for the measure is confirmed.
In addition we show that -- to the considered order -- it coincides
with the Faddeev-Popov measure.
This is also true for the maximal generalization of Polyakov's
measure: none of the additional invariant terms that can be added
contributes to the dimensionally regularized system.

The occurring phenomenological Lagrangian describes for example
2-flavor chiral QCD as well as the classical Heisenberg model,
but there are also points of contact with the
Higgs model, superconductors etc. In addition the finite size corrections
to the susceptibility might improve the interpretation of Monte Carlo
results on the lattice. }
\vspace{2mm}

\tableofcontents

\section{Introduction}
If a continuous symmetry is spontaneously broken, the
Goldstone bosons (GB) dominate the low energy behavior of
the system. The interaction among the Goldstone modes is
strongly constrained by symmetries. This represents a universal
feature of all models exhibiting spontaneous symmetry breaking \cite{sym}.

In the present work we choose $O(N)$ as the symmetry to be broken down
to $O(N-1)$. Of course this can also be applied to symmetry groups
(locally) isomorphic to $O(N)$, such as
$ \ SU(2) \times SU(2) \sim O(4) \ $. Thus
it describes QCD with two flavors and broken chiral symmetry.

The underlying theory will be the non-linear $\sigma$-model, which
includes for dimension $d>2$ all invariant terms.
Thus it is not renormalizable
because it contains an infinite number of coupling constants.
\footnote{Different is the 2-dimensional case: there the model {\em is}
renormalizable. Many recent papers concentrate on this case.
We also refer to it in appendix E.}
To any order in the low energy expansion, however, a
``perturbative renormalization'' can be realized, i.e. the coupling
constants occurring to that order are able to absorb all the singularities.
This property provides us with a non-trivial check of the
results and the applied methods.

We are going to use a ``magnetic'' language, so the model most suitable
to our terminology is the classical Heisenberg model for ferromagnets
below the critical temperature.

Particularly in soft pion physics the method of low energy effective
Lagrangians seems to be more efficient than the historical way (current
algebra, Ward identities, etc.)
\cite{leff,endlkopp,gas/leu84,gas/leu87,gas/leu88,tgnullqcd}.
For unknown reasons
two quark flavors are very light compared to the scale of the theory.
If their masses would vanish (chiral limit) the QCD Lagrangian
would exhibit an $SU(2)_{R}\times SU(2)_{L}$ symmetry. This symmetry
spontaneously breaks down to $SU(2)_{R+L} $, creating GBs which are
identified with the pions ($\pi^{+},\pi^{0},\pi^{-} $).\footnote{For
$SU(3)\times SU(3)$ the GBs are identified with the eight lightest
mesons ($\pi ,K,\eta $). This case is not described by an $O(N)$
symmetry; the groups involved in the breakdown $SU(M) \times SU(M)
\to SU(M)$ are locally isomorphic to an orthogonal group only
for $M=2$.} Their
properties reveal the hidden symmetry. They can be analyzed by replacing
${\cal L}_{QCD}$ with its quark and gluon fields by an effective
Lagrangian with pseudoscalar meson fields \cite{gas/leu84,qcd}.
One constructs the
most general ${\cal L} $ in terms of GB fields consistent with the
symmetry of the model. It is generally
assumed -- although not strictly proved -- that the low energy predictions
do {\em only\/} contain this information \cite{wein} .
Then all quantum field theories
generating this type of GBs are covered. Ward identities constrain the
expansion of the Greens function in powers of the momenta and the
external fields. They also imply that the interaction
among the GB modes of low momenta is weak, and
to any finite order of the low energy expansion there
occur only a finite number of coupling constants \cite{endlkopp,wein}.

In the standard model the influence of the gauge and fermion fields
on the scalar sector is weak.
\footnote{Here only a very heavy top quark could be a trouble-maker
\cite{Lind}.}
Concerning the upper bound of the
Higgs mass $m_{H}$ we can consider the $SU(2)$ sector separately
and study an $O(4)$ scalar field theory \cite{higgso4}.
On the other hand the $O(4)$ model is inadequate for questions
involving very weak scalar interactions, such as the lower bound of $m_{H}$;
there the gauge and Yukawa couplings can not be neglected any more.

In the $O(4)$ model the tree level yields the relation $m_{H}^{2}
\propto \lambda_{r}$ , where $\lambda_{r}$ is the renormalized self coupling.
Thus $m_{H}$ enters as a free parameter. Theoretical information about
it can be gained, however, making use of the fact that the cutoff $\Lambda $
is unremovable. Even if we set the bare $\lambda = \infty $, for any given
ratio $\Lambda / m_{H} > O(1)$ \ $\lambda _{r}$ runs away from infinity
fast enough to fix an upper bound for $m_{H}/m_{W} $ that can be
determined numerically. For decreasing $\Lambda /m_{H}$ the latter rises,
so choosing the smallest, physically acceptable value for
$\Lambda / m_{H}$, we obtain an absolute upper bound for $m_{H}$
\cite{higgs,higgstheo}.
But its numerical evaluation is charged with significant finite
size effects due to massless GB \cite{AHas}. We are going to
present more precise analytical results about finite size effects
of that kind.

Generally our results are suitable for comparison
with data of lattice MC simulations (in particular concerning the
conclusions about infinite volume),
finite size properties of ferromagnets etc. About the
link to bosonic strings, see \cite{string}.\\

But of course the confirmation of the perturbative renormalizability
of the results
we get with the Polyakov method for the measure has not at least
a theoretical and technological meaning. This is particularly of
interest in view of the symmetry groups $SU(N)$, where for $N>2$
no other applicable treatment of the measure is known.

We also show that the leading term, which corresponds to
Polyakov's definition of the functional measure, can be completed
by an arbitrary linear combination of further terms obeying the
symmetries of the system: it turns out that in dimensional
regularization -- to the considered order -- all the contributions
of non-leading measure terms vanish. Only if we include
power divergences such a generalization becomes necessary, e.g.
for the invariance of the partition function under some field
transformations.\\

In {\em section~2} we introduce the model and its parameters and
derive the effective action up to the third order in the derivative
expansion for dimensions $d=3$ and $d=4$.

In {\em section~3} we describe Polyakov's definition of the measure
occurring in the path integral and apply it to determine
the measure to the second order. We also confirm the result of
this method with the measure of Faddeev and Popov.

The 1-loop calculation of the partition function is given comprehensively
in {\em section~4}. The extension of this calculation to 3 loops
is described for $d=3$ and $d=4$ in {\em sections 5} and {\em 6}, respectively.

In {\em section~7} we show explicitly that the results of sections
5 and 6 can be renormalized perturbatively; we determine the
constraints on the counter terms.

{\em Conclusions} and five {\em appendices} about rather technical aspects
are added. There we discuss the results with a generalized singularity
structure that includes the leading power divergences and show that
perturbative renormalizability still holds. In finite volume this is
highly non-trivial, hence it provides us with a sensitive consistency check.
We also discuss the explicit form of
the generalized measure, the link to a massive
expansion and the conclusions about the renormalizable
case $d=2$ as well as the reduction to quantum mechanics ($d=1$).

\section{The non-linear $\sigma $-model}
In two dimensions, the non-linear $\sigma $-model can be characterized
by the Lagrangian
\begin{equation}
\label{Lsy}
{\cal L}^{(sy)} = \frac{F^{2}}{2} \partial_{\mu} \vec S \partial_{\mu}
\vec S
\end{equation}
where $\vec S (x) $ is an $N$-component scalar field subject to the
constraint
\begin{equation}
\label{normS}
\sum_{\alpha =0}^{N-1} S^{\alpha }(x) S^{\alpha }(x) = 1
\end{equation}
The model represents a renormalizable, asymptotically free two-dimensional
field theory
which is invariant under
global $O(N)$ rotations of the
vector $\vec S (x)$. One may introduce a term which explicitly breaks
the $O(N)$ symmetry by coupling the system to an external ``magnetic
field'' $\vec H $,
\begin{equation}
\label{Lsb}
{\cal L}^{(sb)} = - \Sigma ( \vec H \vec S )
\end{equation}
For dimension $d>2$, the Lagrangian specified in eqs
(\ref{Lsy}) and (\ref{Lsb})
does, however, not make sense as it stands because the constraint
(\ref{normS}) generates derivative couplings which are not renormalizable
(the coupling constant $F$ carries the dimension [mass]$^{d/2-1}$).
Accordingly, for $d>2$, the term ``non-linear $\sigma $-model'' does not
refer to the Lagrangian (\ref{Lsy}), (\ref{Lsb}) but to the following
more general construction. One considers the set of all possible
$O(N)$-invariant couplings of the field $\vec S (x)$, allowing for
arbitrarily many derivatives. Ordering the infinite series of vertices
according to the number of derivatives, the first few terms are
\footnote{The index of $g$ is the number of derivatives.}
\begin{eqnarray}
{\cal L}^{(sy)}& = &\frac{F^{2}}{2} \partial_{\mu} \vec S \partial_{\mu}
\vec S + \frac{1}{2} g^{(1)}_{4} \partial^{2} \vec S  \partial^{2}
\vec S  + \frac{1}{4} g^{(2)}_{4} (\partial_{\mu} \vec S \partial_{\mu}
\vec S )^{2} \nonumber \\
& & \frac{1}{4} g_{4}^{(3)} (\partial_{\mu} \vec S \partial_{\nu} \vec S)^{2}
+ \frac{1}{2} g_{6}^{(1)}\partial_{\mu}\partial^{2}\vec S
\partial_{\mu}\partial^{2}\vec S + \dots \label{Lsy2}
\end{eqnarray}
where the dots stand for further terms involving six or more derivatives
of the fields. The corresponding generalization of the symmetry breaking
term (\ref{Lsb}) involves derivatives of the field $\vec S (x)$ as well
as higher powers of the magnetic field. Assuming $\vec H $ to be constant,
the first terms are
\footnote{The indices of $h$ are the power of $H$ and
the number of derivatives.}
\begin{eqnarray}
\label{Lsb2}
-{\cal L}^{(sb)} &=& \Sigma (\vec H \vec S) + h_{2,0}^{(1)} (\vec H
\vec S )^{2} + h_{2,0}^{(2)} (\vec H \vec H ) + h_{1,2}^{(1)}(\vec H
\vec S)(\partial_{\mu}\vec S \partial_{\mu}\vec S) + h_{1,2}^{(2)}
(\vec H \partial^{2} \vec S) \qquad \nonumber \\
&& +h_{1,4}^{(1)}(\vec H \vec S)(\partial_{\mu} \vec S \partial_{\mu}
\vec S )^{2} + h_{1,4}^{(2)}(\vec H \vec S)(\partial_{\mu} \vec S
\partial_{\nu} \vec S)^{2} + h_{1,4}^{(3)}(\vec H \vec S) (\partial^{2}
\vec S \partial^{2} \vec S) \nonumber \\
&& +h_{2,2}^{(1)}(\vec H \vec H)(\partial_{\mu}\vec S \partial_{\mu}
\vec S ) + h_{2,2}^{(2)}(\vec H \vec S)^{2}(\partial_{\mu}\vec S
\partial_{\mu}\vec S) + h_{2,2}^{(3)} (\vec H \vec S)(\vec H \partial^{2}
\vec S ) \dots
\end{eqnarray}
In the following, we study the model defined by
\begin{equation}
{\cal L} = {\cal L}^{(sy)} + {\cal L}^{(sb)}
\end{equation}
in $d=3$ and $d=4$. More specifically, we consider the properties of
the corresponding partition function Z in a finite volume, which can
also be understood as an IR regularization.
We introduce a rectangular box
$L_{1} L_{2} \dots L_{d} = V$ ($L_{i}/L_{k}$ not large) imposing
periodic boundary conditions :
\begin{eqnarray*}
{\rm Z} &=& {\bf N} \int [d\vec S ] e^{-\int_{V} {\cal L} dx} \\
\vec S (x) &=& \vec S (x+ \tilde \en) \ , \quad \tilde \en = (\en_{1}L_{1},
\en_{2}L_{2}, \dots ,\en_{d}L_{d}) \ , \quad \en_{\mu } \in \ganz
\end{eqnarray*}
$[d\vec S ]$ is the ordinary measure of the path integral and ${\bf N}$
is an $\vec H $-independent normalization constant (which also requires
renormalization, see below).
One component can also be taken to be imaginary time-like, so the corresponding
side of the Euclidean box, say $L_{d}$, represents the inverse, finite
temperature of the system: $L_{d}= \frac{1}{T}$. From this interpretation
we see that all the coupling constants must be independent of $L_{d}$,
and due to the permutational symmetry of the Euclidean axis they can't
depend on $L_{1}, \dots , L_{d-1}$ either \cite{gas/leu88,has/leu}.

As shown in \cite{gas/leu87,has/leu}, the partition function can be
expanded in inverse powers of the size of the box
\begin{displaymath}
L \doteq V^{1/d}
\end{displaymath}
In our consideration of the free energy, $L^{-1}$ takes the role of the
energy in soft scattering amplitudes.

We consider a small magnetic field of the magnitude
\begin{equation}
\label{H}
H \doteq \vert \vec H \vert = O(V^{-1})
\end{equation}
so the leading term of the symmetry breaking part of the action is of
order $O(1)$.

It has been shown that for any $d>2$,
the leading order in the large volume expansion of the partition function
only involves the two coupling constants $\Sigma $ and $F$, called
``magnetization'' and ``pion decay constant'', respectively. (The latter
denotation can be understood by noting that $F^{2}$ is the residue of the
GB pole in the current correlation function at zero external field.)

Statistical physics sometimes introduces a ``helicity modulus'' $\Upsilon $
defined as the increase of the free energy when the external field is
slowly rotated: $\Delta f = \frac{1}{2} \Upsilon \alpha^{2}$, ($\alpha
\doteq $ rotation angle/distance). It turns out that $\Upsilon $
coincides with $F^{2}$ \cite{has/leu}.

The remaining
terms in eqs (\ref{Lsy2}) and (\ref{Lsb2}), which involve additional
derivatives or higher powers of the magnetic field, only show up at
non-leading order. Moreover, the large volume expansion can be worked
out perturbatively: to a given order in the expansion in powers of
$1/L$, only graphs involving a limited number of loops contribute.
In the present work, we extend the results of \cite{gas/leu87,has/leu}
by considering the large volume expansion of the partition function
up to and including terms of order $(L^{2-d})^{3}$ \
\footnote{Actually we expand in powers of the small dimensionless
quantity $L^{2-d}/F^{2}$. This is what we really mean when --
for brevity -- we just count the powers of $L^{2-d}$.}. As we will see, this
requires a perturbative evaluation to three loops (the loop propagator
fixes the ordering in magnitudes of $L^{2-d}$). Higher dimensions require
more coupling constants from the second order on, e.g. the terms in
(\ref{Lsb2}) with coefficients \ $h_{2,0}^{(i)}$ \
contribute to the action as $L^{-d}$,
so they are classified differently for $d=3$ and $d=4$.

The GB mass is given by $m^{2} = \frac{\Sigma H}{F^{2}}$,
\footnote{This is not the case for $d=2$, where the GB mass exhibits a gap
at $H \to 0$.} such that
$mL << 1$ , i.e. the GB modes feel the boundary conditions strongly.
\footnote{$ m << 1/L $ -- that corresponds to rule (\ref{H}) and $d>2$ --
characterizes the so-called ``$\epsilon $-expansion'', in contrast to the
``chiral expansion'' where $mL=O(1)$.}
On the other hand this must not be true for other mass scales as they
are given here e.g. by the mass of the $\sigma $-particle
(or the $\rho $-particle in QCD):
there $L$ is much larger than the Compton wave length.

We note that for small $H$, $\Sigma $ and $F$ control the finite
size effects, the GB mass and the correlation functions.

In the limit of purely spontaneous symmetry breaking, $ H \to 0$,
the model contains zero modes.
They correspond to space-independent fields, $\vec S (x) = $ const. ,
for which only the symmetry breaking part of the action is different
from zero -- for weak magnetic fields the action reduces to
$\Sigma \vec H \vec S V$. In the region $ H = O(V^{-1})$
we are studying here, the direction of the vector $\vec S $ does therefore
not strongly favor the direction of $\vec H $; in the absence of a
magnetic field, all directions become equally likely. The standard
perturbative expansion of the model -- where the field $\vec S $
is expanded around the direction of the magnetic field -- is not
applicable here. As pointed out in \cite{gas/leu87}, the problem can
be solved by using collective variables. The general field configuration
is represented as
\begin{equation}
\label{rot}
\vec S (x) = \Omega ^{-1} \vec \pi (x) \doteq
\Omega^{-1} (\pi^{0}, \pif ) \quad ,
\quad \vec H = (H,{\underline 0})
\end{equation}
where $\Omega \in O(N)$ is a global rotation associated with the
zero modes (or quasi zero modes for $H > 0 $)
while the vector $\vec \pi (x)$ describes the non-zero
modes. The condition
\begin{equation}
\label{fluc}
\int_{V} \pi^{i}(x) dx = 0 \ ; \qquad i=1, \dots ,N-1
\end{equation}
insures that this vector fluctuates around the direction $(1,{\underline 0})$,
such that $\pi^{0}$ can be expanded as
\footnote{In the literature the component $\pi^{0}$ is often called
$\sigma $, naming the model.}
\begin{displaymath}
\pi^{0} = 1-\frac{1}{2} \pif^{2} -\sum_{k=2}^{\infty} \frac{(2k-3)!!}
{2^{k}k!} (\pif^{2})^{k}
\end{displaymath}
Thus for $H \to 0$ the expectation values are:
\begin{displaymath}
<\pi^{0} > = 1  \qquad , \qquad < \pif > = {\underline 0}
\end{displaymath}
i.e. the $N-1$ vector field $\pif $ represents (small) transversal
excitations of $\vec S$ around its longitudinal component $\pi^{0}$
(parallel to $\vec H $).

It turns out that also the leading term of ${\cal L}^{(sy)}$
-- the kinetic term $\frac{F^{2}}{2} \partial_{\mu} \pif \partial_{\mu}
\pif $ -- contributes to the action in order $O(1)$
(see \cite{has/leu}), so the power counting rule (\ref{H}) is completed by
\begin{equation} \label{crule}
\partial_{\mu} \propto L^{-1} \ , \quad
\pi^{i}(x) \propto L^{1-d/2} \ ; \quad i=1, \dots ,N-1
\end{equation}
Inserting the decomposition (\ref{rot}) in the action and using the rules
(\ref{H}), (\ref{crule}), we obtain a series of the form
\begin{displaymath}
S = \int_{V} {\cal L} dx = \sum_{\ell =0}^{\infty } \{ S_{\ell}+
\tilde S_{\ell}(H) \}
\end{displaymath}
where $S_{\ell}$ and $\tilde S_{\ell}$ stem from ${\cal L}^{(sy)}$
and ${\cal L}^{(sb)}$, respectively, and the sum
$S_{\ell} + \tilde S_{\ell}(H)$
collects all contributions of order $L^{(2-d)\ell }$.
Accordingly, the partition function becomes
\begin{displaymath}
{\rm Z} = {\bf N} \int [d \vec S ] e^{-\{S_{0}+\tilde S_{0}(H)\}}
\left[ 1- \sum_{\ell \geq 1} (S_{\ell} +\tilde S_{\ell}(H)) + \frac{1}{2}
\{ \sum_{\ell \geq 1} (S_{\ell} +\tilde S_{\ell}(H)) \}^{2} \dots \right]
\end{displaymath}
In this work,
we are interested only in the $H$-dependence of Z, so we absorb
in the normalization constant ${\bf N}$ an overall
$H$-independent factor. Therefore, to calculate the partition function
to $O((L^{2-d})^{3})$, we need to evaluate $S_{0},S_{1},S_{2}$ and
$\tilde S_{0}(H) \dots \tilde S_{3}(H) $.

First we simplify ${\cal L}^{(sy)}$ given by (\ref{Lsy2}):
the transformation
\begin{equation} \label{feldtr1}
\vec S \to \frac{\vec S + \alpha F^{-4/(d-2)} \partial^{2}\vec S}
{\vert \vec S + \alpha F^{-4/(d-2)} \partial^{2}\vec S \vert }
\end{equation}
does not change the
form of the Lagrangian but only the coupling constants \cite{has/leu}. This
allows us to choose the (dimensionless) parameter $\alpha $ in such
a manner that $g^{(1)}_{4}$ vanishes. Then only the terms with
$F^{2},g_{4}^{(2)},g_{4}^{(3)}$ and $g_{6}^{(1)}$ contribute
to the order we want to consider.

If we apply the counting rule (\ref{H}) in ${\cal L}^{(sb)}$,
\vspace*{1mm}
only the terms with the coupling
\vspace*{1mm}
constants $\Sigma, h_{2,0}^{(1)},
h_{2,0}^{(2)}, h_{1,2}^{(1)}, h_{1,2}^{(2)}$ and $h_{1,4}^{(3)}$
remain relevant in eq. (\ref{Lsb2}).

The transformation
\begin{equation} \label{feldtr2}
\vec S \to \frac{\vec S + \beta \Sigma F^{-2d/(d-2)} \vec H}{\vert \vec S +
\beta \Sigma F^{-2d/(d-2)} \vec H \vert}
\end{equation}
enables us to put also $h_{1,2}^{(2)} = 0$ \
without disturbing $g_{4}^{(1)} = 0$, since the latter is
independent of $\vec H$.
\footnote{Actually the elimination of $h_{1,2}^{(2)}$ is motivated only
by the possibility of a space dependent magnetic field $\vec H (x)$.
In our case where $\vec H = const.$ this term does not contribute
to the action and transformation (\ref{feldtr2}) can be used to make
$h_{2,0}^{(1)}$ or $h_{2,0}^{(2)}$ vanish. We will recall this
remark when counting the degrees of freedom of the counter terms
associated with the non-leading coupling constants in section 7
and appendix B.} \\
In addition with the transformation
\begin{equation} \label{feldtr3}
\vec S \to \frac{\vec S + \lambda \Sigma F^{-2(d+2)/(d-2)}
(\vec H \vec S ) \partial^{2} \vec S }{\vert \vec S +
\lambda \Sigma F^{-2(d+2)/(d-2)} (\vec H \vec S ) \partial^{2} \vec S
\vert }
\end{equation}
we also succeed in making $h_{1,4}^{(3)}$ vanish. Else it
would occur for $d=4$. There, this last transformation
manipulates $\vec S$ only in the $3^{rd}$ order, so it maintains the
coupling constants up to the $2^{nd}$ order, in particular \
$g_{4}^{(1)} = h_{1,2}^{(2)} = 0 $ .

All the three transformations performed above do not change $\vec S$ in the
leading order, so the leading couplings $F^{2}$ and
$ \Sigma $ remain untouched. The rest of the coupling constants,
which correspond to our special choice of $\alpha, \beta$ and $\lambda$,
we denote by $g_{4}^{(2)}{'} $ etc. In appendix B we give the explicit
transformation formula (\ref{lagtrans}), which prove that these
simplifications are possible.

Now \ $S = \int {\cal L} (\vec \pi , \Omega \vec H ) dx $ \ shall be
expanded:
\footnote{The region of spatial integration is always $V$, unless we indicate
something else.}
\footnote{Throughout this work $\sim ,\simeq ,\cong $ mean:
to an accuracy of $1^{st},2^{nd},3^{rd}$ order in $(L^{2-d})$, respectively.}
\begin{eqnarray*}
S \cong \int \Big[ \frac{F^{2}}{2} (\partial_{\mu} {\underline \pi}
\partial_{\mu} {\underline \pi} + \partial_{\mu} \pi^{0} \partial_{\mu}
\pi^{0}) -\Sigma H \Omega ^{00} \pi^{0} \qquad \qquad \qquad \qquad
\qquad \\
+ h_{1,2}^{(1)}{'}
H ( \Omega^{00} \pi^{0}+ \Omega^{0i} \pi^{i}) (\partial_{\mu}
{\underline \pi} \partial _{\mu} {\underline \pi} + \partial _{\mu} \pi ^{0}
\partial _{\mu} \pi ^{0} )
- h_{2,0}^{(1)}{'} H^{2} (\Omega^{00} \pi ^{0}+ \Omega^{0i}\pi^{i} )^{2} \\
-h_{2,0}^{(2)}{'} H^{2}
+ \frac{1}{4} g_{4}^{(2)}{'}(\partial_{\mu} {\underline \pi} \partial_{\mu}
{\underline \pi} )^{2} + \frac{1}{4} g_{4}^{(3)}{'}
(\partial_{\mu} {\underline \pi}
\partial _{\nu} {\underline \pi} )^{2}
+\frac{1}{2} g_{6}^{(1)}{'} ( \partial_{\mu} \partial^{2} \pif
\partial_{\mu} \partial^{2} \pif ) \Big] dx
\end{eqnarray*}

If we insert $\pi ^{0}$, omit once more the $O(N)$ symmetrical
terms of $3^{rd}$ order and choose dimensionless coupling constants
$k_{1} \dots k_{6}$, the expansion of the action becomes:

\begin{displaymath}
\begin{array}{lclr}
S_{0} & = & \frac{F^{2}}{2} \int \partial _{\mu} {\underline \pi} \partial
_{\mu} {\underline \pi} dx - \Sigma HV \Omega ^{00} & \\
& & & \\
S_{1} & = & \frac{F^{2}}{2} \int ({\underline \pi} \partial _{\mu} {\underline
\pi} )^{2} dx + \frac{\Sigma H \Omega ^{00}}{2} \int \pif ^{2} dx & \\
& & & \\
S_{2} & = & \frac{F^{2}}{2} \int {\underline \pi} ^{2} ({\underline \pi}
\partial _{\mu} {\underline \pi} )^{2} dx + \frac{\Sigma H \Omega ^{00}}{8}
\int ({\underline \pi} ^{2} )^{2} dx & d=3 \\
& & & \\
S_{2} & = & \frac{F^{2}}{2} \int {\underline \pi} ^{2} ({\underline \pi}
\partial _{\mu} {\underline \pi} )^{2} dx + \frac{\Sigma H \Omega ^{00}}{8}
\int ({\underline \pi} ^{2} )^{2} dx & \\
&&& \\
 & & + \frac{\Sigma k_{1} H \Omega^{00}}{F^{2}}
\int \partial _{\mu} {\underline \pi} \partial _{\mu} {\underline \pi} dx
- \frac{\Sigma ^{2} H^{2} V [ k_{2} (\Omega ^{00} )^{2}+k_{3} ] }{F^{4}}
& \\ &&& \\ && +
\frac{1}{4} \int [k_{4} (\partial _{\mu} {\underline \pi} \partial_{\mu}
{\underline \pi} )^{2} + k_{5} (\partial _{\mu} {\underline \pi} \partial
_{\nu} {\underline \pi} )^{2} + 2 \frac{k_{6}}{F^{2}}
(\partial_{\mu}\partial^{2} \pif )^{2}] dx & d=4 \\
& & & \\
S_{3}(H) & = & \frac{\Sigma H \Omega ^{00}}{16} \int ({\underline \pi}^{2}
)^{3} dx + \frac{\Sigma k_{1} H \Omega ^{00}}{F^{4}} \int \partial _{\mu}
{\underline \pi} \partial _{\mu} {\underline \pi} dx - \frac{\Sigma ^{2}
H^{2} V}{F^{6}} \left[ k_{2} ( \Omega^{00})^{2} + k_{3}) \right]
& d=3 \\
& & & \\
S_{3}(H) & = & \frac{\Sigma H \Omega ^{00}}{16} \int ({\underline \pi}^{2})^{3}
dx + \frac{\Sigma k_{1} H \Omega ^{00}}{F^{2}} \Big[ \int ({\underline \pi}
\partial _{\mu} {\underline \pi} )^{2} dx -\frac{1}{2} \int \pif^{2}
(\partial_{\mu} \pif )^{2} dx \Big] & \\
& & & \\
&& + \frac{\Sigma^{2} H^{2} k_{2}} {F^{4}}
\Big[ (\Omega^{00})^{2}\int {\underline \pi}^{2} dx
- \Omega^{0i} \Omega^{0k} \int \pi^{i} \pi^{k} dx \Big] & d=4
\end{array}
\end{displaymath}
\vspace*{5mm}

Concerning the sign flip of $k_{1}$, we follow the convention
given in \cite{has/leu}. We define the non-leading coupling constants,
however, consequently dimensionless, like the transformation parameters
$\alpha,\ \beta,\ \lambda $ before.
This is achieved by multiplying with the suitable power
of $F$, which is the only dimension-carrying coupling constant.

In addition the magnetic field is always accompanied by the
(dimensionless) constant $\Sigma $. \\

Actually for \ $d=4$ \ there is also one term of order
 \ $5/2 $ , namely:
\begin{displaymath}
\frac{\Sigma H \Omega^{0i}}{F^{2}} k_{1} \int \pi^{i}
(\partial_{\mu} \pif )^{2} dx \quad .
\end{displaymath}
But from the contraction rules it is obvious that this term will not
contribute to our three loop result. So we omit it, although
this is not justified in general: if we wanted to calculate
to five loops we would have to include this term, since there its square in
the exponential expansion {\em does} contribute.

\section{The measure}

In the presence of derivative couplings, the step from the classical
Lagrangian to the quantum theory is not straightforward.
In the framework of the canonical quantization procedure,
the problem manifests itself, e.g., in the fact that the
interaction Hamiltonian does not coincide with the negative
interaction Lagrangian.

In the path integral formulation of quantum theory, the issue
concerns the measure, i.e. the volume element in the space of
field configurations over which we are to integrate.
In particular the finite spatial volume we consider here causes
non-trivial contributions.

In the following, we construct the measure by means of the
method introduced by Polyakov in his analysis of the path
integrals for bosonic strings \cite{poly}. Polyakov introduces
a metric on the space of classical field configurations and defines
the measure as the volume element induced by this metric.

In our case, where the classical configurations are characterized
by the field $\vec S (x)$, the metric is a quadratic form
involving the difference $d\vec S $ between two neighboring
configurations,
\begin{displaymath}
ds^{2} = \int \int dx dy \sum_{i,k} K_{i,k}(x,y; \vec S, \vec H )
 \ dS_{i}(x)dS_{k}(y)
\end{displaymath}
The most important requirement to be imposed on the metric is
locality: the support of the kernel $K_{i,k} $
must be concentrated at $x=y$. In addition, the metric must
respect the symmetries of the Lagrangian -- Euclidean invariance
as well as the invariance under global $O(N)$ rotations in the
isospin space.

The ansatz
\begin{equation} \label{met}
ds^{2} \doteq \frac{1}{V} \int  d\vec S(x) d\vec S (x) dx
\end{equation}
certainly satisfies these requirements
\footnote{The factor $1/V$ is unimportant here, i.e. a question of
normalization. Polyakov did not include such a factor; he considers a
{\em curved} space where it would have violated locality.}
, but by no means represents the general form of a local metric.
In particular, locality allows also derivatives to occur.

To find the maximal generalization allowed by the symmetry requirements,
we perform a derivative expansion and include the magnetic field,
as we did for the Lagrangian. This leads to the form:
\begin{eqnarray}
ds^{2}_{g} &=& \frac{1}{V} \int \Big\{ (d\vec S )^{2}
+ \frac{a_{1}}{F^{4/(d-2)}} (\partial_{\mu} d\vec S )^{2} +
\frac{a_{2}}{F^{4/(d-2)}} (d\vec S )^{2}
(\partial_{\mu}\vec S )^{2} \label{Massallg} \\
 && + \frac{a_{3}}{F^{8/(d-2)}} (\partial ^{2} d\vec S )^{2}
 + \dots  \nonumber \\
 & & + b_{1} \frac{\Sigma}{F^{2d/(d-2)}} (d\vec S )^{2}(\vec H \vec S)
 + b_{2}\frac{\Sigma}{F^{2(d+2)/(d-2)}} (\partial_{\mu}
 d \vec S )^{2}(\vec H \vec S)
 \dots \Big\} dx \nonumber
\end{eqnarray}
where we introduce new coupling constants for the non-leading
contributions.
The explicit discussion of this general measure
is given in appendix B. It turns out that all
its non-leading terms only yield physically irrelevant
power divergences,
i.e. they do not contribute to the dimensionally regularized
-- more generally: not to the renormalized -- free energy.

In the same appendix we also give the
transformation rules of the couplings in the measure
under the field transformations (\ref{feldtr1}) - (\ref{feldtr3}).
They show in particular that
the form (\ref{met})
does not permit such substitutions: starting from the measure
(\ref{met}), they cause non-trivial
changes of all the coupling constants in (\ref{Massallg}) (which are,
however, not relevant in dimensional regularization, since they only
affect power divergences).
At last we demonstrate the invariance of the entire partition
function to 3 loops -- including all possible couplings in the Lagrangian
and in the measure as well as the leading power divergences --
under those transformations with arbitrary coefficients.

It looks as if the quantization of theories including derivative
couplings contain a considerable number of degrees of freedom, here
represented by the unspecified constants $a_{i}, \ b_{i}$.
The essential point of appendix B is, however, that the freedom
of choice for
the coupling constants in the measure does not mean an
increased number of degrees of freedom of the system.

So it is justified to consider in the main part of this work
the simplified measure (\ref{met}), which yields relevant
contributions, as we will see. In this section we
give its evaluation in terms
of the collective variables introduced in section 2.

Without consideration of a magnetic field, the set of functions $\{ \vec S (x)
\}$ is parame-trized by the direction of the mean magnetization
\begin{displaymath}
\vec M = \frac{1}{V} \int_{V} \vec S (x) dx
\end{displaymath}
-- which plays the role of the collective variable introduced in section 2 --
and by a set of coordinates associated with the remaining degrees of
freedom, which has to be regularized.

The collective variable may be identified with a subset of the rotation
group, namely the rotations in the planes containing a fixed vector, say
$\vec e = (1,{\underline 0})$. There is a rotation $\Omega $ of this subset
that takes the direction of $\vec M $ into $\vec e$ ,
\begin{equation} \label{m}
\vec m \doteq \frac{\vec M }{\vert \vec M \vert } = \Omega^{T} \vec e
\end{equation}
This is the rotation needed to represent the general element $\vec S (x)$
of our space of functions by eq. (\ref{rot}).

The non-zero modes we parametrize by a decomposition into a complete
orthonormal system of periodic functions $u_{n}^{i}(x)$ :
\begin{equation} \label{zerleg}
\pi^{i}(x) = \Sigma_{n}{'} u_{n}^{i}(x) q^{n}
\end{equation}
where $n=(k, \bar n )$ runs over the flavours $k=1 \dots N-1$ and over
the modes $\bar n =(\bar n_{1}, \dots , \bar n_{d}), \quad \bar n_{\mu}
\in \ganz $.
The form $u_{n}^{i}(x) = \delta^{i}_{k} u_{\bar n}(x)$ insures
the remaining $O(N-1)$ symmetry and the prime
indicates that the zero mode is excluded, in accordance with eq.
(\ref{fluc}). The mode functions obey
\begin{eqnarray}
\frac{1}{V} \int u_{n}^{i~*}(x) u_{m}^{i} (x) dx & =& \delta_{n m} \\
\frac{1}{V} \Sigma_{nm}' u_{n}^{i}(x)u_{m}^{k}(y) &=& \delta^{ik}
\left( \delta (x-y)-\frac{1}{V} \right) \label{del}
\end{eqnarray}
where $1/V$ has to be subtracted in eq. (\ref{del}) because of the
missing zero mode.
{}From relation (\ref{crule}) we see that the coefficients $q^{n}$
are of order $O(L^{1-d/2})$.\\

In terms of the variables $\Omega ,\ \vec \pi (x)$ the metric takes the form:
\begin{displaymath}
ds^{2} = \frac{1}{V} \int_{V} [d\vec \pi (x) - \omega \vec \pi (x)]^{2} dx
\qquad (\omega = d\Omega \cdot \Omega^{T} = -\omega^{T} = {\rm ~inf.~rotation
{}~matrix).}
\end{displaymath}
Inserting the decomposition (\ref{zerleg}) this becomes:
\begin{eqnarray*}
ds^{2} & = & g_{mn} \delta q^{m} \delta q^{n} + 2g_{m i}
\delta q^{m} \omega^{0i} + g_{ik} \omega^{0i} \omega^{0k}\\
{\rm where} \quad & & \delta q^{n} = dq^{n} - \frac{1}{V} \omega^{ik} \int
u^{i}_{n}(x) \pi^{k}(x) dx\\
& &  g_{m n} = \delta_{m n} + \frac{1}{V} \int \frac{u^{i}_{m}(x)
u^{k}_{n}(x) \pi^{i}(x) \pi^{k}(x)}{[\pi^{0}(x)]^{2}} dx\\
& & g_{n i} = \frac{1}{V} \int u^{k}_{n}(x) \left( \delta_{ik} \pi^{0}(x) +
\frac{\pi^{i}(x) \pi^{k}(x)}{\pi^{0}(x)} \right) dx\\
& & g_{ik} = \delta_{ik} + \frac{1}{V} \int [\pi^{i}(x) \pi^{k}(x) -
\delta_{ik} \pif^{2}(x)] dx
\end{eqnarray*}
(with summation over double indices).
Accordingly, the regularized volume element is given by $[d\vec S] =
\sqrt{{\bf g}} \ [d\pi ]' \Pi^{N-1}_{i=1} \omega^{0i}$,
where $ [d\pi ]' \doteq \prod_{n}' dq^{n} $ and
\begin{displaymath}
{\bf g} = det \left( \begin{array}{ccc} g_{m n} & & g_{m i}\\ g_{k n}
& & g_{ki}\\
\end{array} \right) \quad . \end{displaymath}

Note that the volume element only involves the components $\omega^{0i}$
of $\omega$. \vspace*{1mm}
{}From definition (\ref{m}) we get  $ \quad
d\vec m ^{.} d\vec m = \Sigma^{N-1}_{i=1} (\omega^{0i})^{2}$. This shows
that their product is the volume element on the unit sphere:
$\Pi^{N-1}_{i=1} \omega^{0i} = d\mu (\vec m )$.

The determinant ${\bf g}$ can be evaluated perturbatively.
$\sqrt{{\bf g}}$ is an $H$-independent factor, so
we need it only to the $2^{nd}$ order, i.e.
to $(q^{n})^{4}$.
\footnote{In order to be very precise we had to say that actually in
$0^{th}$ order the integral over $[d\pi ]'$ does not even occur,
so ${\bf g}$=1 is
already the first order and the numeration of the orders is shifted
by one. But such a notation would cause considerable confusion.}

We first write it in the form \begin{displaymath} {\bf g} =
det (1 + \varepsilon )
\qquad \varepsilon = \left( \begin{array}{ccc} \varepsilon_{m n} & &
\varepsilon_{m i}\\ \varepsilon_{k n} & & \varepsilon_{ki} \end{array} \right)
\quad  \doteq (\varepsilon_{ab}) \end{displaymath}
where all the matrix elements \ $\varepsilon_{ab} = g_{ab} - \delta_{ab}$
\ have the magnitude $L^{2-d}$. Hence:
\begin{displaymath}
\sqrt{{\bf g}} = (e^{tr \ ln (1+\varepsilon )})^{1/2}
= e^{\frac{1}{2}tr \varepsilon - \frac{1}{4} tr \varepsilon^{2}}
+ O((L^{2-d})^{3})
\end{displaymath}
This is the form of \ $\sqrt{{\bf g}}$ \ we need in order
to add \ $-ln \sqrt{{\bf g}}$ \ to the action.\\

Gaussian integration yields the contraction rule
\begin{equation}
< \pi^{i}(x) \pi^{k}(y) > = \delta^{ik} \frac{1}{F^{2}} G (x-y) \ ,
\end{equation}
where $G (x-y)$ is the propagator, which satisfies
\begin{equation}
-\partial^{2} G (x) = \delta (x) - \frac{1}{V} \ .
\end{equation}
The measure can be expressed in terms of $G $.
Hence {\em all} the contributions to the partition
function Z are determined by the propagator $G $ and its derivatives.

The most obvious choice for the mode functions $u_{n}^{i}(x)$ is plane waves
(Fourier decomposition). We are going to refer to this decomposition in
the forthcoming when we deal with momenta $p_{n}: \ \ p_{n} u_{n}^{k} =
-i \nabla u_{n}^{k}$. This implies the form
\begin{equation}
G(x) = \frac{1}{V} \sum_{\bar n}{'} \frac{e^{ip_{\bar n}x}}{p_{\bar n}^{2}}
\end{equation}
(again the prime at the summation sigma indicates that we omit the zero mode).
The choice of the complete orthonormal
system is unimportant and we will often refer to
plane waves only for convenience.

The traces in $ln \sqrt {\bf g}$, however, contain UV divergences -- e.g.
the undefined term $\delta (0)$ -- so the system needs regularization.

Many regularization schemes are known.
They all violate important
physical properties -- if this could be avoided, regularization would
not be needed. Since the violated properties vary from one regularization
scheme to an other, the question of their equivalence in the final limit
is highly non-trivial.

Our model is very sensitive for the type of regularization;
e.g. a sharp cutoff in momentum space or the Pauli-Villars
regularization turn out to be unsuitable, see subappendix A.1.

A better possibility is {\em dimensional regularization}
\cite{dimreg}. It has been used in \cite{ge/leu} and
\cite{has/leu}, and also in this work we apply it.

Its essential peculiarity can be demonstrated if we decompose the two-point
Green function into the limit at infinite volume
and a volume-dependent correction:
\begin{equation} \label{propzerleg}
G(x-y) = G(x-y)\vert_{V \to \infty} + g(x-y) \ .
\end{equation}
The first term is a distribution that takes in Fourier decomposition
the form \vspace{1mm} \\ \vspace{1mm} $ \frac{1}{(2\pi )^{d}}
\int d^{d}p \frac{e^{ip(x-y)}}{p^{2}}$. Regularization reduces
it to a function $G^{\Lambda}(x-y)$,
whose Laplacian is a regularized $\delta $-function :
\begin{equation} \label{deltadelta}
\delta^{\Lambda}(x-y) \doteq -\partial^{2}G^{\Lambda}(x-y)
\end{equation}
At \ $x=y$ \ this is in general a function of the regularization parameter
that diverges when the parameter is removed, whereas in the
particular case of dimensional regularization it simply vanishes.
This happens to all the singularities, which -- in case of a momentum cutoff
-- contain powers of the cutoff (such as $\gl (0) $ for $d>2$);
there remain, however, logarithmic singularities (like $\gl (0) $ for $d=2$).
For a discussion of dimensional regularization on compact manifolds,
see \cite{dimreg2}.

Whenever we deal with the singularity structureref{propzerleg}) of the
propagator
and refer to dimensional regularization.
For technical reasons, however, we are going to
consider a more general construction which keeps track also of
the leading power divergence of each term
occurring in other regularization schemes.
We will see that the essential properties of the result still
hold for this generalized singularity structure:
we show in appendix A that
perturbative renormalization still works
and in appendix B we verify the invariance
of the entire partition function
under the field transformations (\ref{feldtr1}) $\dots $
(\ref{feldtr3}).
As far as the power divergences are concerned, we consider only
those with the highest possible power of the UV cutoff (that we denote
by $\Lambda $, see appendix A).
In divergences of the second order, e.g., we only include the term
$\propto \Lambda^{2(d-2)}$ and drop the one $\propto \Lambda^{d-2}$
which is sensitive to the used regularization. Of course the
logarithmic divergences that also occur in dimensional regularization
are not discarded.

As a consequence, throughout the main part of this work we assume $\dlf (x) $
to act under the integral like an exact  $\delta $-function.
This means an interchange of limits,
which is justified for dimensional regularization, but which would
be dangerous in other regularizations (the results without of this
assumption are discussed in subappendix A.1).

So we will permit ourselves the luxury of including power divergences
such as $\dl $;
they will reveal some aspects hidden by dimensional
regularization, in particular it provides us with
very significant consistency tests.\\

Using the relation,
\begin{equation} \label{delta}
\frac{1}{V} \sum_{n}{'}u_{n}^{i}(x) u_{n}^{k}(y) \vert_{reg.}
= \delta^{ik} \left( \dlf (x-y) - \frac{1}{V} \right) \ ,
\end{equation}
a lengthy calculation leads to:
\begin{eqnarray} \label{polymass}
ln \sqrt{{\bf g}}  & \simeq  & \frac{V\delta^{\Lambda}(0)-N+1}{2V}
\int \pif^{2}(x) dx + \frac{2V\delta^{\Lambda}(0)-N+1}{8V}
\int ( \pif^{2}(x))^{2} dx  \nonumber\\ \label{determ}
& & - \frac{N-1}{8} \left( \frac{1}{V} \int \pif^{2}(x) dx \right) ^{2}
\end{eqnarray}

As a consistency test we compare our result (\ref{polymass}) to the one of
\cite{has/leu} (section 6), where -- also in the framework of
dimensional regularization -- the Faddeev-Popov method yields
\footnote{We discuss here the analogue to the Polyakov measure (\ref{met}).
Of course a generalization with non-leading terms is possible for
the Faddeev-Popov measure too.} :
\begin{displaymath} {\rm Z} = {\bf N_{0}} \int d\mu (\vec m )
\int [d\pi ]' \vert \vec M \vert ^{N-1} e^{-S} \quad . \end{displaymath}
So $\vert \vec M \vert
^{N-1}$ should be the measure for $\delta^{\Lambda}(0) = 0$.
With $ \vert \vec M \vert = \frac{1}{V} \int \pi^{0} dx$ we get:
\begin{eqnarray}
\label{FPmd}
\vert \vec M \vert ^{N-1}
&=& e^{(N-1) \ ln ( \frac{1}{V} \int \sqrt{1-\pif^{2}} dx)} \\
& \simeq & exp \left( (N-1)\Big[ - \frac{1}{2V} \int
\pif^{2}dx - \frac{1}{8} \{ \frac{1}{V} \int \pif^{2}
dx \}^{2} - \frac{1}{8V} \int \{ \pif^{2} \}^{2} dx \Big] \right) \nonumber
\end{eqnarray}
This is indeed the $\dl $-independent part of $\sqrt{{\bf g}}$,
so we can affirm the consistency and write:
\begin{displaymath}
\sqrt{{\bf g}} \ [d\pi ]' = [dq] \vert \vec M \vert ^{N-1} \ ;
\quad [dq] \simeq
e^{ \frac{1}{4} \delta ^{\Lambda} (0) [2 \int  \pif ^{2} dx +
\int (\pif^{2} )^{2} dx]} [d\pi ]' .
\end{displaymath}

In contrast to the Faddeev-Popov method, the integration over the
collective variable $\Omega$ here only extends over a subset of the
rotation group. This difference is inessential, however, for the following
reason. Perform the change of variables $ \pi ^{i} (x) \to R^{ik} \pi ^{k}
(x)$, where $R$ is an element of the little group belonging to $ \vec e
= (1,{\underline 0})$, and replace Z by an average over the little group.
Both, the magnetization and the measure $[dq]$ are invariant under this
transformation. Since every $\Omega \in O(N)$ can be decomposed as
$\Omega = R \Omega _{\vec m}$ (where $\Omega_{\vec m}$ belongs to the subset
specified above), an integral over all elements of the little group followed
by an integral over all directions $ \vec m $ amounts to an integral of the
full group. Finally, exploiting also $O(N)$ - invariance of the action:
$S (\Omega ^{T} \vec \pi , \vec H ) = S (\vec \pi , \Omega \vec H)$
we arrive at
\begin{equation}
\label{partfun}
{\rm Z} = {\bf N} \int d\Omega \int [d\pi ]' \sqrt{{\bf g}} \
e^{- \int {\cal L} (\vec \pi , \Omega \vec H ) dx}
 \end{equation}
where $d\Omega $ is the Haar measure on $O(N)$.

The procedure of \cite{has/leu} was first applied in \cite{fadpop} where
Hasenfratz finds for lattice regularization the measure
\begin{displaymath}
[d\pi ]' \prod_{i=1}^{N-1} \delta \Big( \sum_{x} \pi_{x}^{i}\Big) \
exp \Big( -\sum_{x} ln \pi^{0}_{x} \Big) \ exp \Big( (N-1)
ln \sum_{x} \pi^{0}_{x} \Big)
\end{displaymath}
($\sum_{x} $: sum over lattice points).

In the continuum limit, the last factor becomes $const. \cdot
\vert \vec M \vert ^{N-1}$. The factor $exp(-\sum_{x} ln \pi_{x}^{0})$
has been omitted in \cite{has/leu} because for dimensional regularization
the exponent vanishes.
In the present case, the leading power divergence corresponds to
\begin{equation} \label{latticedelta}
\dl = \frac{1}{a^{d}} \qquad \qquad \qquad
(a \doteq  {\rm lattice~ constant})
\end{equation}
so this factor becomes:
\begin{displaymath}
exp\left( -\dl \int ln \pi^{0} dx\right) = exp \left( \frac{1}{2} \dl \
\left[ \int \pif^{2} + \frac{1}{2} \int (\pif ^{2})^{2}dx \dots \right] \right)
= \frac{[dq]}{[d \pi]'}
\end{displaymath}

The only difference between the lattice regularization in \cite{fadpop}
and the regularized expansion in a complete orthonormal system
of periodic functions used here is, as far
as the functional measure is concerned, the representation for the
regularized volume element associated with the non-zero modes.\\

We have seen explicitly to the second order that the measures of
Polyakov and Faddeev-Popov coincide.

An important motivation for testing the Polyakov method is its
applicability to a larger class of symmetry groups than it is the case
for the FP measure. In particular it can be applied to every
symmetry breaking $SU(N) \times SU(N) \to SU(N) $,
in contrast to the FP method; for those symmetry breakings
it is not known how one could
apply the FP method if $N>2$ (as we mentioned in section 1, the case
$N=3$ is of interest for chiral QCD with three flavors).
\footnote{That the transition to the unitary groups is not
straightforward can be seen from the fact that for
$SU(N) \times SU(N)$ there is not such a simple invariant
as $\vert \vec S \vert $ in the case of $O(N)$, but there is a set
of invariants, which is not easy to handle.}

\section{1-Loop calculation of the partition function}
For the 1-loop calculation of the partition function
we have to consider:\\
$\sqrt{\bfg }$ to the $0^{th}$ order ($=1$), $S_{0}, \ \tilde S_{0}(H)$
and $\tilde S_{1}(H)$.
Inserting this in (\ref{partfun}) we get:
\footnote{Here, we clearly recognize the GB mass \
$m^{2} = \frac{\Sigma H}{F^{2}}$ .}
\begin{displaymath}
{\rm Z} \approx {\bf N} \int d\Omega \int [d \pi ]'
e^{-\frac{F^{2}}{2} \int \partial _{\mu} {\underline \pi} \partial
_{\mu} {\underline \pi} dx + \gamma \Omega ^{00} -\frac{\gamma
\Omega ^{00}}{2V} \int {\underline \pi } ^{2} dx }
\end{displaymath}
where we defined:
\begin{displaymath}
\gamma \doteq \Sigma HV \propto O(1) \quad .
\end{displaymath}
The evaluation of the generating functional is more convenient
than the study of individual Green functions \cite{gas/leu84}.
$d=3$ and $d=4$ needn't be distinguished yet.

With the substitution $q_{1}^{n} \doteq  \sqrt{\frac{F^{2}}{2}V
p_{n}^{2} + \frac{\gamma \Omega^{00}}{2}} \ q^{n}
\propto O(1) $ ,
the second integral only contributes to the normalization constant, hence
\begin{displaymath}
{\rm Z} \approx {\bf N}\int d\Omega e^{\gamma \Omega^{00}}
\int [d\pi ]' e^{-\pif_{1}^{2}}
= {\bf N_{1}} \int d\Omega e^{\gamma \Omega^{00}} \left( \prod_{n}{'} \Big[
\frac{F^{2}}{2}Vp_{n}^{2} + \frac{\gamma \Omega^{00}}{2} \Big]
\right)^{-\frac{1}{2}}
\end{displaymath}

Now we have to expand the Jacobian: $\frac{F^{2}}{2}V(p_{n})^{2} +
\frac{\gamma \Omega^{00}}{2} \doteq \frac{1}{2} L^{d-2} (\alpha_{n} +
\frac{\beta}{L^{d-2}}) \ $, where $\alpha_{n}, \beta \propto
O(1) , \quad \alpha_{n} $ being independent of $H$.
\begin{displaymath}
\prod_{n}{'} \Big\{ \frac{1}{2} L^{d-2} (\alpha_{n} + \frac{\beta }{L^{d-2}})
\Big\} = {\bf N'} \Big[ 1+ \frac{\beta}{L^{d-2}} \Sigma_{n}'
\frac{1}{\alpha_{n}} + O((L^{2-d})^{2}) \Big]
\end{displaymath}
Inserting this, we obtain:
\begin{equation} \label{einloop}
{\rm Z} \approx {\bf N_{2}} \int d\Omega e^{\gamma \Omega^{00} (1-\frac{1}
{2F^{2}} \frac{1}{V} \sum_{n}' [1/p_{n}^{2}])} =
{\bf N_{3}} Y_{N} \left( \gamma \Big[ 1- \frac{N-1}{2F^{2}} G_{1} \Big]
\right)
\end{equation}
The last step is an identity
of the `modified Bessel function'
\footnote{The identity is $\int d\Omega e^{x\Omega^{00}} \equiv \Gamma
(\frac{N}{2}) Y_{N}(x) $ for $\Omega \in O(N) $. The modified Bessel function
has the expansion $Y_{N}(x) = \sum_{k=0}^{\infty} \frac{1}{k! \Gamma (k+
\frac{N}{2})} \bigl( \frac{x}{2} \bigr) ^{2k} \ $, which is not
oscillating, in contrast to the common Bessel function. }
, obeying
\begin{equation} \label{diffY}
Y''_{N}(x) + \frac{N-1}{x}Y'_{N}(x) - Y_{N}(x) \equiv 0 \quad ,
\end{equation}
and $G_{1} \doteq G(0) = \frac{1}{V} \sum_{\bar n}'
\frac{1}{(p_{\bar n})^{2}} $\ .
We expand this definition to $\ G_{k} \doteq \frac{\Gamma (k)}{V}
\sum_{\bar n}' \frac{1}{(p_{\bar n}^{2})^{k}}$ , see figure 1.\\
\begin{figure}[hbt]
 \centerline{\ \psfig{figure=F1.ps,height=2cm}\ }
 \caption{Diagrams representing the terms: \ a) $G_{1}$ \ \ \ b) $G_{2}$
 \ \ \ c) $G_{3}$ \ \ \ etc. }
\end{figure}

Eq. (\ref{einloop}) is the result Hasenfratz and Leutwyler
found with a different method \cite{has/leu}. But the
`Jacobian method' demonstrated here is not applicable
beyond the 1-loop level.
So for the 3-loop calculation we will follow the
straightforward procedure (expansion of the exponential, Wick contractions).

\section{The partition function to 3 loops in 3 dimensions}
We first consider {\bf $d=3$}. If we insert the results of section 2
and 3, the partition function takes the form :
\begin{eqnarray}
{\rm Z} & \cong & {\bf N} \ \int d\Omega \int \sqrt{\bfg } \ [d \pi ]'
e^{-S_{0} -S_{1} -S_{2} -S_{3}(H)} \nonumber \\
& \cong & {\bf N} e^{\frac{\gamma^{2}}{F^{6}V} k_{3}} \int d\Omega
e^{\gamma \Omega^{00} +\frac{\gamma^{2}}{F^{6}V} k_{2} (\Omega^{00})^{2}}
\int [d\pif ]' e^{- \frac{1}{2} F^{2} \int
(\partial_{\mu}\pif )^{2} dx } \ \cdot  \nonumber \\
& & \{1 - \frac{F^{2}}{2} \int ({\underline \pi} \partial_{\mu} {\underline
\pi})^{2} dx - \frac{\beta}{2V} \int {\underline \pi}^{2} dx \nonumber \\
& & + \frac{F^{4}}{8} \bigl( \int ({\underline \pi} \partial_{\mu}
{\underline \pi})^{2} dx \bigr) ^{2} + \frac{\beta^{2} -N+1}{8V^{2}}
\bigl( \int {\underline \pi}^{2} dx \bigr) ^{2} +\frac{F^{2} \beta}{4V}
\bigl( \int ({\underline \pi} \partial_{\mu} {\underline \pi})^{2}dx
\bigr) \bigl( \int {\underline \pi}^{2} dy \bigr) \nonumber \\
& & -\frac{F^{2}}{2} \int {\underline \pi}^{2} ({\underline \pi} \partial_{\mu}
{\underline \pi})^{2}dx - \frac{\rho }{8V} \int ({\underline \pi}^{2})^{2}
dx \nonumber \\
& & -\frac{F^{4}\gamma \Omega^{00}}{16V} \bigl( \int ({\underline \pi}
\partial_{\mu} {\underline \pi})^{2} dx \bigr) ^{2} \bigl( \int {\underline
\pi}^{2}dy \bigr) - \frac{F^{2} \beta^{2}}{16V^{2}} \bigl( \int ({\underline
\pi} \partial_{\mu} {\underline \pi})^{2}dx \bigr) \bigl( \int {\underline
\pi}^{2}dy \bigr) ^{2} \nonumber \\
&& -\frac{\beta(\beta^{2} -3(N-1))}{48 V^{3}} \bigl(
\int {\underline \pi}^{2} dx \bigr) ^{3}
+\frac{F^{2} \gamma \Omega^{00}}{16 V} \bigl( \int ({\underline \pi}
\partial_{\mu} {\underline \pi})^{2}dx \bigr) \bigl( \int ({\underline \pi}
^{2})^{2} dy \bigr) \nonumber \\
&& + \frac{F^{2}\gamma \Omega^{00}}{4 V} \bigl( \int
{\underline \pi}^{2} dx \bigr) \bigl( \int {\underline \pi}^{2} (
{\underline \pi} \partial_{\mu} {\underline \pi})^{2} dy \bigr)
+\frac{\beta \rho }{16 V^{2}} \bigl( \int {\underline \pi}^{2} dx \bigr)
\bigl( \int ({\underline \pi}^{2})^{2} dy \bigr) \nonumber \\
&& - \frac{\gamma \Omega^{00}}
{16V} \int ({\underline \pi}^{2})^{3} dx - \frac{\gamma \Omega^{00} k_{1}}
{F^{4}V} \int \partial_{\mu} {\underline \pi} \partial_{\mu} {\underline
\pi} dx \} \label{lang}
\end{eqnarray}
where
\begin{equation}
\label{def}
\beta \doteq \gamma \Omega^{00} +N-1-V\delta^{\Lambda}(0) \qquad {\rm and}
\qquad \rho \doteq \beta -V\delta^{\Lambda}(0) \quad .
\end{equation}
In the expansion $\{ \dots
\}$ only the $\gamma-$ (i.e. $H-$) dependent contributions to the $3^{rd}$
order have to be included. The rest has been omitted, rsp. absorbed by
the normalization constant. Accordingly for the coefficients $\beta^{2},
\quad \beta (\beta^{2}-3(N-1))$ and $\beta \rho$ in the $3^{rd}$ order
-- that is: in the last four lines -- only the $\gamma $-dependent part
needs to be included.

We denote: $ < \dots > \doteq \frac{\int [d\pi ]' e^{-
\frac{1}{2} F^{2} \int (\partial_{\mu} \pif )^{2} dx } (~ \dots ~)}
{\int [d\pi ]' e^{- \frac{1}{2} F^{2} \int (\partial_{\mu}
\pif )^{2} dx }} \ $ \ .

With $< 1 > = 1$ there remain 15 terms to be evaluated. To this end we use:
\begin{eqnarray*}
< \frac{1}{V} \int \pif \cdot \pif \ dx > & = &  \frac{N-1}{F^{2}} G_{1}\\
< \int \partial_{\mu} \pif \partial_{\mu} \pif \ dx > & = &- \frac{N-1}{F^{2}}
V \partial^{2} G(0) = \frac{N-1}{F^{2}} \ \Big( V \dl -1 \Big)
\end{eqnarray*}

Together with the contraction rules this enables us to
calculate the terms in eq. (\ref{lang}).
We repeat that
throughout these calculations (i.e. throughout section 5 and also
section 6) $\delta^{\Lambda}(x)$ is treated like an exact $\delta $-function
under the integrals.
For the moment the difference between the dimensionally regularized
system we actually refer to and the more general singularity
structure we also want
to consider manifests itself only in the presence of the term $\dl $.

{}From the contraction rules we obtain:
\begin{equation}
\label{fosi}
1)\qquad <-\frac{F^{2}}{2} \int (\pi^{i} \partial_{\mu} \pi^{i} )
(\pi^{k} \partial_{\mu}  \pi^{k}) dx > \ = \
-\frac{N-1}{2F^{2}} (V\dl -1)G_{1} \qquad \qquad
\end{equation}
\begin{equation}
2)\qquad < -\frac{\beta }{2V} \int {\underline \pi}^{2} dx > =
-\frac{\beta }{2F^{2}} (N-1) G_{1} \qquad \qquad \qquad \qquad
\qquad \qquad \quad
\end{equation}
Here we recognize the 1-loop result again.
\begin{eqnarray}
3) & & <\frac{F^{4}}{8} \Big( \int ({\underline \pi} \partial_{\mu}
{\underline \pi})({\underline \pi} \partial_{\mu} {\underline \pi} ) dx
\Big) \Big( \int ({\underline \pi} \partial_{\nu} {\underline \pi})
({\underline \pi} \partial_{\nu} {\underline \pi}) dy \Big) >
\doteq (N-1) \cdot {\cal I} \qquad \qquad \qquad \label{Idef}
\end{eqnarray}
We leave it like this because -- as we will see --
we don't need to know ${\cal I}$ explicitly.
We just mention that it contains the term
\begin{equation} \label{j2}
J_{2} \doteq \int (\partial_{\mu} G \partial_{\mu} G)^{2}dx \quad ,
\end{equation}
which can not be expressed in terms of $G$-functions at $x=0$.
\footnote{For completeness we give the result nevertheless. The
calculation yields: \\
${\cal I}= \frac{1}{8F^{4}} \Big[ \{ (N-1)(V\dl )^{2} + 8V\dl -N-9
\} G_{1}^{2}
+ \{ 2(V\dl )^{2} -4V \dl +4 \} \frac{1}{V}G_{2}
+ 2(N-2) VJ_{2} \Big] $ }
\begin{eqnarray}
4) & & < \frac{\beta^{2}-N+1}{8V^{2}} \bigl( \int {\underline \pi}^{2}
dx \bigr) ^{2} > =
\frac{\beta^{2} -N+1}{8F^{4}} (N-1) \Big[ (N-1)G_{1}^{2} + \frac{2}{V} G_{2}
\Big]  \qquad \quad \\
5) & & \frac{F^{2}\beta }{4V} < [\int (\pi^{i}\partial_{\mu} \pi^{i})
(\pi^{j}\partial_{\mu} \pi^{j}) dx] \ [\int \pi^{k} \pi^{k}dy] > \nonumber
\end{eqnarray}
There are two types of pairing that contribute: we can either
contract the two derivated fields -- this situation is similar
to 4) -- or contract each of them
with one $\pi ^{k}(y)$. The sum of these two contributions is :
\begin{eqnarray}
&=& \frac{\beta }{4F^{4}} (V\delta^{\Lambda}(0)-1) (N-1) \Big[ (N-1)
G_{1}^{2} + \frac{2}{V} G_{2} \Big]  + \frac{\beta }{2F^{4}} (N-1)
G_{1}^{2} \qquad \qquad \\
6) & & -\frac{F^{2}}{2} < \int (\pi^{i} \pi^{i})( \pi^{j} \partial_{\mu}
\pi^{j})( \pi^{k}\partial_{\mu} \pi^{k}) dx >  \nonumber\\
& = & -\frac{V\dl -1}{2} < (\delta_{ii}\delta_{jj}+2\delta_{ij}) \int (
\pi^{i}\pi^{i})(\pi^{j}\pi^{j}) dx> \nonumber\\
& = & -\frac{V\delta^{\Lambda}(0)-1}
{2F^{4}} (N^{2}-1) G^{2}_{1} \\
7) & & -\frac{\rho}{8V} < \int (\pi^{i} \pi^{i})( \pi^{k} \pi^{k})dx> =
 -\frac{\rho }{8F^{4}} (N^{2}-1) G^{2}_{1}
\end{eqnarray}
Now we have finished the first and second order. (A term of order
$\ell $ can easily be recognized by the factor $F^{-2\ell }$.)
Except for $G_{1},\ G_{2}$ that were already represented by diagrams in
fig. 1, there occurs \ $G_{1}^{2}$ , see fig. 2.
\begin{figure}[hbt]
 \centerline{\ \psfig{figure=F2.ps,height=2cm}\ }
 \caption{Diagrams to: \ a) $G_{1}^{2}$ \ \ \ b) $G_{1}^{3}$ \ \ \ etc. }
\end{figure}
\begin{figure}[hbt]
 \centerline{\ \psfig{figure=F3.ps,height=2cm}\ }
 \caption{Graph for the term $J_{2} $.\ Here and in the following diagrams,
  greek letters denote partial derivatives. }
\end{figure}
$J_{2}$, which is included in ${\cal I}$ is represented in fig. 3.

The remaining 8 terms are of third order and require more effort. We show
the treatment of the most difficult one that includes all the steps we used
for the third order. Then we satisfy ourselves for the rest by
quoting the result of each term.
\begin{displaymath}
8) \qquad
-\frac{F^{4}\gamma \Omega^{00}}{16V} < \int (\pi^{i}\partial_{\mu}
\pi^{i})(\pi^{j}\partial_{\mu}\pi^{j})dx \int (\pi^{k} \partial_{\nu}
\pi^{k})(\pi^{\ell}\partial_{\nu}\pi^{\ell})dy \int (\pi^{w}\pi^{w})dz > \qquad
\end{displaymath}
We can reduce the effort remarkably by inserting the ${\cal I}$
of definition (\ref{Idef}).
\begin{equation} \label{not}
= -{\cal I} \cdot \frac{(N-1)^{2}}{2F^{2}} G_{1} - \frac{F^{4}\gamma
 \Omega^{00}}
{16V} <[(a{\bf b})(c{\bf d})][(e{\rm f})(g{\rm h})][(r\vert s)]>
\end{equation}
in an obvious notation: $a\dots s$: momenta, [ ]: same space-point, ( )
same flavor, bold/roman: $ \partial_{\mu},\partial_{\nu},\quad
 \vert $ : pairing $r=s$ is excluded .\\
Thus $r,s$ have to be paired with $a \dots h$. For this there are 7
equivalence classes:\\
\begin{displaymath}
\begin{array}{lllllll}
i) & a=r, \ c=s \quad {\rm (class~with}& 4{\rm ~variants)} & \qquad
&v)& a=r, \ e=s \qquad  & (8) \\
ii) & a=r, \ b=s & (8) & & vi)& a=r, \ f=s & (16) \\
iii) & a=r, \ d=s & (8) & & vii)& b=r, \ f=s & (8) \\
iv)& b=r, \ d=s & (4) & & & &
\end{array}
\end{displaymath}
$i)$ We further distinguish 3 subclasses and get:
\begin{eqnarray}
-\frac{\gamma \Omega^{00}(N-1)}{4F^{6}V} \Big[ (N-1)(V\delta^{\Lambda}(0))^{2}
+ 2(V\delta^{\Lambda}(0)-1) + 2 \int \partial_{\mu \nu}G \partial_{\mu \nu}
G du \Big] G_{1} G_{2} = \nonumber\\
 -\frac{\gamma \Omega^{00}(N-1)}{4F^{6}V}
\Big[ (N-1)(V\delta^{\Lambda}(0))^{2}-2(N-3)V\delta^{\Lambda}(0)+N-5 \Big]
 G_{1}G_{2} \qquad
\end{eqnarray}
$ii)$ and $iii)$ include the factor \ $ \partial_{\mu}G_{2} = 0$ .\\
$iv)$ Again there are 3 subclasses.
Using $\ \int \partial_{\mu}G \partial_{\mu} G dx = G_{1} $ we find:
\begin{equation}
-\frac{\gamma \Omega^{00}(N-1)}{4F^{6}} \left( (N-1)(V\delta^{\Lambda}(0)-1)
G^{3}_{1} + 2(V\delta^{\Lambda}(0)-1) \frac{G_{1}G_{2}}{V} + 2G^{3}_{1}\right)
\end{equation}
In the remaining classes, $r$ and $s$ are connected in a mixed way to the
$x-$ and $y-$block, i.e. the factor $\int G(x-z) G(y-z) dz$ occurs:
\begin{displaymath}
\frac{1}{V^{2}} \int \left( \Sigma_{\bar n }'
\frac{e^{ip_{\bar n }(x-z)}}{p_{\bar n }^{2}}
\right) \left( \Sigma_{\bar m}'\frac{e^{ip_{\bar m}(y-z)}}{p_{\bar m}^{2}}
\right)
dz = \frac{1}{V} \Sigma_{\bar n}'
\frac{e^{ip_{\bar n}(x-y)}}{(p_{\bar n}^{2})^{2}}
\doteq \dot G (x-y)
\end{displaymath}
Explication of the notation: $G(x) = \sum_{\bar n}'
\frac{e^{ip_{\bar n}x}}{m^{2}+
p_{\bar n}^{2}} \vert_{m=0} \quad ; \qquad \dot G(x) \doteq -\frac{d}{dm^{2}}
G(x) \quad \Rightarrow \quad -\partial^{2}\dot G = G ,\quad \dot G(0) = G_{2}
, \quad \ddot G(0) = G_{3}\ $ etc. (see appendix C). \\
As in $i)$ we denote $u=x-y$. Then in $v) \dots vii)$ there occur integrals
over 4 $G$-functions of $u$, whereby one dot and four partial derivatives
(two by $\mu$ and two by $\nu$) are distributed in all possible ways.
So we have to deal with the terms:
\begin{equation} \label{gammas}
\begin{array}{ll}
\Gamma_{0} \doteq \int G (\partial_{\mu \nu} G)^{2} \dot G du &
\Gamma_{1} \doteq \int \partial_{\mu} G \partial_{\nu} G \partial_{\mu \nu}
G \dot G  du \\
\Gamma_{2} \doteq \int G \partial_{\nu}G \partial_{\mu \nu}G \partial_{\mu}
\dot G du \qquad &
\Gamma_{3} \doteq \int \partial_{\mu}G \partial_{\nu}G \partial_{\nu}G
\partial_{\mu} \dot G du \\
\Gamma_{4} \doteq \int G^{2} \partial_{\mu \nu}G \partial_{\mu \nu}\dot G du &
\Gamma_{5} \doteq \int G \partial_{\mu}G \partial_{\nu}G
\partial_{\mu \nu}\dot G du \\
\end{array}
\end{equation}
With partial integrations all these quantities can be expressed
in terms of one of them; we choose $\Gamma_{3} \ (= \dot J_{2} /4 )$.
\begin{equation} \label{gam3}
\begin{array}{lll}
\Gamma_{0} & = & -\frac{1}{6} G_{1}^{3} + (\delta^{\Lambda}(0)-\frac{1}{2V})
G_{1}G_{2} - \frac{1}{4V^{2}}G_{3} - \frac{1}{12V} J_{3} + \Gamma_{3} \\
\Gamma_{1} & = & \frac{1}{2} (-\frac{1}{V}G_{1}G_{2} + \frac{1}{2V^{2}}G_{3}
+ \frac{1}{2V} J_{3} + \Gamma_{3}) \\
\Gamma_{2} & = & \frac{1}{6} (G^{3}_{1} - \frac{1}{V}J_{3}) -
\frac{1}{2}\Gamma_{3} \\
\Gamma_{4} & = & \frac{1}{3} (2G^{3}_{1} + \frac{1}{V}J_{3}) + \Gamma_{3} \\
\Gamma_{5} & = & -\frac{1}{6} (G^{3}_{1}
+\frac{2}{V}J_{3})-\frac{1}{2}\Gamma_{3}
\end{array}
\end{equation}
where $\quad J_{3} \doteq \int G^{3}(u) du $ .

\begin{figure}
 \centerline{\ \psfig{figure=F4.ps,height=2cm}\ }
 \caption{Diagrams for : \ a) $J_{3}$ \ \ \ b) $\Gamma_{3}$ }
\end{figure}
Thus we can express everything through $G_{1},G_{2},G_{3}, \ J_{3}$ and
$\Gamma_{3}$, represented in figures 1, 2 and 4.
The occurrence of precisely these terms in the $3^{rd}$
order is consistent with the massive expansion in \cite{ge/leu},
as we show in appendix C.\\

We continue to decompose the equivalence classes in subclasses. Each of
them contributes a summand to:
\begin{displaymath}
\begin{array}{rl}
v)& -\frac{\gamma \Omega^{00}}{2F^{6}} (N-1)[(V\delta^{\Lambda}(0)-1)^{2}
\frac{G_{3}}{2V^{2}} + N\Gamma_{0} + (N+2)\Gamma_{1} ] \\
vi)& -\frac{\gamma \Omega^{00}}{F^{6}} (N-1) [(V\delta^{\Lambda}(0)-1)
\frac{1}{V}
G_{1}G_{2} + \frac{N+2}{6} (G^{3}_{1}-\frac{1}{V}J_{3}) + (\frac{N}{2}-1)
\Gamma_{3})] \\
vii)& -\frac{\gamma \Omega^{00}}{2F^{6}} (N-1) [G^{3}_{1} + N\Gamma_{4} +
(N+2)\Gamma_{5}]
\end{array}
\end{displaymath}
If we insert the identities of the $\Gamma $'s and add up we arrive at
\begin{eqnarray}
&& - \frac{F^{4}\gamma \Omega^{00}}{16} <[(a{\bf b})(c{\bf d})]
[(e{\rm f})(g{\rm h})][(r\vert s)]> =
-\frac{\gamma \Omega^{00}}{4F^{6}} (N-1) \ \cdot
\qquad \qquad \qquad \qquad \qquad  \nonumber\\
&& \Big\{ \Big[(N-1)V\delta^{\Lambda}(0) +
\frac{N}{3} + \frac{17}{3}\Big] G_{1}^{3} +
\Big[ (N-1)(V\delta^{\Lambda}(0))^{2}
+ 12V\delta^{\Lambda}(0)-N-13\Big] \frac{1}{V} G_{1}G_{2}  \nonumber\\
&& + \Big[ (V\delta^{\Lambda}(0))^{2} - 2V\delta^{\Lambda}(0) + 2 \Big]
\frac{1}{V^{2}}G_{3} - \frac{N+5}{3V} J_{3} + 4(N-2) \Gamma_{3} \Big\}
\end{eqnarray}
$9) \dots 15)$ can be treated in the same way following Wick's
theorem: determination of the
classes that avoid summation over linear momenta, decomposition of each
class in subclasses corresponding to the possibilities to pair the
remaining fields. No further term is as lengthy as 8), and exept for 11)
the results include only propagators at 0. We write them down such
that each line stems from one equivalence class.
\begin{eqnarray*}
9)& & -\frac{\beta^{2}}{16F^{6}} (V\delta^{\Lambda}(0)-1)(N-1)
\Big[ (N-1)^{2}G^{3}_{1} + 6(N-1)\frac{1}{V}G_{1}G_{2} +
 \frac{4}{V^{2}}G_{3} \Big] \\
 & & - \frac{\beta^{2}}{4F^{2}} (N-1) \Big[ (N-1)G^{3}_{1} +
  \frac{2}{V} G_{1}G_{2} \Big] \\
 & & - \frac{\beta^{2}(N-1)}{2F^{6}V} G_{1}G_{2} \\
 & &{\rm (There~is~a~fourth~class~but~its~contribution~vanishes.)} \\
10) & & -\frac{\beta (\beta^{2}-3(N-1))}{48V^{3}} (N-1)
\Big[ (N-1)^{2} G^{3}_{1} + (N-1)\frac{6}{V} G_{1}G_{2}
 + \frac{4}{V^{2}} G_{3} \Big] \\
11) & & \frac{\gamma \Omega^{00}}{16F^{6}} (V\delta^{\Lambda}(0)-1)
(N^{2}-1) \Big[ (N-1)G^{3}_{1} + \frac{4}{V}G_{1}G_{2} \Big] \\
 & & + \frac{\gamma \Omega^{00}}{12F^{6}} (N^{2}-1)
 \Big[ 5G_{1}^{3}-\frac{2}{V} J_{3} \Big] \\
&& {\rm For~the~latter~we~used:} \int G^{2} \partial_{\mu} G \partial_{\mu} G
du = \frac{1}{3} (G^{3}_{1}-\frac{1}{V}J_{3}) \\
12) & & \frac{\gamma \Omega^{00}}{4F^{6}} (V\delta^{\Lambda}(0)-1)
(N^{2}-1) \Big[ (N-1)G^{3}_{1} + \frac{4}{V} G_{1}G_{2} \Big] \\
 & & + \frac{\gamma \Omega^{00}}{2F^{6}} (N^{2}-1) G^{3}_{1} \\
13) & & \frac{\beta \rho}{16F^{6}} (N^{2}-1) \Big[ (N-1)G^{3}_{1} +
\frac{4}{V}G_{1}G_{2} \Big] \\
&& 11 \ i), \  12 \ i),\ {\rm and}\ 13) \quad
 {\rm are~essentially~based~on~the~same~calculation.}\\
14) & & -\frac{\gamma \Omega^{00}}{16F^{6}} (N^{2}-1)(N+3) G^{3}_{1} \\
15) & & -\frac{\gamma \Omega^{00}k_{1}}{F^{6}V} (N-1)(V\delta^{\Lambda}(0)
-1) \\
\end{eqnarray*}
Having completed the evaluation, we add everything up and the factor
$ <\{ \dots \} > $ in eq. (\ref{lang}) takes the form:
\begin{displaymath}
<\{ \dots \} >  =  1 + (N-1)\Big[ \mu_{1} + \mu_{2} + \gamma \Omega^{00}
(\nu_{1}+\nu_{2}+\nu_{3}) + (\gamma \Omega^{00})^{2} (\rho_{2}+\rho_{3})
+ (\gamma \Omega^{00})^{3} \sigma_{3} \Big] \\
\end{displaymath}
where the indices correspond to the order of magnitude.
If we lift this up to the exponent we get
\begin{eqnarray*}
<\{ \dots \} > & \cong & const. \cdot exp\{ (N-1)[(\alpha_{1}+\alpha_{2}
+\alpha_{3})\gamma \Omega^{00} + (\beta_{2}+\beta_{3})(\gamma \Omega^{00})^{2}
+ \gamma_{3}(\gamma \Omega^{00})^{3} ] \} \\
&&{\rm where~the~}const.{\rm ~is~} H {\rm -independent~and} \\
\alpha_{1}  =  \nu_{1},&& \alpha_{2} = \nu_{2}-(N-1)\mu_{1}\nu_{1},
\quad \alpha_{3} = \nu_{3} - (N-1)(\nu_{1}\mu_{2}+\mu_{1}\nu_{2}) +
(N-1)^{2} \mu_{1}^{2}\nu_{1} \\
\beta_{2} & = & \rho_{2} - \frac{N-1}{2} \nu_{1}^{2}, \quad
\beta_{3} = \rho_{3} - (N-1)(\mu_{1}\rho_{2}+\nu_{1}\nu_{2}) + (N-1)^{2}
\mu_{1}\nu_{1}^{2} \\
\gamma_{3} & = & \sigma_{3} -(N-1)\nu_{1}\rho_{2} + \frac{(N-1)^{2}}{3}
\nu_{1}^{3}
\end{eqnarray*}
The most complicated contributions, $\mu_{2}$ and $\nu_{3}$ only occur
in $\alpha_{3}$. There the ugly term ${\cal I}$ cancels, so it
was justified not to insert its evaluation.

Now we actually have a representation of Z, but we prefer to have
the same form as in section 4, i.e. we want to transform the integral
on the unit sphere in the iso-space, \
$ \int d\Omega e^{\gamma \Omega^{00} + \frac{\gamma^{2}}{F^{6}V}
k_{2} (\Omega^{00})^{2}} exp\{ \dots \} \ $, to the form
$\ e^{\delta_{1}' + \delta_{2}' + \delta_{3}'} \int d\Omega e^{\gamma
\Omega^{00} (1+\varepsilon_{1}'+\varepsilon_{2}'+\varepsilon_{3}')} $.

This can be realized making use of the differential equation (\ref{diffY}),
as we outline in appendix D. The result is
\begin{displaymath}
\begin{array}{l}
\delta_{1}' = 0 , \quad \delta_{2}' = (N-1)\beta_{2}, \quad \delta_{3}' =
(N-1)(\beta_{3}-(N-1)\gamma_{3})+\frac{k_{2}}{F^{6}} \\
\varepsilon_{1}' = (N-1)\alpha_{1} , \quad \varepsilon_{2}' = (N-1)
(\alpha_{2}-(N-1)\beta_{2}) \\
\varepsilon_{3}' = (N-1)[\alpha_{3}-(N-1)\beta_{3}+(\gamma^{2}+N(N-1))
\gamma_{3} + (N-1)^{2}\alpha_{1}\beta_{2} - \frac{k_{2}}{F^{6}} ]
\end{array}
\end{displaymath}\\

Inserting everything we arrive at the final result:
\begin{eqnarray}
\hline
\label{Z3}
{\rm Z} & \cong & {\bf N}\ e^{\frac{\gamma^{2}}{F^{4}}
 (\delta_{2}+\frac{1}{F^{2}}
\delta_{3})}\quad Y_{N} \left( \gamma \Big[ 1+\frac{\varepsilon_{1}}{F^{2}}+
\frac{\varepsilon_{2}}{F^{4}}+\frac{\varepsilon_{3}}{F^{6}} \Big] \right) \\
\hline
\label{E1}
\varepsilon_{1} & = & -\frac{N-1}{2} G_{1} \\
\delta_{2} & = & \frac{N-1}{4V} G_{2} \\
\varepsilon_{2} & = & \frac{(N-1)(N-3)}{8} \left( -G_{1}^{2} +
\frac{2}{V} G_{2} \right) \\
\delta_{3} & = & \frac{N-1}{4} \left( \frac{N-3}{V} G_{1}G_{2} -
\frac{2N-5}{3V^{2}} G_{3} \right) + \frac{k_{2}+k_{3}}{V} \\
\label{E3}
\varepsilon_{3} & = & (N-1) \Big\{ \frac{(N-3)(3N-7)}{48} (- G_{1}^{3} +
\frac{6}{V} G_{1}G_{2} )
 - ((N-3)(N-4)+\gamma^{2})\frac{G_{3}}{12V^{2}} \nonumber\\
 & & - \frac{N-3}{12V}J_{3} - (N-2)\Gamma_{3}
- \frac{(V\dl -1)k_{1} + k_{2}}{V} \Big\}
\end{eqnarray}
\vspace*{5mm}

It is not surprising that all the
contributions are proportional to the number of flavors, $N-1$, except
for the terms with $k_{2}$ and $k_{3}$, which do not stem from a
coupling of ${\underline \pi }$-fields.
On the other hand we notice
a repeated appearance of the factors $(N-3)$ and $(3N-7)$ that can only
be interpreted in the context of renormalization.
Even more striking is that there are many terms occurring in the
course of the calculation that cancel at the end. Already in the first
order the ($H$-independent) $\dl $-contributions of the measure and
the action just compensate each other. Further examples for such terms are
in $\delta_{2}: NG_{1}^{2}$ and $G_{1}^{2}$. We write
$ \quad \delta_{2} : \quad (N,1)G_{1}^{2} \quad $ etc.

\begin{displaymath}
\begin{array}{ll}
\delta_{2}: & (N,1)G_{1}^{2} \qquad \varepsilon_{2}:\quad
(N,1)V\dl G^{2}_{1},\quad (N,1)\dl G_{2},\quad (N^{2},N,1)VJ_{2}\\
\delta_{3}: & (N^{2},N,1)V\dl G_{1}^{3},\quad (N^{3},N^{2},N,1)G_{1}^{3},
\quad (N,1)\dl G_{1}G_{2}, \quad N^{3} G_{1}G_{2}/V \\
\varepsilon_{3}: & (N^{2},N,1)(V\dl )^{2}G_{1}^{3},\quad (N^{2},N,1)V\dl
G_{1}^{3}, \quad (N,1)(\dl )^{2} V G_{1}G_{2} \\
 & (N^{2},N,1)\dl G_{1}G_{2}, \quad (N,1)\dl G_{3}/V,\quad
(N,1)VJ_{2}G_{1}
\end{array}
\end{displaymath}
This long list of canceled terms exhibits a remarkable property
of the system. The vast part of it would have been ignored
if we would have restricted ourselves to the terms that really appear in
dimensional regularization.
We will see that all these
cancellations of measure and Lagrangian are strictly required
by the perturbative renormalizability of the structure that
includes the leading power divergences.

\section{3-Loop expansion of the partition function for d=4}
The difference in the expansion of $d=3$ and $d=4$ is due to the
magnitude of the terms with the coupling constants $k_{1} \dots k_{6}$.
Generally, the terms with $k_{1} \dots k_{5}$
are all $\propto \frac{1}{V} $, which means of $3^{rd},2^{nd}$ order for
$d=3,4$ respectively, and the $k_{6}$-term is $\propto L^{-4}$, i.e.
of $4^{th},2^{nd}$ order for $d=3,4$.
Thus for $d=3$ only the $H$-dependent terms with
$k_{1},k_{2},k_{3}$ had to be taken into account. For $d=4$, however,
those terms contribute to the $2^{nd}$ and $3^{rd}$ order and additionally
contribute mixed terms with the first order in the exponential expansion.
The latter is also true for the $k_{4}\dots k_{6}$-terms, so we have
to include three coupling constants more than in section 5.

We write down the expansion of the partition function as in (\ref{lang}):
\begin{eqnarray}
{\rm Z} & \cong & {\bf N} e^{\frac{\gamma^{2}k_{3}}{F^{4}V}} \int d\Omega
e^{\gamma \Omega^{00} + \frac{(\gamma \Omega^{00})^{2} k_{2}}{F^{4}V} }
\int [d\pi ]' e^{-\frac{F^{2}}{2}\int (\partial_{\mu}\pif )^{2} dx } \cdot
\nonumber \\
& & \{ 1-\frac{\beta }{2V} \int \pif^{2}dx - \frac{F^{2}}{2} \int (\pif
\partial_{\mu} \pif)^{2} (1+\pif^{2})dx - \frac{\rho}{8V} \int (\pif^{2})^{2}
dx - \frac{N-1-\beta^{2}}{8V^{2}} \left( \int \pif^{2}dx \right) ^{2}
\nonumber \\
& & + \frac{F^{4}}{8} \left( \int (\pif \partial_{\mu} \pif )^{2} dx
\right) ^{2} - \frac{\gamma \Omega^{00} k_{1}}{F^{2}V} \int \partial_{\mu} \pif
\partial_{\mu} \pif dx + \frac{\beta F^{2}}{4V} \int \pif^{2}dx \int
(\pif \partial_{\mu} \pif )^{2} dy \nonumber \\
& & - \frac{1}{4} \int [k_{4}(\partial_{\mu} \pif \partial_{\mu} \pif
)^{2} + k_{5} (\partial_{\mu} \pif \partial_{\nu} \pif )^{2}
+k_{6}\frac{2}{F^{2}} (\partial_{\mu}\partial^{2} \pif )^{2}] dx \nonumber \\
& & - \frac{\beta^{3}}{48V^{3}} \left( \int \pif^{2}dx \right) ^{3} -
\frac{\beta^{2}F^{2}}{16V^{2}} \left( \int \pif^{2} dx \right) ^{2}
\int (\pif \partial_{\mu} \pif )^{2}dy - \frac{\beta F^{4}}{16V} \left(
\int \pif^{2}dx \right) \left( \int (\pif \partial_{\mu} \pif )^{2}dy
\right) ^{2} \nonumber \\
& & + \frac{\beta F^{2}}{4V} \int \pif^{2}dx
\int (\pif \partial_{\mu} \pif )^{2}
\pif ^{2}dy + \frac{\beta \rho}{16V^{2}} \int \pif^{2}dx \int (\pif^{2})^{2}dy
+ \frac{\beta (N-1)}{16V^{3}} \left( \int \pif^{2}dx \right) ^{3} \nonumber \\
& & + \frac{\beta \gamma \Omega^{00}k_{1}}{2F^{2}V^{2}} \int \pif^{2}dx
\int \partial_{\mu} \pif \partial_{\mu} \pif dy + \frac{\gamma \Omega^{00}}
{8V} \int \pif^{2}dx \int [k_{4}(\partial_{\mu} \pif \partial_{\mu} \pif )^{2}
+ k_{5} (\partial_{\mu} \pif \partial_{\nu} \pif )^{2} ]dy \nonumber \\
& & + \frac{F^{2}\rho }{16V} \int (\pif \partial_{\mu} \pif )^{2}dx \int
(\pif ^{2})^{2}dy + \frac{\gamma \Omega^{00} k_{1}}{2V}  \int (\pif
\partial_{\mu} \pif )^{2}dx \int ( \partial_{\nu} \pif )^{2} dy \nonumber \\
&& + \frac{\gamma \Omega^{00}k_{1}}{2F^{4}V^{2}} \int \pif^{2}
(\partial_{\mu} \pif )^{2}dx
- \frac{\gamma \Omega^{00}}{16V} \int (\pif^{2})^{3}dx - \frac{\gamma
\Omega^{00}k_{1}}{F^{2}V} \int (\pif \partial_{\mu} \pif )^{2}dx -
\frac{(\gamma \Omega^{00})^{2}k_{2}}{F^{4}V^{2}} \int \pif^{2}dx \nonumber \\
&& +\frac{\gamma^{2} k_{2}}{F^{4}V^{2}} \Omega^{0i} \Omega^{0k} \int
\pi^{i} \pi^{k} dx + \frac{\gamma \Omega^{00}k_{6}}{4VF^{2}}
\int \pif^{2}dx \int (\partial_{\mu} \partial^{2} \pif )^{2} dy \}
\end{eqnarray}
where $\beta $ and $ \rho $ keep the meaning given in
definition (\ref{def}).

Most of these terms have already been evaluated in section 5 (identically
or only with different coefficients). We just discuss the new ones.

First we define two new distributions :
\begin{equation} \label{strongsing}
D^{\Lambda}(x) \doteq -\partial^{2} \dlf (x) =
\partial^{2} \partial^{2} \gl (x) = \partial^{2}\partial^{2}G(x)
\quad {\rm and} \quad \Delta^{\Lambda}(x)
\doteq - \partial^{2} D^{\Lambda}(x) \quad .
\end{equation}
In the framework of our decomposition $G=\gl +g$, where $-\partial^{2}
G = \dlf -1/V$, we find $\partial^{2}g(x) \equiv 1/V$ and \ $D^{\Lambda}(x),
\ \Delta^{\Lambda}(x)$ \ are volume independent.

At $x=0$ they are pure power divergences, stronger than those
introduced before:
\begin{displaymath}
D^{\Lambda}(0) = \frac{1}{V} \sum_{\bar n} p_{\bar n}^{2} \quad , \qquad
\Delta^{\Lambda}(0) = \frac{1}{V} \sum_{\bar n} (p_{\bar n}^{2})^{2}
\end{displaymath}
They vanish in dimensional regularization but as in the case of
$\dl $ we do not omit them.

In addition we define:
\begin{displaymath}
G_{\mu \nu} \doteq \partial_{\mu} \partial_{\nu} G(x) \vert _{x=0}
\end{displaymath}
Then we find to the
\begin{displaymath}
\begin{array}{l}
2^{nd} {\rm ~order} \\
- \frac{k_{4}}{4} < \int (\partial_{\mu} \pif \partial_{\mu} \pif )(
\partial_{\nu} \pif \partial_{\nu} \pif )dx > = -\frac{(N-1)k_{4}}{4F^{4}V}
\Big[ (N-1)(V\dl -1)^{2} + 2V^{2}(G_{\mu \nu})^{2} \Big] \\
\quad \\
-\frac{k_{5}}{4} < \int \partial_{\mu} \pi^{i} \partial_{\nu} \pi^{i}
\partial_{\mu} \pi^{k} \partial_{\nu} \pi^{k} dx > =
-\frac{(N-1)k_{5}}{4F^{4}V} \Big[ (V\dl -1)^{2} + NV^{2} (G_{\mu \nu})^{2}
\Big] \\
\quad \\
- \frac{k_{6}}{2F^{2}} < \int \partial_{\mu} \partial^{2} \pif
\partial_{\mu} \partial^{2} \pif dx > =  k_{6} \frac{N-1}{2F^{2}} V
\Delta^{\Lambda}(0) \\
\quad \\
3^{rd} {\rm ~ order} \\
< \frac{\beta \gamma \Omega^{00} k_{1}}{2F^{2}V}  \int \pif^{2}dx \int
\partial_{\mu} \pif \partial_{\mu} \pif dy > = \frac{\beta \gamma
\Omega^{00} k_{1}}{2F^{6}V} (N-1) \Big[ (N-1)(V\dl -1) + 2 \Big] G_{1} \\
\quad \\
\frac{\gamma \Omega^{00}k_{4}}{8V} < \int \int \pi^{i}(x) \pi^{i}(x)
\partial_{\mu} \pi^{j}(y) \partial_{\mu} \pi^{j}(y) \partial_{\nu} \pi^{k}(y)
\partial_{\nu} \pi^{k}(y) dx dy > \\
\quad = \frac{\gamma \Omega^{00}k_{4}}{8F^{6}V} (N-1) \Big[ \{ (N-1)
 (V\dl -1)^{2}
+ 4(V\dl -1) + 2V^{2}(G_{\mu \nu})^{2} \}(N-1)G_{1}  \\
\qquad \qquad \qquad \qquad \qquad \qquad \qquad \qquad \qquad \qquad
\qquad \qquad \qquad \qquad \qquad \qquad
+ 8VG_{\mu \nu} \dot G_{\mu \nu} \Big] \\
{\rm where~we~have~used:}~\int \partial_{\mu}G\partial_{\nu}G du = \dot G_{\mu
\nu} \quad . \\
\end{array}
\end{displaymath}
\begin{displaymath}
\begin{array}{l}
\frac{\gamma \Omega^{00}k_{5}}{8V} < \int \int \pi^{i}(x) \pi^{i}(x)
\partial_{\mu}\pi^{j}(y) \partial_{\nu}\pi^{j}(y) \partial_{\mu}\pi^{k}(y)
\partial_{\nu}\pi^{k}(y) dx dy > \\
\quad = \frac{\gamma \Omega^{00}k_{5}}{8F^{6}V} (N-1) \Big[
\{ (N-1)(V\dl -1)^{2}
+ 4(V\dl -1) +N(N-1)V^{2}(G_{\mu \nu})^{2} \} G_{1} \\
\qquad \qquad \qquad \qquad \qquad \qquad \qquad \qquad \qquad \qquad
\qquad \qquad \qquad \qquad \qquad \qquad
+ 4NG_{\mu \nu} \dot G_{\mu \nu} \Big] \\
\frac{\gamma \Omega^{00}k_{1}}{2V} < \int (\pif \partial_{\mu} \pif)
(\pif \partial_{\mu} \pif )dx \int \partial_{\nu} \pif \partial_{\nu} \pif
dy > \\
\quad = \frac{\gamma \Omega^{00}k_{1}}{2F^{6}V} (N-1) \Big[ (N-1)(V\dl -1)^{2}
+ 4(V\dl -1) \Big] G_{1}
\end{array}
\end{displaymath}
where we inserted $\int \partial_{\mu \nu } G \partial_{\mu \nu } G du
= \dl -1/V $.
\begin{displaymath}
\begin{array}{l}
\frac{\gamma \Omega^{00}k_{1}}{2F^{2}V}<\int \pif^{2} (\partial_{\mu} \pif
\partial_{\mu} \pif ) dx > = \frac{\gamma \Omega^{00}k_{1}}{2F^{6}V}
(N-1)^{2}(V\dl -1) \  G_{1} \qquad \qquad \qquad \qquad \\
\quad \\
\frac{\gamma^{2} k_{2}}{F^{4} V^{2}} < \int [\Omega^{0i} \Omega^{ok}
\pi^{i} \pi^{k} -(\Omega^{00})^{2} \pif^{2} ] dx > =
\frac{\gamma^{2} k_{2}}{F^{6}V} [1-N(\Omega^{00})^{2}] \ G_{1} \\
\quad \\
\frac{\gamma \Omega^{00} k_{6}}{4VF^{2}} < \pif^{2}dx \int
\partial_{\mu} \partial^{2} \pif \partial_{\mu} \partial^{2} \pif dy > =
\frac{\gamma \Omega^{00} k_{6}}{4F^{6}} (N-1) \Big[ -(N-1)G_{1}V
\Delta^{\Lambda}(0) + 2 D^{\Lambda}(0) \Big]
\end{array}
\end{displaymath}

So we found in connection with the new coupling constants $k_{4}\dots k_{6}$
also new kinds of terms:
$(G_{\mu \nu})^{2}$ and $G_{\mu \nu} \dot G_{\mu \nu}$ -- see fig. 5 --
are associated with $k_{4},k_{5}$ and include a regular contribution.
Their occurrence is again consistent with \cite{ge/leu}, see appendix C.
\begin{figure}[hbt]
 \centerline{\ \psfig{figure=F5.ps,height=2cm}\ }
 \caption{Diagrams for \ $G_{\mu \nu} G_{\mu \nu}$ \ \ and \ \
 $G_{\mu \nu} \dot G_{\mu \nu}$}
\end{figure}

$k_{6}$ contributes
the remaining new terms $V\Delta^{\Lambda}(0), \ G_{1}V\Delta^{\Lambda}(0)$
and $D^{\Lambda}(0) $ that vanish in
dimensional regularization (i.e. they can not be found in \cite{ge/leu});
so the coupling constant $k_{6}$ is physically irrelevant.
(It can easily be seen that this is true to all orders of magnitude.)\\

But the new terms don't prevent us from following exactly the same procedure
as in section 5. We add all the summands of the integrand of Z, heave it
in the exponent and remove the parts with $(\Omega^{00})^{2}$ and
$(\Omega^{00})^{3}$ according to appendix D. Thus we arrive at the result:
\begin{eqnarray}
{\rm Z} & \cong & {\bf N} e^{\frac{\gamma^{2}}{F^{4}}(\delta_{2} +
\frac{1}{F^{2}}
\delta_{3})} \quad Y_{N}\left( \gamma [1 + \frac{\varepsilon_{1}}{F^{2}}
+ \frac{\varepsilon_{2}}{F^{4}} + \frac{\varepsilon_{3}}{F^{6}} ] \right) \\
& & \nonumber \\
\varepsilon_{1} & = & - \frac{N-1}{2F^{2}} G_{1} \\
\delta_{2} & = & \frac{N-1}{4V} G_{2} + \frac{k_{2}+k_{3}}{V} \label{4E2} \\
\varepsilon_{2} & = & - \frac{(N-1)(N-3)}{8} \left( G^{2}_{1} - \frac{2}{V}
G_{2} \right) - \frac{N-1}{V} \Big[ (V\dl -1)k_{1} + k_{2} \Big]
\label{e2d4} \\
\delta_{3} & = & (N-1) \left( \frac{N-3}{4V} G_{1}G_{2} - \frac{2N-5}
{12V^{2}} G_{3} + \frac{k_{1}-k_{2}}{V} G_{1} \right) \label{4D3} \\
\varepsilon_{3} & = & (N-1) \Big\{ -\frac{(N-3)(3N-7)}{48} (G_{1}^{3}
- \frac{6}{V} G_{1}G_{2}) \label{4E3} \nonumber\\
 & & - \frac{(N-3)(N-4) + \gamma^{2}}{12V^{2}} G_{3} - \frac{N-3}{12V}
J_{3} - (N-2) \Gamma_{3} \nonumber\\
 & & + \frac{1}{2V} \Big[ (N-1)V\dl k_{1} - (N+1)(k_{1}-k_{2})
 +(V\dl -1) \{ (N-1)k_{4}+k_{5} \} \Big] G_{1} \nonumber\\
 & & + (k_{4} + \frac{N}{2} k_{5}) G_{\mu \nu } \dot G_{\mu \nu }
 + \frac{1}{2}k_{6}D^{\Lambda}(0) \Big\}
\label{e3d4}
\end{eqnarray}
Now we have completed the evaluation of the partition functions
for $d=3$ and $d=4$ to the $3^{rd}$ order.
We repeat that for dimensional regularization we can omit
$\dl $ and $\Dl $, and as a consequence the coupling constant $k_{6}$.

If we want the 2-loop result we simply neglect $\delta_{3}$ and
$\varepsilon_{3}$. Then the difference of $d=3$ and $d=4$ are only the
additional terms of the latter with $k_{1},k_{2},k_{3}$. The 2-loop
result has been given already in \cite{has/leu}.
There one finds in the appendices also a description of
methods and results for the numerical determination of $g_{1}$ and
$g_{2}$, the regular parts of $G_{1}, \ G_{2}$.\\

As {\em new} cancellations for $d=4$ we can report:
\begin{displaymath}
\begin{array}{ll}
\delta_{3}: & (N^{2},N,1) \dl k_{1} G_{1} \\
\varepsilon_{3}: & (N^{2},N,1)\dl k_{1}G_{1},\quad (N^{2},N)k_{1}G_{1}/V \\
 & (N^{2},N,1)V(\dl )^{2}G_{1}(k_{4},k_{5}), \quad
N^{2}(\dl ,\frac{1}{V})k_{4}G_{1},\quad N(\dl ,\frac{1}{V})k_{5}G_{1} \\
 & (N,1)k_{4}V(G_{\mu \nu})^{2}G_{1},\quad (N^{2},N,1)k_{5}(G_{\mu \nu})^{2}
G_{1}, \quad (N^{2},N,1)G_{1}k_{6} V\Delta^{\Lambda}(0)
\end{array}
\end{displaymath}
Again the power singularities are strongly represented in this list,
in particular there are very few $\dl $ in the final result -- even though
we met whole polynomials of $V\dl $ in the calculation --
and $\DL $ does not show up at all.

All this is part of a sensitive consistency check, because only
due to this cancellations even the generalized singularity structure
we consider is perturbatively renormalizable, see appendix A.

\section{Perturbative renormalization}
Now we are going to verify that the results obtained in sections
5 and 6 fulfill the constraints imposed by renormalizability
\footnote{For brevity here we write sometimes ``renormalizability''
where we actually
mean ``perturbative renormalizability'', as specified in section 1.}.
We repeat that for the singularity structure
we refer to the decomposition (\ref{propzerleg}).
Most singularities of our system of leading divergences
are power divergent (i.e. physically irrelevant). In this
section we restrict the consideration to the relevant singularities
that really occur in the dimensional regularized system.
The remarkable property that even the generalized structure we
dealt with so far is perturbatively renormalizable is discussed
in appendix A.

We apply the ``mass independent'' renormalization prescription,
that sets all the finite parts of the counter terms zero \cite{ren}.
This prescription is preferable especially in view of the
$\beta $-functions \cite{vorschriften}. In the path integral
formalism, the rescaling of the fields to be integrated over can
be absorbed by rescaling the source accordingly \cite{vorschriften}.
But since $H$ only occurs in the product $\Sigma H$, it suffices
to renormalize all the coupling constants, where $\Sigma_{r}H$
actually means $(\Sigma H)_{r}$ (the subscript $r$ denotes the
renormalized quantities). On each of the three levels, the
couplings should be able to absorb the divergences
without picking up a $V$-dependence (see section 2). The volume
independence of the counter terms is the constraint that
provides us with the non-trivial check of our results.

For the subsequent discussion we introduce a
measure for the degree of divergence. Let $\Lambda $ be a characteristic
regularization parameter with the dimension of momentum,
so the non-regularized system corresponds to
$^{lim}_{\Lambda \to \infty }$. (In the naive momentum cutoff
``regularization'',
$\Lambda $ would be the cutoff,
or more generally a characteristic length of the support
in momentum space, e.g. for a smooth cutoff. In a lattice regularization
$\Lambda $ would be proportional to the inverse lattice constant, etc.)
Then we can express the degree of divergence in powers of $\Lambda $.
We recall that for dimensional regularization only the
singularities $\propto ln \Lambda $ remain and that we only
consider them in this section.

It is required by the concept of low
energy expansion and included in our Bessel representation that
after renormalization the leading coupling
constants $\Sigma $ and $F$ coincide with the bare couplings,
so we don't need to renormalize them here.

This is not true, however, for the
non-leading coupling constants $k_{j}$. There we make the ansatz:
\begin{equation} \label{ctk}
k_{jr} = k_{j} +\kappa_{0,j}
\qquad \qquad \kappa_{0,j} \propto ln \Lambda \ , \
\end{equation}
where in the counter terms $\kappa_{0,j}$
the ``mass independent'' renormalization prescription
excludes additional (finite) terms.

The partition function has the form \
$ {\rm Z} = {\bf N} e^{\frac{(\Sigma HV)^{2}}{F^{4}}
\rho_{2}} \ Y_{N}(\Sigma HV\rho_{1})\ .$ Hence the renormalization
has to provide:
\begin{eqnarray} \label{eins}
\Sigma_{r} \cdot \rho_{1r} & = & \Sigma \cdot \rho_{1} \\
\label{zwei}
\frac{\Sigma^{2}_{r}}{F^{4}_{r}} \cdot \rho_{2r} & = &
\frac{\Sigma^{2}}{F^{4}} \cdot \rho_{2}
\end{eqnarray}
where in $\rho_{1r},\ \rho_{2r} $ all the singularities are removed.
We are going the evaluate these two equations order by order.\\

In the first order there are no counter terms available since
only the leading coupling constants are involved.
This is in accordance with the fact that the 1-loop result
does not contain singularities; $\gle \propto \Lambda^{d-2}$
vanishes in dimensional regularization (both, for $d=3$ and $d=4$).\\

For the second and third order, $d=3$ and $d=4$ have to be discussed
separately. We start with ${\bf d=3}$ :

$G_{n}\vert_{V\to \infty}$ diverges if $2n \leq d$. Then we
regularize it to
\footnote{For $d=2n$ \ \ $G_{n}^{\Lambda}$ becomes $\propto ln \Lambda $,
i.e. relevant for our discussion.
This is also the meaning of $\Lambda $ with vanishing power in the following.}
$G_{n}^{\Lambda} \propto \Lambda ^{d-2n} $
and \ $G_{n} = G^{\Lambda}_{n}+g_{n}(V)$\ .
This concerns for $d=3$ just
$G_{1}$, whereas $G_{2},G_{3} \dots$ are regular $V$-dependent functions.

As a consequence the partition function has no divergences even to
the second order, in agreement with the observation that there are
still no counter terms available.

For the remaining terms that contain integrations over $V$,
a corresponding decomposition is more complicated. Let $J$
be such a term: \ $J=\int_{V} T(x) d^{d}x \ $, which shall
be brought to the form \ $J= j(V) +
\sum_{\ell} J^{\Lambda}_{\ell}j_{\ell }(V)$ \ ,
$J_{\ell}^{\Lambda}$ being $V$-independent divergences (that can be
absorbed by the renormalized coupling constants), and $j,j_{\ell }$
being regular.
$T$ is some combination of $G$-functions. Their decomposition
into $G^{\Lambda}(x) + g(x,V)$ yields the form \
$T = \sum_{k} T_{k}^{\Lambda}(x) \cdot t_{k}(x,V)$\ , where
the $t_{k}$ are regular functions, whereas the $T_{k}^{\Lambda}$
depend on $\Lambda $ but not $V$. If $T_{k}^{\Lambda} \propto
\Lambda^{a_{k}}$ and $t_{k} \propto L^{-b_{k}}$, then
$(a_{k}+b_{k})$ will be
fixed for all $k$, since $\Lambda $ has the dimension of a momentum.
This also means that close to the origin
$T^{\Lambda}_{k}(x) \propto x^{-a_{k}} $ .
Thus only the summands with $a_{k}\geq d$ are really
singular for $\Lambda \to \infty $ (if \ $t_{k}(0,V) \neq 0$)
\footnote{The generalization for the case $t_{k}(0,V) =0$ is straightforward.}
, the rest contributes to $j$.

For the treatment of those singularities, let's call them
$\int_{V} T_{\ell}^{\Lambda}(x)t_{\ell}(x,V) d^{d}x$, we apply
a technique that was similarly used in \cite{ge/leu}.
Let $S$ be a sphere around the origin inside the box $V$.
If we decompose $\int_{V} \dots $ into $\int_{S} \dots + \int_{V-S}\dots $,
only the first integral is singular (at $x=0$); the second
one can be added to $j$.
Let $t^{(0)}_{\ell}$ be the Taylor expansion of $t_{\ell }(x)$ around
$x=0$ to the order $a_{\ell}-d$ (its coefficients depend on $V$);
then we write the singular term as:
\begin{displaymath}
\int_{S} T^{\Lambda}_{\ell}(x) [t_{\ell}(x,V)-t^{(0)}_{\ell}(x,V)] d^{d}x
+  \int_{S} T^{\Lambda}_{\ell}(x) t^{(0)}_{\ell}(x,V) d^{d}x \quad .
\end{displaymath}
The first term is regular and contributes to $j$. Finally: \\
$ \int_{S} T^{\Lambda}_{\ell}(x) t^{(0)}_{\ell}(x,V) d^{d}x
=  \int_{ \reell ^{d}} T^{\Lambda}_{\ell}(x) t^{(0)}_{\ell}(x,V) d^{d}x
-  \int_{ \reell ^{d}-S} \dots $\ . We include
the last term in $j$ again, and the integral over the entire Euclidean space
is the desired
$J^{\Lambda}_{\ell}j_{\ell}(V)$  ;
$J^{\Lambda}_{\ell} $\ is independent of $V$,
with a leading divergence $\propto \Lambda ^{a-d}$.
\footnote{We introduce the sphere $S$ instead of just writing
$\int_{V} T^{\Lambda}_{\ell}t^{(0)}_{\ell} dx =
\int_{ \reell ^{d}} T^{\Lambda}_{\ell}t^{(0)}_{\ell} dx
- \int_{ \reell ^{d}-V} T^{\Lambda}_{\ell}t^{(0)}_{\ell} dx $ because it
permits an additional selection of the singularities, as we will see.}

That this procedure corresponds to the decomposition of the $G_{k}$
described above can be confirmed if we apply it on $\int_{V} (G(x))^{2} d^{d}x
= G_{2}$; we arrive at the same $G^{\Lambda}_{2}$ (for $d\geq 4$).

With this concept, we investigate the structure of
\begin{displaymath}
\frac{1}{V} J_{3} = \frac{1}{V} \int_{V} [\gl{^{3}}+3\gl{^{2}}g+3\gl g^{2}
+g^{3}] d^{3}x
\end{displaymath}
We look for singularities close to the origin where $\gl (x) \propto 1/x $.
Only the first term is singular.
We find:
\begin{displaymath}
J_{3} = J^{\Lambda} + j_{3} \ , \quad
{\rm where} \ \  J^{\Lambda} \propto ln \Lambda \ .
\end{displaymath}
The singularity $J^{\Lambda}$ survives dimensional regularization, so it
must be renormalized, see below.\\
Our most complicated term is \vspace*{1mm} $
\Gamma_{3} = \int \partial_{\mu}(\gl +g)\partial_{\mu}(\gl +g)
\partial_{\nu}(\gl +g)\partial_{\nu}(\dot \gl +\dot g) \ d^{3}x $.
$g(x)$ is an even
function, so $\partial_{\mu}g = c x_{\mu}+c^{\alpha \beta
\gamma}_{\mu} x_{\alpha}x_{\beta}x_{\gamma} + \cdots \ $
(from section 3 \vspace{1mm} we know: $\partial_{\mu \nu}g \vert_{x=0} =
\delta_{\mu \nu}/V \ \to \ c=\frac{1}{V}$). \vspace{1mm} \\
About $\dot g(x) $ we know: \vspace{1mm}
$\ \partial_{\mu}\dot g (x)\vert_{x=0} = 0 ,\quad
\partial_{\mu \nu}\dot g (x)\vert_{x=0} = -\delta_{\mu \nu}g_{1} $ , so
$\partial_{\nu}\dot g$ has the expansion: \ \
$\partial_{\nu}\dot g = -g_{1}x_{\nu} + c'^{\alpha \beta \gamma}_{\nu}
x_{\alpha}x_{\beta}x_{\gamma} + \cdots $ .
Applying this we find:
\begin{displaymath}
\Gamma_{3} = \Gamma_{a}^{\Lambda}+
\Gamma_{b}^{\Lambda}g_{1}+\Gamma_{c}^{\Lambda}\frac{1}{V}+\gamma_{3}
\quad
{\rm where:~} \Gamma_{a}^{\Lambda} \propto \Lambda^{3},~\Gamma_{b}^{\Lambda}
\propto \Lambda^{2},~\Gamma_{c}^{\Lambda} \propto ln \Lambda,~\gamma_{3}
{}~{\rm regular}
\end{displaymath}
It might seem that there is also a singularity $\propto \Lambda $
associated with $c'$, but the corresponding
volume-dependent factor vanishes as we see when we integrate over
$S$. This happens to all the non-covariant terms.
In addition only $\Gamma_{c}^{\Lambda}$ is relevant for the
present discussion.

The $3^{rd}$ order of $\rho_{1}, \ \rho_{2}$ is denoted by
$\frac{1}{F^{6}} \varepsilon_{3}, \ \frac{1}{F^{2}} \delta_{3}.$
Inserting the result of section 5 we see that in $\delta_{3}$
there are no logarithmic singularities, so we find for the
counter term the constraint:
\begin{equation} \label{ct3kd}
\kappa_{0,2} + \kappa_{0,3} = 0 \quad .
\end{equation}

Exploiting in the same way the $3^{rd}$ order of (\ref{eins}),
we arrive at:
\begin{equation} \label{oben7}
\kappa_{0,2} - \kappa_{0,1}  =  \frac{N-3}{12}J^{\Lambda}
+(N-2)\Gamma^{\Lambda}_{c} \label{unten7}
\end{equation}
We conclude that
the set of counter terms $\{ \kappa_{0,1}, \ \kappa_{0,2}, \
\kappa_{0,3} \}$ is submitted to the two (independent)
constraints (\ref{ct3kd}) and (\ref{unten7}), so the
counter terms keep one degree of freedom. But as we mentioned
in section 2 we could have used the transformation (\ref{feldtr2})
to eliminate either $k_{2}$ or $k_{3}$: if we do so, i.e. if
we exploit maximally the freedom of choice of the fields to reduce the number
of coupling constants in the Lagrangian (for $\vec H = const.$),
then the two remaining counter terms are uniquely determined.\\

Now we carry out the procedure for ${\bf d=4}$.\\

In the final result of section 6 there are three terms
that contain a singularity in dimensional regularization:
\begin{eqnarray*}
G_{2} &=& \glz + g_{2} \\
J_{3} &=& J_{b}^{\Lambda} g_{1} + j_{3} \\
\Gamma_{3} &=& \Gamma_{d}^{\Lambda} \frac{1}{V}g_{1} + \gamma_{3}
\end{eqnarray*}
and the renormalization involves already the second order of
eqs. (\ref{eins}) and (\ref{zwei}). We insert the results for $\delta_{2}$
and $\varepsilon_{2}$ :
\begin{eqnarray} \label{drct1}
\kappa_{0,2} + \kappa_{0,3} &=& \frac{N-1}{4}\glz \\
\kappa_{0,1}-\kappa_{0,2} &=& \frac{N-3}{4} \glz \label{drct2}
\end{eqnarray}
The 2-loop results of $d=3$ and 4 are alike, only the $k_{j}$-terms
are one order higher for $d=3$. It is remarkable that on the other
hand the mechanism of renormalization is much different. Here
$G_{2}$ is divergent, but the counter terms $\kappa_{0,j}$
allow it to occur in $\delta_{2}$ and $\varepsilon_{2}$ .\\

{\em Third order} \\

If we insert $\delta_{3}$ in (\ref{zwei}) we make the interesting
observation that the constraint (\ref{drct2}) is identically repeated.

In the third order of eq. (\ref{eins}) also the counter terms associated
with $k_{4}, \ k_{5}$ occur
\footnote{Here and to all order no counter terms of $k_{6}$ are
involved in the renormalization.}.
If we insert $\varepsilon_{3}$ and
apply eq. (\ref{drct2}), only the new counter terms remain and
and can be determined:
\begin{equation} \label{ctcon4}
\left( \begin{array}{c} \kappa_{0,4} \\ \kappa_{0,5} \\ \end{array}
\right) = \frac{2}{N(N-1)-2} \left( \begin{array}{c} -N/2 \\
1 \\ \end{array} \right) \Big\{ \frac{(N-3)(2N-3)}{2} \glz -
\frac{N-3}{6} J_{b}^{\Lambda} - 2(N-2) \Gamma_{d}^{\Lambda} \Big\}
\end{equation}

Actually there are three constraints imposed on
$\{ \kappa_{0,1}, \ \kappa_{0,2}, \ \kappa_{0,3} \} $,
but only two of them are independent, so
the set $\{ \kappa_{0,1} \dots \kappa_{0,5} \} $ keeps
one degrees of freedom. However, if we eliminate $k_{2}$ or $k_{3}$,
then all the counter terms are uniquely determined, as for $d=3$.\\

In summary we repeat that perturbative renormalizability can be affirmed
on all the three levels of magnitude, for \ $d=3$ \ and \ $d=4$.
This we could demonstrate in the framework of dimensional regularization
without determining the singularities (nor the regular terms) explicitly.

If we reduce the number of coupling constants in the Lagrangian by
means of field transformations to its minimum, then renormalization
can only be realized due to the coincidence of various constraints
imposed on the counter terms, which are associated with the remaining coupling
constants. In this case, all the counter terms are determined uniquely,
both for $d=3$ and $d=4$.

\section{Conclusions}
We have investigated the non-linear $\sigma $-model in 3 and 4 dimensions,
describing a system of Goldstone bosons
in a large but finite volume and in presence of a weak magnetic field
of the order of the inverse volume
(such that the Goldstone bosons feel the finite size
strongly).
The corresponding partition function is perturbatively
renormalizable as we have shown explicitly to 3 loops.

We can also
confirm the applicability of Polyakov's functional measure that
contains relevant contributions in terms of the finite size.
Referring to dimensional regularization,
an arbitrary linear combination of further invariant terms can be
added to this measure without yielding any contribution to the action.\\

The explicit three loop results for the large volume expansion
of the source dependent part of the free energy
are given for spatial dimension $d=3, 4$ in section 5, 6, respectively,
without specification of the isospin space dimension $N$.
They take a particularly simple form for $N=3$ (Heisenberg model).
They provide a basis for the interpretation of Monte Carlo results,
in particular for their extrapolation to infinite volume.

\vspace*{3cm}

{\bf Acknowledgement} \\

This work represents my PhD. thesis under
the direction of Prof. H. Leutwyler.
I am very grateful to him for
suggesting an interesting subject as well as patient and absolutly
crucial help during all the stages of this work.\\

I am also deeply indept to Prof. P. Hasenfratz for
his advice of great importance in particular concerning
the regularization scheme.\\

In addition I am grateful for the interest of Prof. J. Gasser,
Prof. H. Kleinert, Prof. M.V. Marinov and Prof. P. Minkowski
in my work, particularly in the treatment of the functional measure.

\appendix
\section{Perturbative renormalization with power divergences}
In this appendix we are first going to show that the results of
sections 5 and 6 are perturbatively renormalizable, even if we include
the leading power divergences of all the singular terms,
as specified in section 3.
In the second part we add some
remarks about other attempts to regularize our model.\\

The property
of perturbative renormalizability with leading power divergences
is much more general than the one shown in
section 7 where we restricted the discussion to the terms that occur
in dimensional regularization. The latter could be deduced from
this appendix as a corollary. The generalization, however,
suffers from the problem that we do not know a regularization scheme that
leaves us exactly with the divergences we include here.
What we consider are the terms that all regularizations displaying
power divergences have in common. So we are not very surprised
that this structure is meaningful. On the other hand
the generalizability is remarkable, since there are many divergences involved
with many different volume dependent prefactors.
Hence the constraints imposed on the counter terms become very narrow;
in particular various counter terms are actually overdetermined
by the number of constraints and renormalizability only holds
due to the coincidence of several of them. This coincidence requires
very special relations among the coefficients of the terms occurring
in our results, which provide us with the highly non-trivial check
announced in the introduction.

Like in section 7 we use the ``mass independent renormalization
prescription'' which does, however, not affect the additional
counter terms we include here. They will be expressed in positive powers
of $\Lambda^{d-2}/F^{2}$, where $\Lambda $ is the characteristic
regularization parameter (of dimension of a momentum) introduced
in section 7.

We will proceed in a manner that does not only show the renormalizability
by determining the counter terms explicitly, but also the significance
of this property. Starting from the $\ell $ loop result ($\ell =0,1,2$)
we make a quite general ansatz for the ($\ell +1$) level -- including
all the terms occurring in the calculation to this level -- and examine,
which constraints renormalizability imposes on its coefficients.
Although some coefficients are arbitrary from this point of view,
a lot of them are determined exactly: from the beginning many of
them must be zero (in accordance with the long list of canceled
terms given at the end of section 5 and 6) and for $\ell =2$ we can
even ``predict'' non-vanishing coefficients.
\footnote{As we will see, the renormalizability does not depend
on any special ratio between different singularities of the same power.
This justifies the predictions mentioned above. If we could assume
this independence of singularity ratios also to higher orders,
we could immediately predict the coefficients of various multiloop
terms, such as $G_{1}^{k}, \ G_{1}^{k-2}G_{2}/V, \ (k=4,5 \dots )$ etc.
However, we have no prove for this assumption.

Hence for exact predictions we would need a general prove of this assumption
as well as  the assumption that the singularity structure we consider is
really
renormalizable to all orders. So we write ``predictions'' in inverted commas.}
This clarifies the meaning of the result
and gives a sound basis for the conclusions about the multiloop terms
($\ell >3$).

As a general ansatz for the renormalized
leading coupling constants we write:
\begin{eqnarray} \label{ctsf}
\Sigma_{r} &=& \Sigma ( 1 + \sigma_{1} + \sigma_{2} + \sigma_{3} \dots ) \\
F_{r} &=& F (1+f_{1} +f_{2} +f_{3} \dots ) \qquad \qquad
\sigma_{n},\ f_{n}  \propto  \frac{\Lambda ^{(d-2)n}}{F^{2n}} \nonumber
\end{eqnarray}
where $\Sigma,F$ are the bare couplings and
$\sigma_{n},f_{n}$ their associated counter terms.
The renormalized non-leading coupling constants can also contain
logarithmic counter terms (known from section 7), so there we write:
\begin{equation}
k_{jr} = k_{j} +\kappa_{0,j} + \kappa_{1,j}+\kappa_{2,j}+ \dots  \qquad
 \qquad \kappa_{0,j} \propto ln \Lambda \ , \
\kappa_{n,j} \propto \frac{\Lambda^{(d-2)n}}{F^{2n}} \ .
\end{equation}
Again we discuss order by order eqs. (\ref{eins}) and (\ref{zwei}).\\

First we consider the 1-loop result of section 4; here only (\ref{eins})
is involved. It takes the form:
\begin{displaymath}
\Sigma (1+\sigma_{1}+\dots )\left( 1-\frac{N-1}{2(F[1+f_{1}+\dots ])^{2}}
\ g_{1} \right)=
\Sigma \left( 1-\frac{N-1}{2F^{2}}(G^{\Lambda}_{1}+g_{1})\right)
\end{displaymath}
This may be exploited to the first order $(\propto F^{-2})$\ :
\begin{equation} \label{renone}
\sigma_{1} = -\frac{N-1}{2F^{2}}
G^{\Lambda}_{1}
\end{equation}
So the 1-loop result {\em is} renormalizable and fixes $\sigma_{1}$.
Now we wonder,
how significant this is, i.e. which subset of the possible 1-loop results
can be renormalized. \\
A general ansatz is:
\begin{displaymath}
\rho_{1} = 1 + \frac{a_{1}+P_{\alpha ,1}}{F^{2}} \ G_{1} \
, \ {\rm where} \quad
P_{\alpha ,1} \doteq \sum_{k \geq 1} \alpha_{k,1} (V\dl )^{k}
\end{displaymath}
$a_{1}$ and $\alpha_{k,1}$ being arbitrary constants.
The $V$-independent part yields: \\
$\sigma_{1} = \frac{a_{1}}{F^{2}} \gle $ , and the $V$-
dependent part: $ \alpha_{k,1}=0 \quad (\forall~k) $ .
Thus we already have non-trivial constraints, although the renormalizable
class is quite large up to now. \\

We consider ${\bf d=3}$ and list the singularity structures
of the divergent terms occurring in the 3-loop
calculation:
\begin{eqnarray*}
G_{1} &=& \gle +g_{1} \ ; \qquad \gle \propto \Lambda \ , \quad g_{1} \
{\rm regular} \\
J_{2} &=& I_{a}^{\Lambda} + I_{b}^{\Lambda} \frac{1}{V} + j_{2} \ ; \qquad
I_{a}^{\Lambda} \propto \Lambda^{5}, \ I_{b}^{\Lambda}\propto \Lambda ^{2},
\quad j_{2} \ {\rm regular} \\
J_{3} &=& J^{\Lambda} + j_{3} \ ; \qquad J^{\Lambda} \propto ln \Lambda ,
\quad j_{3} \ {\rm regular} \\
\Gamma_{3} &=& \Gamma_{a}^{\Lambda} + \Gamma_{b}^{\Lambda} g_{1}
+ \Gamma_{c}^{\Lambda} \frac{1}{V} + \gamma_{3} \ ; \qquad
\Gamma_{a}^{\Lambda} \propto \Lambda^{3} , \ \Gamma_{b}^{\Lambda}
\propto \Lambda^{2} , \ \Gamma_{c}^{\Lambda} \propto \ln \Lambda , \quad
\gamma_{3} \ {\rm regular}
\end{eqnarray*}
Presuming the already exploited 1-loop result, a general ansatz
for the 2-loop result is:
\begin{displaymath}
\begin{array}{l}
\rho_{1} = 1 - \frac{N-1}{2F^{2}}G_{1} + \frac{1}{F^{4}} \{ (a_{2}+
P_{\alpha ,2}) G_{1}^{2} + \frac{b+P_{\beta}}{V} G_{2} +
(t +P_{\tau})VJ_{2}\} \\
\rho_{2} = (\tilde a_{2} + P_{\tilde \alpha ,2})G^{2}_{1} + \frac{\tilde b
+ P_{\tilde \beta}}{V}G_{2} + (\tilde t +P_{\tilde \tau})VJ_{2}
\end{array}
\end{displaymath}
where the $P$'s are again polynomials in $V\dl $.
For $\rho_{1}$ we have to consider the second order of (\ref{eins}) :
\begin{displaymath}
\sigma_{2} - \frac{N-1}{2F^{2}}g_{1}(\sigma_{1}-2f_{1}) = \qquad \qquad
\qquad \qquad \qquad \qquad \qquad \qquad \qquad \qquad \qquad
\end{displaymath}
\begin{displaymath}
\frac{1}{F^{4}} \Big\{ (a_{2}+P_{\alpha ,2})
(\gle{^{2}} + 2\gle g_{1}) + P_{\alpha ,2}g_{1}^{2} + \frac{P_{\beta}}{V}G_{2}
+(t +P_{\tau})(VI^{\Lambda}_{a}+I^{\Lambda}_{b}) + P_{\tau}j_{2} \Big\}
\end{displaymath}
We conclude: $P_{\alpha ,2}=P_{\beta}= t =P_{\tau}=0,\quad \sigma_{2} =
a_{2} \gle{^{2}}/F^{4},~f_{1}=(\frac{a_{1}}{2}-\frac{a_{2}}{a_{1}})\gle
.\  $
$a_{2}$ and $b$ are arbitrary. The required cancellations are achieved,
thanks to the compensation of the $\dl $-contributions of the measure
and the Lagrangian. Inserting $a_{1},\ a_{2}$ we get:
\begin{equation} \label{cts2f1}
\sigma_{2} = -\frac{(N-1)(N-3)}{8F^{4}} \ \gle{^{2}} \ , \quad
f_{1} = -\frac{N-2}{2F^{2}} \ \gle
\end{equation}

Now also (\ref{zwei}) must hold to the $2^{nd}$ order ($\propto F^{-4}$).
The renormalized side can not compensate any divergent terms, so
the general
{\em renormalizable} ansatz is just: $\rho_{2} = \tilde b  \ G_{2}/V $.
This is indeed what we found, with \ $\tilde b = (N-1)/4 $ . \\

{\em Third order} \\

Here the constraints imposed by renormalizability develop for the first
time their full power. They no longer only exclude certain terms and
cut off the polynomials in ($V\dl $) in the coefficients, but also
``predict'' the exact values of non-vanishing coefficients.

The $3^{rd}$ order of $\rho_{1}, \ \rho_{2}$ is denoted by
$\frac{1}{F^{6}} \varepsilon_{3}, \ \frac{1}{F^{2}} \delta_{3}.$
As a general ansatz for $\varepsilon_{3}, \ \delta_{3}$ we take
an arbitrary linear combination of the terms : \\
$G_{1}^{3}, \ \frac{1}{V}G_{1}G_{2}, \ \frac{1}{V^{2}} G_{3}, \
\frac{1}{V} k_{j} \ (j=1,2,3), \ \frac{1}{V}J_{3}, \ \Gamma_{3}, \
{\rm and} \ VG_{1}J_{2} $ \ , \\
where again the coefficients include polynomials in $(V\dl )$.
Presupposing the 2-loop result, the $3^{rd}$ order of (\ref{zwei})
requires that in $\delta_{3}$ a lot of contributions vanish and
fixes the form:
\begin{displaymath}
\delta_{3} = \frac{(N-1)(N-3)}{4V} G_{1}G_{2} + \frac{\tilde c }{V^{2}}G_{3}
+ \frac{1}{V} (\tilde r J_{3} + \sum_{j} \tilde d _{j} k_{j} )
\end{displaymath}
which is
in accordance with section 5. $\tilde c $ is arbitrary and
if we insert the result in the remaining constraint: \
$\sum_{j} d_{j} \kappa_{0,j} = \tilde r J^{\Lambda} $ \ we arrive at
eq. (\ref{ct3kd}).

Exploiting in the same way the $3^{rd}$ order of eq. (\ref{eins}),
we find for $\varepsilon_{3}$ the form :
\begin{eqnarray*}
\varepsilon_{3} &=& \frac{(N-1)(N-3)(3N-7)}{48} \Big( -G_{1}^{3}
+ \frac{6}{V}G_{1} G_{2} \Big) + \frac{c}{V^{2}} G_{3} \\
&& + \frac{1}{V} \sum_{j} \Big( d_{j} +V\dl d_{1,j} \Big) k_{j}
+\frac{r}{V} J_{3} + s \Gamma_{3}
\end{eqnarray*}
where $c, \ d_{j}, \ d_{1,j}, \ r \ {\rm and} \ s $ are almost
arbitrary. Again the specified coefficients are in accordance with
the result of section 5.

So to the 34 terms that had to vanish non-trivially for $d=3$ we can
add 3 more coefficients that precisely take the only value that
permits renormalization of the structure we consider.

The counter terms are :
\begin{eqnarray}
\sigma_{3} & = & -\frac{N-1}{F^{6}} \left( \frac{(N-3)(3N-7)}{48}
\gle{^{3}} + \dl k_{1}  + (N-2)\Gamma_{a}^{\Lambda}
\right) \label{sig3d3} \\
f_{2} & = & -\frac{N-2}{F^{4}} \left( \frac{N-2}{4}\gle{^{2}} +
\Gamma^{\Lambda}_{b} \right) \\ \label{last}
\sum_{j=1}^{3} d_{j} \kappa_{0,j} &=& rJ^{\Lambda} + s\Gamma_{c}^{\Lambda}
\end{eqnarray}
Inserting the explicit result in eq. (\ref{last}) yields
constraint (\ref{unten7}). We note that for $N=2$ and for $N=3$
the counter terms take a particularly simple form.

\vspace*{8mm}

Now we carry out the procedure for ${\bf d=4}$.\\

Since no confusion is possible,
we denote the parameters equally as for $d=3$, as we already
started to in section 6 and 7. So $a_{k},b,c$ etc. have a local meaning
for the dimension we are discussing at present.

The singularity structures of the terms occurring up to the $3^{rd}$ order
take the following form:
\begin{displaymath}
\begin{array}{rl}
G_{1} &= \gle + g_{1}~,\qquad \gle \propto \Lambda^{2} \qquad {\rm
(first~degree ~of~divergence)} \\
G_{2} &= \glz + g_{2}~,\qquad \glz \propto ln\Lambda \\
J_{2} &= I^{\Lambda}_{a} + I^{\Lambda}_{b}\frac{1}{V} + I^{\Lambda}_{c}
\frac{1}{V^{2}} + j_{2}~,\qquad I^{\Lambda}_{a} \propto \Lambda^{8},
I^{\Lambda}_{b} \propto \Lambda^{4},I^{\Lambda}_{c} \propto ln\Lambda \\
J_{3} &= J^{\Lambda}_{a} + J^{\Lambda}_{b}g_{1} +
j_{3}~, \quad J^{\Lambda}_{a} \propto \Lambda^{2},J^{\Lambda}_{b}
\propto ln\Lambda \\
\Gamma_{3} &= \Gamma^{\Lambda}_{a} + \Gamma^{\Lambda}_{b}g_{1} +
\Gamma^{\Lambda}_{c} \frac{1}{V} + \Gamma^{\Lambda}_{d}
\frac{1}{V}g_{1} + \gamma_{3} \\
& \qquad \qquad \Gamma^{\Lambda}_{a},\Gamma^{\Lambda}_{b},\Gamma^{\Lambda}
_{c},\Gamma^{\Lambda}_{d} \propto \Lambda^{6},\Lambda^{4},\Lambda^{2},
ln \Lambda \quad {\rm (respectively)} \\
G_{\mu \nu } &= - \delta_{\mu \nu} \ \dl + g_{\mu \nu } \ , \qquad
\dl \propto \Lambda^{4} \\
\dot G_{\mu \nu} &= -\delta_{\mu \nu} \ \gle + \dot g_{\mu \nu} \ , \qquad
\gle \equiv \dot \dl \propto \Lambda^{2} \\
D^{\Lambda}(0) & \propto \Lambda^{6} \ ;
\quad \Delta^{\Lambda}(0) \propto \Lambda^{8}
\end{array}
\end{displaymath}
where the last term is regular everywhere.

The general ansatz for $\varepsilon_{2}$ has to be extended to:
\begin{eqnarray*}
\varepsilon_{2} &=& (a_{2}+P_{\alpha ,2})G_{1}^{2} + \frac{b+P_{\beta}}{V}
G_{2} + \frac{1}{V} \sum_{j=1}^{3}(d_{j}+P_{d_{j}})k_{j}
+(t +P_{\tau})VJ_{2} + (r+P_{\rho}) V \Delta^{\Lambda} (0)
 \end{eqnarray*}
and again the same for $\delta_{2}$, with $\tilde a_{2}$ etc.

{}From the $2^{nd}$ order of (\ref{zwei}) we see: \\
$\tilde a_{2} = P_{\tilde \alpha_{2}} = P_{\tilde \beta} =
P_{\tilde d_{j}} = \tilde t = P_{\tilde \tau} =\tilde r =
P_{\tilde \rho}= 0 ,\quad
\sum_{j} \tilde d_{j} \kappa_{0,j} = \tilde b \glz $ \\
(except for this, $\tilde d_{j}$ and
$\tilde b$ are free).
Inserting what we found in section 6 we arrive at eq. (\ref{drct1}).

{}From eq. (\ref{eins}) we can only conclude: \
$P_{\alpha_{2}} = P_{\beta} = d_{k>1,j} = t = P_{\tau } = r = p_{\rho}= 0 $.\\
This time already in the $2^{nd}$ order $\dl $ can occur, and indeed it does.
The rest is almost free, and:
\begin{eqnarray}
\sigma_{2} &=& \frac{1}{F^{4}} \Big\{ a_{2} \gle{^{2}} +
 \dl \sum_{j} d_{1,j}k_{j} \Big\} \nonumber \\ &=& \label{sig2d4}
-\frac{(N-1) (N-3)}{8F^{4}} \gle{^{2}} - \frac{N-1}{F^{4}}\dl k_{1} \\
f_{1} &=& \frac{1}{F^{2}} \Big\{
\frac{\sigma_{1}}{2} + \frac{2a_{2}}{N-1}\gle \Big\} =
- \frac{N-2}{2F^{2}} \gle \\
\sum_{j=1}^{3} d_{j}\kappa_{0,j} &=& b\glz \label{geringbeding}
\end{eqnarray}
Eq. (\ref{geringbeding}) contains the slight restriction
of the freedom announced above and leads to constraint (\ref{drct2}).\\

{\em Third order} \\

This time the  ansatz for $\varepsilon_{3}, \ \delta_{3}$ is a
linear combination of the following terms:
$G_{1}^{3}, \ \frac{1}{V}G_{1}G_{2}, \ \frac{1}{V^{2}}G_{3}, \
\frac{G_{1}}{V} k_{j} \ (j=1 \dots 5), \ \frac{1}{V}J_{3}, \ \Gamma_{3}, \
VG_{1}J_{2}, \ G_{1}V(G_{\mu \nu})^{2}k_{j} \ (j= 4,5), \\
G_{\mu \nu} \dot G_{\mu \nu} k_{j} \ (j=4,5), \ k_{6} D^{\Lambda}(0), \
k_{6} G_{1}V\Delta^{\Lambda}(0) $ .

In $\delta_{3}$ once more most terms have to vanish. Omitting them
it just remains:
\begin{displaymath}
\begin{array}{l}
F^{2}(2\sigma_{1}-4f_{1}) (\frac{N-1}{4} g_{2} + k_{2} +\kappa_{2,0}
+ k_{3} + \kappa_{3,0}) +\kappa_{1,2} + \kappa_{1,3}
+ \sum \tilde e_{j}\kappa_{0,j} = \\
\tilde b_{3}(\gle \glz +\gle g_{2} + g_{1}\glz ) +
\gle \sum \tilde e_{j}k_{j} \\
\end{array}
\end{displaymath}
Inserting the counter terms known from the $2^{nd}$ order, we find :
\begin{eqnarray}
\tilde b_{3} & = & \frac{(N-1)(N-3)}{4} \nonumber \\
\sum \tilde e_{j}\kappa_{0,j} & = & \tilde b_{3} \glz \label{unorig} \\
\kappa_{1,2}+\kappa_{1,3} & = & \frac{1}{F^{2}} (N-1)(k_{1}-k_{2}) \gle
\end{eqnarray}
As we mentioned in section 7, eq. (\ref{unorig}) reproduces identically
the constraint (\ref{drct2}). In addition we see now that
also constraint (\ref{drct1}) has been repeated.

Eq. (\ref{eins}) causes more work; generally
there we can exclude less quantities because it starts
from the tree-level. A lengthy book-keeping yields in the pure
$G_{n}$ sector what we got for $d=3$ before, but beyond that some novelties:
\begin{displaymath}
\begin{array}{ll}
\varepsilon_{3} = & \frac{(N-1)(N-3)(3N-7)}{48} (-G_{1}^{3} + \frac{6}{V}
G_{1}G_{2}) + \frac{c}{V^{2}}G_{3} + \frac{G_{1}}{V} \sum_{j}
(e_{j} + e_{1,j} V\dl )k_{j} \\
 & + \frac{r}{V}J_{3} + s \Gamma_{3} + G_{\mu \nu} \dot G_{\mu \nu}
\sum_{j} w_{j}k_{j} + p k_{6} D^{\Lambda}(0)
\end{array}
\end{displaymath}
$c,e_{j},e_{1,j},r,s,w_{j}$ and $p$ are almost arbitrary,
as we see if we compute the counter terms:
\begin{equation} \label{ctd4}
\begin{array}{l}
\sigma_{3} = \frac{N-1}{F^{6}} \Big[ - \frac{(N-3)(3N-7)}{48} \gle {^{3}}
+ \frac{1}{2} ([N-1]k_{1} +[N+1][k_{4} + k_{5}])\gle \dl \\ \qquad \qquad
- (N-2)\Gamma^{\Lambda}_{a} +\frac{1}{2}k_{6}D^{\Lambda }(0) \Big] \\
\quad \\
f_{2} = \frac{1}{F^{4}} \left[ -\frac{(N-2)^{2}}{8} \gle{^{2}}
+ \frac{1}{2}([N+3]k_{1}+[N+1][k_{4}+k_{5}]) \dl
-(N-2)\Gamma^{\Lambda}_{b} \right] \\
\quad \\
\kappa_{1,2}-\kappa_{1,1} = \frac{1}{2F^{2}} \Big[
\{ 2(N-1)k_{1} - 2(N-3)k_{2} + (N+1)(k_{4}+k_{5}) \} \gle \\
\qquad \qquad \qquad \qquad
+\frac{N-3}{6} J^{\Lambda}_{a} +2(N-2)\Gamma^{\Lambda}_{c} \Big]
\end{array}
\end{equation}
where on the logarithmic order eq. (\ref{ctcon4}) has to be added.

In summary we repeat that perturbative renormalizability can be affirmed
even for the singularity structure with leading power divergences
on all the three levels of magnitude, for \ $d=3$ \ and \ $d=4$.
This we could demonstrate without determining the power singularities
explicitly.
This property imposes very narrow constraints on the
coefficients of the 3 loop result.

The treatment of power divergences becomes in part applicable when we
discuss conclusions about the renormalizable case $d=2$ in appendix E.
There, e.g. the singularity $\gle $ is logarithmic, so it must be
included in the renormalization.

\subsection{Constraints on the regularization}
In this subappendix we add some remarks about the problems that occur
if we try to regularize our model in a way different from dimensional
regularization.

In the main part of this work we have treated the regularized
$\delta $-function \ $\dlf (x)$
like an exact $\delta $-function under the spatial integral without worrying.
If we don't choose dimensional regularization,
this is risky since we anticipate a limit which we ought
to take only at the very end, after renormalization.
First we are going to
give generalized results for the measure and the partition function
without any assumptions about $\dlf (x)$ (rsp. $\gl (x)$). Then
we observe which properties have been used in the first part of this
appendix and
how far they are necessary for perturbative renormalizability.
As an example, the Pauli-Villars regularization fails
to fulfil the required properties. Concerning the physical properties,
it maintains covariance but violates
unitarity (the opposite is the case for lattice regularization).
At the end we outline why also a sharp momentum cutoff is unsuitable
for this model. We are not much surprised about this when we consider
that it violates both, unitarity as well as covariance.

First we consider the measure and only use relation (\ref{delta}),
which is now understood as a {\em definition} of $\dlf (x)$. No
further properties of this function are presupposed. Then the measure
takes the generalized form:
\begin{eqnarray}
\label{mgen}
ln \sqrt{{\bf g}} & =& \frac{V\dl -N+1}{2V} \int \pif^{2}(x) dx
+\frac{V\dl -1}{2V} \int (\pif^{2}(x))^{2}dx \\ &&  -\frac{N-1}{8}
\left(\frac{1}{V} \int \pif^{2}(x)dx \right)^{2}
 -\frac{N-5}{8V} \int \int \dlf (x-y) \pif^{2}(x) \pif^{2}(y) dx dy
 \nonumber \\
&& -\frac{1}{4} \sum_{i,k} \int \int (\dlf (x-y))^{2} \pi^{i}(x) \pi^{k}(x)
\pi^{i}(y) \pi^{k}(y) dx dy \nonumber
\end{eqnarray}
Here we already observe modifications in the second order.\\

Let us consider the partition function for ${\bf d=3}$ and generalize the
result of section 5. We insert eq. (\ref{mgen}) and -- in accordance with
eq. (\ref{delta}) -- we use eq. (\ref{deltadelta}) when performing the Wick
contractions. In the evaluation of the contracted terms,
we always maintain the generality of $\dlf (x)$, except for the
following three assumptions about the regularized system: \\
a) The regularized propagator is translation invariant :
$G^{\Lambda}(x,y) = G^{\Lambda}(x-y)$.\\
b) Partial integrations are permitted everywhere without causing
extraordinary terms. \\
c) The regularization does not require additional terms in the Lagrangian.\\

For example on the lattice all the three assumptions are not
fulfilled: e.g. the non-covariance requires to include terms like
$g_{4}^{(4)} ( \partial_{\mu} \vec S \partial_{\mu} \vec S )
( \partial_{\mu} \vec S \partial_{\mu} \vec S )$ in the Lagrangian.\\

As a consequence of assumption a) an odd number of derivatives of
the propagator at the origin has to vanish. This property has been used
very extensively: \\
Already to the first order, the evaluation (\ref{fosi}) only holds with
the (non-trivial) constraint
\begin{displaymath}
\partial_{\mu} \gl (x) \vert_{x=0} = 0 \quad .
\end{displaymath}
This would be violated for ``regularizations'', which are not symmetric
around the origin in momentum space, e.g. a sharp cutoff
$\vert p-p_{0} \vert \leq \Lambda $.
We see easily that it is required
from the 2-loop renormalization; else it would cause there a non-vanishing
contribution $(\partial_{\mu} \gl (x)\vert_{x=0})^{2} \gl $, which
can not be absorbed. \\

On the 2-loop level we find modifications for the following terms:
\begin{eqnarray*}
I_{1} & \doteq &
< \frac{F^{4}}{8} \Big( \int (\pif \partial_{\mu} \pif )^{2} dx \Big) ^{2} >
 \\
-I_{2} & \doteq & <-\frac{N-5}{8V} \int \int \dlf (x-y) \pif^{2}(x)
\pif^{2}(y) dx dy > \\
- I_{3} & \doteq & <-\frac{1}{4} \sum_{i,k} \int \int (\dlf (x-y))^{2}
\pi^{i}(x) \pi^{k}(x) \pi^{i}(y) \pi^{k} (y) dx dy >
\end{eqnarray*}
\begin{eqnarray*}
<\frac{F^{2}\beta }{4V} \Big( \int (\pif \partial_{\mu} \pif )^{2} dx \Big)
\Big( \int \pif^{2} dy \Big) >& =& \frac{\beta (N-1)}{4F^{4}}
\Big[ (V\dl -1) \Big( (N-1) G_{1}^{2} + \frac{2}{V} G_{2} \Big) \\
 & & +2G_{1} \int \dlf (x) G(x) dx \Big]
\end{eqnarray*}
The source independent  terms $I_{1},I_{2},I_{3}$ do not enter the final
result because their product with $S_{1}(H)$ is cancelled on the
3-loop level. So we don't need to evaluate them. \\

On the 3-loop level we focus our attention again on the troublesome
term
\begin{eqnarray*}
<- \frac{F^{4} \gamma \Omega^{00}}{16V} \Big( \int (\pif \partial_{\mu}
\pif )^{2} dx \Big)^{2} \Big( \int \pif^{2} dz \Big) > &=&
-\gamma \Omega^{00} \frac{N-1}{2F^{2}} G_{1} I_{1} - \\
&&  \frac{\gamma \Omega^{00}}{16F^{6}} <[(a {\bf b})(c{\bf d})]
[(e{\rm f})(g{\rm h})][(r\vert s)]>
\end{eqnarray*}
with the notation introduced in eq. (\ref{not}). In particular the functions
$\Gamma_{0} \dots \Gamma_{5}$  defined in eqs. (\ref{gammas}) can not be
expressed in terms of $\Gamma_{3}$ as easily as in section 5.
Instead of eqs. (\ref{gam3}) we have:
\begin{displaymath}
\begin{array}{ll}
\Gamma_{0} = \Gamma_{3} + \gamma_{2} + \frac{1}{2}(\gamma_{1}-\gamma_{4})
 \qquad \qquad & {\rm where :} \\
\Gamma_{1} = -\frac{1}{2} (\Gamma_{3}+\gamma_{1}) &
\gamma_{1} \doteq  \int (- \dlf +\frac{1}{V}) \partial_{\mu}G \partial_{\mu}G
\dot G dx \\
\Gamma_{2} = \frac{1}{2} (\gamma_{4} - \Gamma_{3}) &
\gamma_{2} \doteq \int \partial_{\mu}G \partial_{\mu} \dlf G \dot G dx \\
\Gamma_{4} = \Gamma_{3} + \gamma_{3} - \gamma_{4} &
\gamma_{3} \doteq \int \partial_{\mu} \dlf G^{2} \partial_{\mu} \dot G dx \\
\Gamma_{5} = \gamma_{4} - \frac{1}{2} (\Gamma_{3}+\gamma_{3}) &
\gamma_{4} \doteq \frac{1}{3} (\int \dlf G^{3} dx - \frac{1}{V} J_{3}) \\
\end{array}
\end{displaymath}

Also in the remaining nine $H$-dependent 3-loop terms there are numerous
modifications. Some of them include again the quantities $\gamma_{1}
\dots \gamma_{4} $. Following all the steps of section 5 with these
generalized terms we arrive at a partition function of the form
(\ref{Z3}) with:
\begin{eqnarray*}
\varepsilon_{1} &=& - \frac{N-1}{2} \ G_{1} \\
\delta_{2} &=& \frac{N-1}{4V} \ G_{2} \\
\varepsilon_{2} &=& \frac{N-1}{8} \Big[ -(N+1)G_{1}^{2} + \frac{2(N-3)}{V}
G_{2} + 4G_{1} \int \dlf (x) G(x) dx \Big] \\
\delta_{3} &=& \frac{N-1}{2V} \Big[ \frac{-2N+5}{3V} G_{3} + \frac{N-1}{2}
G_{1}G_{2} - G_{2} \int \dlf (x) G(x) dx \Big] + \frac{k_{2}+k_{3}}{V} \\
\varepsilon_{3} &=& (N-1) \Big\{ - \frac{1}{16} (N+1)(N+3) G_{1}^{3}
 - \frac{1}{12V^{2}} \Big[ (N-3)(N-4) + \gamma^{2} \Big] G_{3} \\
 && +\frac{1}{8V} \Big[ N^{2} -8N +11 -4V \int (\dlf (x))^{2}dx
 +2(N^{2}-6N+7) \int \dlf (x) dx \Big] G_{1} G_{2} \\
 && + \frac{3N+1}{4} G_{1}^{2} \int \dlf (x) G(x) dx +
 - \frac{N-3}{2V} G_{2} \int \dlf (x) G(x) dx \\
 && + \frac{1}{2} G_{1} \int (\dlf (x))^{2} \dot G (x) dx
 + \frac{N}{2} \int (\dlf (x))^{2} G(x) \dot G (x) dx \\
 && +\frac{N-5}{2V} \int \dlf (x) G(x) \dot G (x) dx -
 \frac{1}{2} G_{1} \Big( \int \dlf (x) G(x) dx \Big)^{2} \\
 && -\frac{1}{4} \Big[ 4(N-2) \Gamma_{3} -2\gamma_{1} +2N \gamma_{2}
 +(N-2) \gamma_{3} +(-N+6) \gamma_{4} \Big] \\
 && -\frac{1}{V} \Big[ (V\dl -1)k_{1} - k_{2} \Big] \Big\}
\end{eqnarray*}

Comparison to the corresponding result of section 5 (eqs (\ref{E1}) $\dots $
(\ref{E3}) ) shows that:

$> \quad \varepsilon_{1}$ and $\delta_{2}$ are unchanged.

$> \quad\varepsilon_{2}$ and $\delta_{3}$ coincide with the corresponding
quantities of section 5 only if
\begin{equation} \label{conG1}
\int \dlf (x) G(x) dx = G_{1} \quad .
\end{equation}

$> \quad $ For $\varepsilon_{3}$ the analogous transition requires eq.
(\ref{conG1}) and 8 further constraints:
\begin{eqnarray}
\int \dlf dx &=& 1 \\
\int (\dlf )^{2} dx &=& \dl \label{con3} \\
\int (\dlf )^{2} \dot G dx &=& \dl G_{2} \label{con4} \\
\int (\dlf )^{2} G \dot G dx &=& \dl G_{1} G_{2} \label{con5} \\
\int \dlf G \dot G dx &=& G_{1} G_{2} \label{con6} \\
\int \dlf \partial_{\mu} G \partial_{\mu} G \dot G dx &=& 0 \\
\int \dlf G \partial_{\mu} G \partial_{\mu} \dot G dx &=& 0 \label{con8} \\
\label{conend}
\int \dlf  G^{3} dx &=& G_{1}^{3}
\end{eqnarray}

So the nine properties (\ref{conG1}) $\dots $ (\ref{conend})
of $\dlf $ rsp. $\gl $ are required
for the transformation of this 3 loop partition function to the former,
simplified form.

Note that in the evaluations of section 5 we have additionally used
two more relations of that kind,
but since they affect only $I_{2}$ and $I_{3}$ (which cancel
separately), they are not needed for the final result. \\

We have, however, not answered the crucial question, which among
those constraints are really necessary for the perturbative renormalizability
of the 3 loop result.

To handle this question we procceed as follows: we assume that we are
dealing with a regularization in the proper sense, i.e. if we remove
the regularization parameters the propagator and all its derivatives
return to the original form.
We consider now the constraints
(\ref{conG1}) $\dots $ (\ref{conend})
for such a proper, but general regularization and
list the additional singular terms that might occur.
Using in particular the symmetry of the regularized
propagator in momentum space,
we find that \\
1. A discrepancy on the level of the leading divergence is possible
in eqs (\ref{conG1}), (\ref{con3}) $\dots $ (\ref{con6}) and
(\ref{conend}).\\
2. Additional (non-leading) divergences can occur in
\begin{displaymath}
\int (\dlf )^{2} dx \ , \quad \int (\dlf )^{2} \dot G dx \ ,
\quad \int (\dlf )^{2} G \dot G dx \quad {\rm and} \quad \int \dlf G^{3} dx
\quad .
\end{displaymath}
Their possible, non-leading contributions are of order \
$\Lambda , \ ln \Lambda , \ \Lambda , \ \Lambda $,
respectively. Thus $\varepsilon_{3}$ can receive an extra term
of the form
\begin{displaymath}
\alpha_{1} \Lambda L^{-2}G_{1}G_{2} + \alpha_{2}ln \Lambda \ L^{-2}G_{1}
+N\alpha_{3} \Lambda L^{-2} + (N-6)\alpha_{4} \Lambda L^{-2}
\end{displaymath}
($\alpha_{i} = const.$).
Obviously all these terms contain a singularity with the volume
dependent prefactor $L^{-2} \propto g_{1}^{2}$, which can
(in general) {\em not}
be renormalized: the only counter terms that can absorb
singularities with the same prefactor in $\varepsilon_{3}$
are \
$\sigma_{1}$ and $f_{1}$. But they are uniquely determined from the 2
loop level and therefore not able to absorb further singularities.

For the analogous reasons all the constraints about the leading
divergences are necessary.

If in a regularization such additional divergences exist,
they have to fulfill very special relations to preserve
perturbative renormalizability. \\

For ${\bf d=4}$ the generalized partition function has again the form
(\ref{Z3}), where
\begin{eqnarray*}
\varepsilon_{1} &=& - \frac{N-1}{2} \ G_{1} \\
\delta_{2} &=& \frac{N-1}{4V} \ G_{2} + \frac{k_{2}+k_{3}}{V} \\
\varepsilon_{2} &=& (N-1) \Big[ -\frac{N+1}{8} G_{1}^{2} + \frac{N-3}{4V}
G_{2} + \frac{1}{2}G_{1} \int \dlf G dx
-\frac{(V\dl -1)k_{1}-k_{2}}{V} \Big] \\
\delta_{3} &=& \frac{N-1}{V} \Big[ \frac{N-1}{4} G_{1}G_{2} -
\frac{1}{2} G_{2} \int \dlf G dx -\frac{2N-5}{12V} G_{3}
+k_{1} \int \dlf G dx -k_{2} G_{1} \Big] \\
\varepsilon_{3} &=& (N-1) \Big\{ - \frac{1}{16} (N+1)(N+3) G_{1}^{3}
 - \frac{1}{12V^{2}} \Big[ (N-3)(N-4) + \gamma^{2} \Big] G_{3} \\
 && +\frac{1}{8V} \Big[ N^{2} -8N +11 -4V \int (\dlf )^{2}dx
 +2(N^{2}-6N+7) \int \dlf dx \Big] G_{1} G_{2} \\
 && + \frac{3N+1}{4} G_{1}^{2} \int \dlf  G dx
 -\frac{N-3}{2V} G_{2} \int \dlf G dx \\
 && + \frac{1}{2} G_{1} \int (\dlf )^{2} \dot G dx
 + \frac{N}{2} \int (\dlf )^{2} G \dot G dx \\
 && +\frac{N-5}{2V} \int \dlf G \dot G dx -
 \frac{1}{2} G_{1} \Big( \int \dlf G dx \Big)^{2} \\
 && -\frac{1}{4} \Big[ 4(N-2) \Gamma_{3} -2\gamma_{1} +2N \gamma_{2}
 +(N-2) \gamma_{3} - (N-6) \gamma_{4} \Big] \\
 && + \frac{1}{V} k_{1} \Big[ \frac{N-3}{2} (V\dl -1) G_{1} -
 \int \dlf Gdx + G_{1} \int \dlf ( V\dlf -1)dx \Big] \\
 && + k_{2} \frac{N+1}{2V} G_{1} + \Big[ (N-1)k_{4}+k_{5}\Big]
 \frac{V\dl -1}{2V} \int \dlf G dx \\
 && +(k_{4} + \frac{N}{2}k_{5} ) G_{\mu \nu} \dot G_{\mu \nu}
 + \frac{1}{2} k_{6} \int \dlf D^{\Lambda} dx  \Big\}
\end{eqnarray*}
In its evaluation, the assumptions a) $\dots $ c) -- and the
comments about them -- remain unchanged. For the simplification
to the result of section 6 we still need eqs (\ref{conG1}) $\dots $
(\ref{conend}) and in addition:
\begin{equation} \label{con10}
\int \dlf D^{\Lambda} dx = D^{\Lambda}(0)
\end{equation}
Concerning the eqs. that have to hold non-trivially on the level of
leading divergences,
eq. (\ref{con10}) has to be added to the list given for $d=3$.

The variety of possible non-leading divergences is much larger here than
in the 3 dimensional case.
\footnote{This can be understood from the fact that in the
Laplacian expansion of $G^{\Lambda} $ every step corres-ponds to
$2/(d-2)$ loop orders.}
The following terms can cause additional contributions, the form
of which is given in the right column:
\begin{eqnarray*}
\int (\dlf )^{2} G \dot G dx \ , \quad
\int \dlf G^{3} dx \ , \quad
\int \dlf D^{\Lambda} dx & \Big\} & \alpha_{1} \Lambda^{4}L^{-2} +\alpha_{2}
\Lambda^{2}L^{-4} + \alpha_{3} ln \Lambda \ L^{-6} \\
\int (\dlf )^{2} dx \ , \quad
\int ( \dlf )^{2} \dot G dx & \Big\} & \alpha_{4} \Lambda^{2}L^{-2}
+\alpha_{5} ln \Lambda \ L^{-4} \\
\int \dlf G dx \ , \quad
\int \dlf G \dot G dx \ , \quad
V \int \dlf G \partial_{\mu} G \partial_{\mu} \dot G dx & \Big\} &
\alpha_{7} ln \Lambda \ L^{-2}
\end{eqnarray*}
($\alpha_{i} = const.$).
To see that we can not permit all these terms to occur with arbitrary
coefficients, it suffices to
look at the contributions $\propto L^{-4} \ \ {\rm i.e.} \ \
\propto g_{1}^{2} $ : like in the 3 dimensional case,
the only counter terms with the same volume dependent prefactor
that enter $\varepsilon_{3}$ are uniquely determined from the 2 loop level.\\

A possibility to regularized the
power singularities without eliminating all of them is provided by the
{\em Pauli-Villars regularization}. Referring to the Fourier
decomposition the propagator
is manipulated to take the form:
\begin{equation} \label{PV}
G_{PV} (x) = \frac{1}{V} \sum_{n}{'} \Big[ \frac{1}{p_{n}^{2}} + \sum_{i}
\frac{c_{i}}{p_{n}^{2}+M_{i}^{2}} \Big] e^{ip_{n}x}
\end{equation}
(the $c_{i}$ are constants and the $M_{i}$ are heavy regularization
masses that go to infinity in the final limit).
We only regularize $G(x)$ and do not try
to construct a correspondingly extended Lagrangian.
\footnote{Such a construction is not to be feasible: the only
way to introduce masses $M_{i}$ is to break again the $O(N)$ symmetry.
After angular integration we do not end up with the desired form.}

This is a proper regularization in the above sense.
To regularize a singularity of power $\Lambda^{2n}$ we have
to introduce at least $n+1$ different
regularization masses. If we choose this minimal number,
the appropriate coefficients are given by:
\begin{displaymath}
\left( \begin{array}{c} c_{1}\\ . \\ . \\ c_{k} \\ \end{array} \right)
= \left( \begin{array}{cccc} 1 & 1 & \dots & 1 \\ M_{1}^{2} & M_{2}^{2}
& \dots & M_{k}^{2} \\ M_{1}^{4} & M_{2}^{4} & \dots & M_{k}^{4} \\
. & . & . & . \\ . & . & . & . \\ M_{1}^{2k-2} & M_{2}^{2k-2} & \dots
& M_{k}^{2k-2} \\ \end{array} \right) ^{-1}
\left( \begin{array}{c} -1 \\ 0 \\ 0 \\ . \\ . \\ 0 \\ \end{array} \right)
\end{displaymath}
($\sum_{i} c_{i} = -1$ is the minimal condition to have
a regularization at all).
Since the strongest divergences (to 3 loops) are of order
$\Lambda^{3},\ \Lambda^{6}$
for $d=3, \ 4$, respectively, we have to introduce at least 2 rsp. 4
different masses.

It turns out, however, that the Pauli-Villars regularization does not obey
the constraints listed above. To illustrate this we consider
eq. (\ref{conG1}) for $d=4$:
we find the difference
\begin{displaymath}
G_{1PV} - \int \dlf_{PV}(x) G_{PV}(x) dx = \frac{1}{V} \sum_{i}
\frac{c_{i}}{M_{i}^{2}} + \frac{1}{16 \pi^{2}} \sum_{i,k} c_{i} c_{k}
M_{k}^{2} \frac{M_{i}^{2} ln M_{i}^{2} -M_{k}^{2} ln M_{k}^{2}}{M_{k}^{2}
-M_{i}^{2}}
\end{displaymath}
which diverges quadratically.
For renormalizability it is not sufficient that this difference
is volume independent: as we saw the counter terms are overdetermined
and have solutions only because their constraints are not independent.
In the presence of such discrepancies, the additional terms would
have to match in a very special manner to keep the regularization
applicable.

\vspace*{5mm}

At last we give a brief illustration, why a {\em sharp cutoff in
momentum space} turns out not to be a suitable regularization.
More precisely: the sharp momentum cutoff does not even
deserve the name ``regularization'' because its limit
$\Lambda \to \infty $ does not always reproduce the
non-regularized quantities.

This can be illustrated by considering the contribution
to a scattering amplitude at low energy, illustrated in figure 6.
\begin{figure}[hbt]
 \centerline{\ \psfig{figure=F6.ps,height=2cm}\ }
 \caption{2 loop contribution to a scattering amplitude at low energy}
\end{figure}

If we apply a sharp momentum cutoff $\Lambda >> \vert k \vert $
and require that the incoming momentum at every vertex vanishes,
this contribution will include
\begin{displaymath}
J= \int dx e^{ikx} \gl (x) \partial^{2} \gl (x) \propto
\int_{B} \frac{1}{p^{2}} d^{d}p
\end{displaymath}
where $B \doteq \{p\vert (\vert p \vert \leq \Lambda ) \wedge
(\vert p-k \vert \leq \Lambda ) \} $ is the intersection of the balls
with radius $\Lambda $ and the centers 0 and $k$.
In the $0^{th}$ approximation ($k=0$) we get
\begin{displaymath}
J_{0} = \frac{2 \pi^{d/2}}{\Gamma (d/2)(d-2)} \Lambda ^{d-2}
\end{displaymath}
The ``moon'' to be subtracted from this is to first order in
$\vert k \vert $ proportional to the sphere of the ball $B\vert_{k=0}$,
hence
\begin{displaymath}
J = J_{0} - \alpha \Lambda^{d-3} \vert k \vert + O(\vert k \vert ^{2})
\qquad \qquad (\alpha > 0)
\end{displaymath}
In particular we find for $d=4$ :
\begin{displaymath}
J = \pi^{2} \Lambda^{2} - \frac{4\pi}{3}\Lambda \vert k \vert -\frac{1}{4}
\pi^{2} \vert k \vert ^{2} + O(\vert k \vert /\Lambda) \quad .
\end{displaymath}
Remembering that the available counter terms have magnitudes
$\Lambda ^{(d-2)n} \quad (n \in \posganz ) $, we see that already
the second term
is not renormalizable for $d > 3 $, i.e. this term is an artifact of
the sharp cutoff (``echo effect''), which destroys locality.

The troublesome difference $J-J_{0}$ stems from the difference
$\dlf (x)- \delta (x) $; the occurrence of odd power differences
in the expansion of $J$ shows the violation of the basic symmetry
properties that the correctly regularized $G^{\Lambda}(x)$ must
have.

For the free energy the situation is a little different since the
diagrams have no external legs. The role of the perturbation
$\vert k \vert $ has to be played by $1/L$.
Note that e.g. eq. (\ref{conG1}) holds in this case; however
we run into trouble of the kind illustrated above on the 3 loop level
when dealing with the integrands that contain more than 2 momenta:
the cutoff acts on each of them, hence also on the sums of them but one.

\section{Generalization of the Polyakov measure}
In section 3 we have calculated the measure according to the
definition (\ref{met}) containing only the simplest invariant
term. The application of this measure in the following sections
has been successful, particularly in view of the the perturbative
renormalization (section 7, appendix A), which could only be performed
due to cancellations of singularities stemming from the measure
and from the Lagrangian.

We repeat that from a physical point of view, it is reasonable to require
the following properties for terms entering the measure: \\
a) locality \\
b) Euclidean translation invariance \\
c) rotational invariance in isospin space \\
-- just like the terms in the Lagrangian.
Now we are going to generalize
the measure by adding an arbitrary linear combination of all
additional terms fulfilling these conditions.
The generalized measure takes the form:
\begin{eqnarray}
\label{Massd3}
d=3: \quad ds^{2}_{g} &=& \frac{1}{V} \int \Big\{ (d\vec S )^{2} [1+
b \frac{\Sigma}{F^{6}} ( \vec H \vec S ) ]
+ \frac{a}{F^{4}} (\partial_{\mu} d\vec S )^{2} + \dots \Big\} dx \\
d=4: \quad ds^{2}_{g} &=& \frac{1}{V} \int \Big\{ (d\vec S )^{2}
+ \frac{a_{1}}{F^{2}} (\partial_{\mu} d\vec S )^{2} +
\frac{a_{2}}{F^{2}} (d\vec S )^{2}
(\partial_{\mu}\vec S )^{2} \label{Massd4} \\
 && + \frac{a_{3}}{F^{4}} (\partial ^{2} d\vec S )^{2} + \dots  \nonumber \\
 & & + b_{1} \frac{\Sigma}{F^{4}} (d\vec S )^{2}(\vec H \vec S)
 + b_{2}\frac{\Sigma}{F^{6}} (\partial_{\mu} d \vec S )^{2}(\vec H \vec S)
 \dots \Big\} dx \nonumber
\end{eqnarray}
where we wrote down the terms that might contribute to our three loop
result, with dimensionless coefficients.
The selection of these terms has to be performed with some
care, paying attention to the relations
\begin{displaymath}
(\vec S d \vec S) = \frac{1}{2} d (\vec S ^{2}) = 0 = (\vec S \partial_{\mu}
\vec S ) \quad
{\rm and} \quad \partial_{\mu}\vec S =\Omega \partial_{\mu}\vec \pi
\propto L^{-d/2} \ .
\end{displaymath}

It appears quite natural to include the further invariants that
can be built purely from $\vec S $ (with $a $ coefficients),
whereas the source dependent terms (with $b $ coefficients)
might surprise a little. To us it seems useful to observe the
consequences of such terms too. We consider it physically
plausible that an external field
might influence the metric in configuration space.

Moreover the reduction to the first term is a priori not acceptable
since we have exploited the freedom of choice of the fields
already when simplifying the Lagrangian in section 2. We recall that
we got rid of three coupling constants by suitable redefinitions
of the fields. Two of the redefinitions included the magnetic field,
which also supports the consideration of the source dependent
terms in the measure.

To be explicit, let us start from the generalized measures
(\ref{Massd3}), (\ref{Massd4}) and perform the substitutions
(\ref{feldtr1}), (\ref{feldtr2}) and (\ref{feldtr3}) of section 2.
Then the coefficients of $ds^{2}_{g}$ change in terms of the
dimensionless parameters \ $\alpha , \ \beta , \ \lambda $ as follows:
\begin{eqnarray}
\{ a,b \} & \to & \{ a-2\alpha , b-2\beta  \}  \\
\{ a_{1},a_{2},a_{3},b_{1},b_{2} \} & \to &
\{ a_{1}-2\alpha , a_{2}+2 \alpha ,
a_{3}+\alpha ^{2}-2a_{1}\alpha , \nonumber \\
&& \ b_{1}-2\beta , b_{2}-2\lambda -2(a_{1}-2\alpha )\beta  \}
\end{eqnarray}
We see that for $d=3$ the full generalization (\ref{Massd3}) has
to be considered. Also for $d=4$ all the parameters of $ds^{2}_{g}$
get activated, but the changes in the three $\vec H $-independent
parameters depend only on $\alpha $, thus they are not independent.
\footnote{If we want to eliminate non-leading couplings in the
maximally generalized measure instead of the Lagrangian, we can reduce it
with these transformations to Polyakov's form for $d=3$, whereas for
$d=4$ there remain two of the $a_{i}$ coefficients.}
Nevertheless we consider the completely generalized form (\ref{Massd4}). \\

In his original paper \cite{poly}, Polyakov did not consider such a
generalization. He had renormalizable models in mind and there it is
not motivated to include additional invariants, not in the Lagrangian
nor in the measure.
The two dimensional non-linear $\sigma $-model corresponds to the
renormalizable case
Polyakov refers to. There he mentions that his measure (for bosonic
strings composed of two terms) is unique, where he requires
locality in a stricter sense than it is done here, i.e. in the sense
that also derivative couplings are excluded.

Here the situation is different: non-leading
couplings are needed and instead of adding them only in ${\cal L}$
(as it is usually done), we can do so in $[d\vec S ]$ as well.
In which way those extensions are related to each other is not
evident and will be discussed explicitly.

Of interest is, how far this
generalization is physically permissible and
to which extent the non-leading coupling constants $k_{i}$
in the Lagrangian can be replaced by new parameters from the measure.
At last, for $d>2$ those terms were included in order to enable a perturbative
renormalization. It will be observed if we need less of them
for this purpose after generalizing the measure.\\

Now we are going to consider the
maximal generalization mentioned above.
If we insert the collective variables introduced in section 2, expand
everything in terms of the transversal fields $\pif (x)$ and consider that
non-diagonal elements of the metric tensor only enter the determinant
quadratically, we arrive at the form:
\begin{eqnarray}
d=3: \quad ds^{2}_{g} & \cong & \frac{1}{V} \int dx
\Big\{ (d\vec \pi - \omega \vec \pi )^{2}
[1+b \frac{\Sigma}{F^{6}} H \Omega^{00}]+ \frac{a}{F^{4}}
(\partial_{\mu} d \pif )^{2} \Big\} \\
d=4: \quad ds^{2}_{g} & \cong & \frac{1}{V} \int dx
\Big\{ (d\vec \pi - \omega \vec \pi )^{2}\Big[ 1+\frac{a_{2}}{F^{2}}
\partial_{\mu} \pif \partial_{\mu} \pif \\
 && + b_{1} \frac{\Sigma}{F^{4}} H ( \Omega^{00} [1-\frac{1}{2}\pif^{2}]
 +\Omega^{0i} \pi^{i})  \Big]
  +\frac{a_{1}}{F^{2}} \Big[ (\partial_{\mu}d\pif )^{2}+
 (\partial_{\mu}(\pif d\pif))^{2} \Big] \nonumber \\
 && + b_{2}\frac{\Sigma}{F^{6}} H \Omega^{00} (\partial_{\mu} d \pif )^{2}
 + \frac{a_{3}}{F^{4}} ( \partial^{2} d \pif )^{2} \Big\} \nonumber
\end{eqnarray}

We begin with ${\bf d=3}$. For the generalized determinant $ {\bf g}_{g}$
we get :
\begin{equation} \label{massallg3}
\sqrt{ {\bf g}_{g}} \cong \sqrt{ {\bf g}} \ exp \{
b\frac{\Sigma}{2F^{6}}H\Omega^{00}(N-1) V\dl
+a \frac{N-1}{2F^{4}} VD^{\Lambda}(0) \}
\end{equation}
where $\sqrt{ {\bf g}}$ is given in (\ref{determ}) and we have defined
$D^{\Lambda}(x)$ in (\ref{strongsing}).

The most important property displayed by eq. (\ref{massallg3}) is that
the non-leading couplings in the measure do not yield any physically
relevant contribution. To be explicit: in dimensional regularization
the determinants simply obey the relation:
\begin{equation} \label{detgl}
{\bf g}_{g} = {\bf g} \quad .
\end{equation}
We recall that there are relevant measure terms, but they are all
included in the leading term that defines Polyakov's measure
(\ref{met}).

If we apply our ansatz for the renormalized non-leading coupling
constants also on the parameters of the measure,
the new counter terms do not enter
the renormalizability conditions because they do not multiply
any regular term. Thus the non-leading coupling constants of
the measure do not contribute new degrees of freedom to the set
of counter terms, constrained by renormalization. \\

For completeness we also show the renormalizability of the leading
power divergences involved in this generalization:

$D^{\Lambda}(0) \propto \Lambda^{d+2}$ can be
absorbed by the normalization constant ${\bf N}$ of the partition function.

The $b$-contribution shifts in the final result (\ref{Z3}) $\varepsilon_{3} $
in the following way:
\begin{equation}
\varepsilon_{3} \to \varepsilon_{3} + \frac{b}{2}(N-1) \dl
\end{equation}
Qualitatively this term is not new: a term $\propto \dl $ was
already found with $k_{1}$. So we can interpret $b$ as a shift
of $k_{1}$, where $k_{2}$ has to perform the same shift in
order to keep the regular term $(k_{1}-k_{2})/V $ unchanged.
We conclude that indeed $a$ and $b$ are completely arbitrary,
even if we include the leading power  divergences.

To be explicit, we just have to replace in eq. (\ref{sig3d3})
\begin{displaymath}
\sigma_{3} \to \sigma_{3} + \frac{b}{2F^{6}} (N-1) \dl \quad .
\end{displaymath}

\vspace*{7mm}

Let us consider the more complicated case ${\bf d=4}$.  We find:
\begin{eqnarray}
ln \sqrt{{\bf g}_{g}} &=& \frac{1}{2} tr \varepsilon_{g} -\frac{1}{4}
tr \varepsilon_{g}^{2} \nonumber \\
&=& \frac{1}{2}tr\varepsilon - \frac{1}{4} tr \varepsilon^{2} \nonumber \\
&& + a_{1} \frac{N-1}{2F^{2}} V \Dl + \frac{a_{1}+(N-1)a_{2}}{2F^{2}}
\dl \int \partial_{\mu} \pif \partial_{\mu} \pif dx \nonumber \\
&& + (2a_{3}-a_{1}^{2}) \frac{N-1}{4F^{4}} V \DL
+ b_{1} \frac{\gamma \Omega^{00}}{2F^{4}} (N-1) \dl \left(
1 - \frac{1}{2V} \int \pif^{2} dx \right) \nonumber \\
&& +(b_{2}-a_{1}b_{1}) \frac{\gamma \Omega^{00}}{2F^{6}} \Dl \label{massallg4}
\end{eqnarray}
$\DLF $ is also defined eq. (\ref{strongsing}), and $\varepsilon_{g}$
is the generalization of the matrix $\varepsilon = g -1$ introduced
in section 3. We note that there occur three cancellations in the considered
order from the trace of the linear and the quadratic matrix
$\varepsilon_{g}$. They concern the terms
\begin{equation} \label{list}
b_{1}\frac{\gamma \Omega^{00}}{2F^{4}V} tr \varepsilon \quad , \quad
\frac{a_{1}}{2F^{2}V} \int (\partial_{\mu} \pif )^{2}dx \qquad
{\rm and} \quad \frac{a_{1}}{2F^{2}} D^{\Lambda}(0) \int \pif^{2} dx
\end{equation}
Eq. (\ref{massallg4}) reveals that our central observation of $d=3$ --
that the non-leading measure couplings do not contribute any physically
relevant term -- still holds for $d=4$. Here eq. (\ref{detgl}) for
the dimensionally regularized system
is confirmed on a highly non-trivial level.
In particular the first two terms of the
list (\ref{list}) would have destroyed this property.

Hence it was justified to evaluate only the leading
measure term in section 3. \\

Since none of the non-leading coupling constants in the measure
multiplies any regular term, its meaning is
already exhausted with their contributions to the (powerful) counter terms.
In particular the measure can not provide any counter terms
that enter the renormalization equations (they behave like
$k_{6}$), so the number of
non-leading coupling constants in the Lagrangian required for
the renormalization of the three loop result remains unchanged
(as we observed for $d=3$ before).

Concerning the power divergences, we note that
the last term of the cancellation list (\ref{list}) would have been
forbidden by perturbative renormalizability.

The factor
\begin{displaymath}
exp \left( \frac{N-1}{2F^{2}}\Big[ a_{1} \Dl +
\frac{2a_{3}-a_{1}^{2}}{2F^{2}} V \DL \Big] \right)
\end{displaymath}
of $\sqrt{{\bf g}_{g}}$ can be absorbed by ${\bf N}$.
The rest changes $\varepsilon_{2},\ \varepsilon_{3}$
in the following way (referring to (\ref{e2d4}), (\ref{e3d4}))
\begin{eqnarray}
\varepsilon_{2} &\to & \varepsilon_{2} + b_{1} \frac{N-1}{2} \dl \\
\varepsilon_{3} & \to & \varepsilon_{3} - \frac{N-1}{2}
\Big[ \{ a_{1}+(N-1) (a_{2}+ b_{1}/2) \} \dl G_{1}
+(a_{1}b_{1}-b_{2}) \Dl \Big]  \nonumber \\
\end{eqnarray}
where again a forbidden term ($\propto V(\dl )^{2}G_{1}$) cancels in
$\varepsilon_{3}$.

The singularities in the additional terms can be renormalized by
generalizing the formulas (\ref{ctcon4}). The counter terms of the
leading coupling constants receive the following additional summands
(with respect to eqs (\ref{sig2d4}) and (\ref{ctd4})) :
\begin{eqnarray*}
\sigma_{2} & \to  & \sigma_{2} + \frac{N-1}{2F^{4}} b_{1} \dl \\
\sigma_{3} & \to & \sigma_{3} - \frac{N-1}{2F^{6}}
\Big[ \{ a_{1}+(N-1)(a_{2}+b_{1}/2) \} \dl \gle
+(a_{1}b_{1}-b_{2}) \Dl \Big] \\
f_{2} & \to & f_{2} - \frac{N-1}{2F^{4}} \{ a_{1}+(N-1)(a_{2}+b_{1}/2) \} \dl
\end{eqnarray*}

\vspace*{5mm}

Since an important motivation for considering $d^{2}s_{g}$ was given
by the transformations (\ref{feldtr1})$\dots $(\ref{feldtr3}),
let us at last take a look at their actual effect,
i.e. we want to observe the outcome if the three terms
\begin{displaymath}
g_{4}^{(1)}(\partial^{2} \vec S)^{2} \quad ; \quad
h_{1,2}^{(2)}(\vec H \partial^{2}\vec S ) \quad ; \quad
h_{1,4}^{(3)}(\vec H \vec S )( \partial^{2} \vec S )^{2}
\end{displaymath}
are included in the Lagrangian. From power counting we see that for
$d=3$ all the three terms can affect our result only to the third
order, unlike $d=4$ where $g_{4}^{(1)}$ and $h_{1,2}^{(2)}$ could
appear already to the second order.

Next we recall that the elimination of the $h_{1,2}^{(2)}$-term is
only motivated by the possibility of a space-dependent magnetic
field $\vec H (x)$. In the case of a constant external field considered
here, this term does not contribute to the action.

Let us discuss the influence of the remaining two terms,
provided with dimensionless coupling constants $K_{1}, \ K_{2}$
(we recall that we choose $F$ to be the only dimension-carrying
coupling). We modify the Lagrangian of section 2 as follows:
\begin{eqnarray*}
{\cal L} & \to & {\cal L} + K_{1} \frac{1}{2F^{2}}
(\partial^{2} \vec S)^{2}  \qquad \qquad \qquad \qquad \qquad \qquad
(d=3) \\
{\cal L} & \to & {\cal L} + \frac{1}{2} K_{1} (\partial^{2} \vec S)^{2}
-K_{2} \frac{\Sigma}{F^{6}} (\vec H \vec S ) (\partial^{2} \vec S)^{2}
\qquad \qquad (d=4)
\end{eqnarray*}

For {\bf $d=3$} only the $K_{1}$-term is relevant to 3 loops.
In the final result, this alters the argument of the Bessel function
such that
\begin{displaymath}
\varepsilon_{3} \to \varepsilon_{3} + \frac{1}{2} K_{1}(N-1)\left(
\dl -\frac{1}{V} \right)
\end{displaymath}
The same kinds of terms were also found with the $k_{i}$ couplings.
In dimensional regularization the singularity structure does not
change due to $K_{1}$. However, in contrast to the non-leading
couplings in the measure, $K_{1}$ creates a regular contribution.

As a consequence, the new counter term introduced by $K_{1}$
rises the degree of freedom of the set
of logarithmic counter terms from 1 to 2. \\

Referring to the renormalization of the leading power divergences
we first note that a forbidden term $\propto (N-1)VG_{1}D^{\Lambda}(0)$
cancels in $\varepsilon_{3}$.
Hence it suffices to replace \
$\sigma_{3} \to \sigma_{3} + \frac{1}{2} K_{1}(N-1) \dl $.\\

For {\bf $d=4$} the action to three loops changes as follows:
\begin{displaymath}
S \to S + \frac{1}{2} K_{1} \int \Big[ (\partial^{2} \pif )^{2} +
(\partial_{\mu} \pif \partial_{\mu} \pif + \pif \partial^{2} \pif )^{2}
\Big] dx
- K_{2} \frac{\gamma \Omega^{00}}{F^{4}V} \int \partial^{2} \pif
\partial^{2} \pif dx
\end{displaymath}
Here the modifications are much more complicated, mainly because
the $K_{1}$-term is of first order. A lengthy calculation --
along the lines of section 6 -- yields:
\begin{eqnarray*}
\varepsilon_{2} & \to & \varepsilon_{2} +\frac{1}{2} K_{1}(N-1)
\left( \dl -\frac{1}{V} \right) \\
\delta_{3} & \to & \delta_{3} - K_{1} \frac{N-1}{2V} G_{1} \\
\varepsilon_{3} & \to & \varepsilon_{3} +(N-1) \Big\{ K_{1}
([N+1]V\dl -3N+5) \frac{1}{4V} G_{1} +2K_{1}G_{\mu \nu } \dot G_{\mu \nu} \\
&& + (K_{1}[k_{1}-\frac{1}{2} K_{1}]+K_{2}) \ \Dl \Big\}
\end{eqnarray*}
The result for $\varepsilon_{3}$ in $d=3$ is shifted down to
$\varepsilon_{2}$ here, since $1/V$ is classified in the second order now.
Also for $d=4$ only $K_{1}$ is relevant in dimensional regularization.
$K_{2}$ does not contribute to the considered
order, i.e. it behaves like $k_{6}$ and all the non-leading couplings
in the measure. Also the observation still holds that $K_{1}$ does not
change the singularity structure but it does change the regular part.
Its associated (logarithmic) counter term rises again the
degree of freedom of the logarithmic counter terms
from 1 to 2. (Note that this counter term enters constraint (\ref{drct2}) --
which was identically imposed by $\varepsilon_{2}$ and $\delta_{3}$ --
both times in such a way that the constraints due to
$\varepsilon_{2}$ and $\delta_{3}$ remain identical.)\\

In the notation used at the end of section 5 and 6, the terms occurring
in the intermediary results and cancelling at the end are:
\begin{eqnarray*}
\varepsilon_{2} & : & (N^{2},N,1)K_{1}G_{1}VD^{\Lambda}(0) \\
\delta_{3} & : & (N^{3},N^{2},N,1)VD^{\Lambda}(0) G_{1}^{2}, \
(N^{2},N,1)D^{\Lambda}(0)G_{2},\ (N^{2},N,1)\dl G_{1},\ N^{2}G_{1}/V \\
\varepsilon_{3} & : & K_{1}: \ (N^{3},N^{2},N,1) \{ G_{1}V(\dl )^{2},
V^{2}\dl D^{\Lambda}(0)G_{1}^{2}, VD^{\Lambda}(0)G_{1}^{2},G_{2}
D^{\Lambda}(0) \},\\
&& \qquad \ (N^{2},N,1)VD^{\Lambda}(0)\dl G_{2},\ N^{3} \{ G_{1}/V,
\dl G_{1} \} \\
&& K_{1}^{2}: \ (N^{3},N^{2},N,1)G_{1}(VD^{\Lambda}(0))^{2}, \
(N^{2},N,1)\{ G_{1}V \DL , V\dl D^{\Lambda}
(0) \}, \\ && \qquad \ N^{2}D^{\Lambda}(0) \\
&& K_{1}k_{1} : \ (N^{2},N,1)V\dl D^{\Lambda}(0),\ N^{2} D^{\Lambda}(0)
\end{eqnarray*}
Most of these cancellations are required by renormalizability
of the structure with power divergences (in
particular there are no counter terms available with volume-dependent
prefactors, whose powers of $L$ are larger than zero;
in the $\delta $-sector even pure singularities are forbidden.) \\
The exceptions are $N^{2}G_{1}/V$ in $\delta_{3}$ and $N^{3}\{ G_{1}/V,
\dl G_{1} \},\ N^{2}D^{\Lambda}(0)\{ K_{1}^{2},K_{1}k_{1} \} $ in
$\varepsilon_{3} $: here the
singularities could be absorbed, but each of these terms would be
a strange novelty
to compare with section 6,
\footnote{Such terms would not permit the field transformation
invariance of Z, pointed out below.}
whereas existing terms just shift the regular and the
power divergent contributions of the coupling constants $k_{i}$.
About the renormalization we remark that since
the singular contributions associated with $K_{1}$ and $K_{2}$
consist of terms we found before in section 6, their renormalization
-- along the lines of appendix A -- works without any problems.

The transformations (\ref{feldtr1})$\dots $ (\ref{feldtr3})
alter the non-leading coupling constants
of the Lagrangian like this:
\begin{eqnarray}
&& k_{1} \to k_{1} - \alpha - \beta \ , \quad k_{2} \to k_{2} - \beta \ ,
\quad k_{3} \to k_{3} + \beta \nonumber \\
&& k_{4} \to k_{4} + 4 \alpha \ , \quad k_{6} \to k_{6} + \alpha^{2}
-2\alpha K_{1} \label{lagtrans} \\
&& K_{1} \to K_{1} - 2 \alpha \ , \quad K_{2} \to K_{2} - \frac{1}{2}
\alpha^{2}  + 2\alpha k_{1} + \beta K_{1} - 2\alpha \beta + \lambda
\nonumber
\end{eqnarray}
and we can confirm our claim of section 2 that $K_{1}$ and $K_{2}$
can be chosen to be zero, as well as $k_{2}$ or $k_{3}$.

If we include in the partition function (for $d=3$ and $d=4$)
{\em all} possible couplings in the Lagrangian and in the measure
and apply the transformation rules for both sets of couplings,
we observe that \ Z$(\vec S , \vec H , k_{i},K_{i},a_{i},b_{i})$ \
-- including the leading power divergences --
is invariant under the transformations (\ref{feldtr1}) $\dots $
(\ref{feldtr3}) for
arbitrary $\alpha , \beta $ and $\lambda $.
\footnote{To the considered order there occur no mixed
products of non-leading coupling constants from the
measure and from the Lagrangian.
So to verify this invariance one just has to insert the transformation rules
in the results given above.}
This is a very sensitive
consistency test for our results.

Thus we have found an other remarkable aspect that supports that
the structure with all the leading divergences still fulfills
the important properties.
This invariance is due to an exchange of terms among all the
non-leading coupling constants, except for $k_{5}$.\\

As a corollary we can conclude that the overall invariance
of Z under the discussed field transformations also holds for
the dimensionally regularized system. There, however, the consistency
test is less sensitive; only $k_{1} \dots k_{4}$ and $K_{1}$
participate in an exchange of regular terms and transformation
(\ref{feldtr3}) is irrelevant.

\vspace*{5mm}

The conclusion of this appendix is that the maximal generalization
of the measure
-- including all the terms fulfilling the three physical properties
listed in the beginning of this appendix -- is permitted
by perturbative renormalizability and only affects
the power divergences in the Lagrangian, i.e. in dimensional
regularization they do not yield any contribution at all.
Hence they do not reduce the number
of required non-leading coupling constants in the Lagrangian.

Since the coefficients for the
non-leading terms are completely arbitrary, there is an infinite
set of equivalent measures of the path integral for our model --
each measure corresponding to a particular type of quantization --
that all yield the same physically relevant contributions.
The Polyakov measure belongs to this set and has the advantages
of its simplicity and its compatibility to renormalizable theories.

Without the transformations eliminating some terms in the Lagrangian,
in the regular part the coupling constants $k_{1} \dots k_{5}$ would
have been shifted.
The singularities, which are present in dimensional
regularization, however, are not affected by these transformations.
The leading power singularity structure would have been altered without
destroying its perturbative renormalizability.

\section{Identification of the terms by massive expansion}

If we assume the GB to be massive, their Green functions take the form
\begin{displaymath}
G_{m}(x) = \frac{1}{V} \sum_{\bar n } \frac{e^{ip_{\bar n }x}}{m^{2}+
p_{\bar n }^{2}}
\quad .
\end{displaymath}
For small masses we can expand this in terms of the massless $G(x)$:
\begin{displaymath}
G_{m}(x) = \frac{1}{Vm^{2}} + G(x) - m^{2}\dot G (x)
+ \frac{m^{4}}{2} \ddot G (x) + \cdots
\end{displaymath}
where \ $\dot G \doteq -\frac{d}{dm^{2}} G_{m} \vert_{m=0} $ \ etc.

In our case we have to insert $m^{2} = \frac{\Sigma H}{F^{2}} =
\frac{\gamma }{F^{2}V}$ , so: $-m^{2}\frac{d}{dm^{2}}\vert_{m=0}
\propto \frac{\gamma }{F^{2}} L^{2-d} $ lowers the magnitude by one unit.\\
Thus we get e.g. the series \
$ \frac{1}{F^{2}}G_{1},\ \frac{1}{F^{4}V} G_{2},\
\frac{1}{F^{6}V^{2}}G_{3}
 \dots $ (index = order). \\

Gerber and Leutwyler carried out a 3-loop calculation with $G_{m}$ for
$d$=4 \cite{ge/leu}.
There only the temperature was finite, not the spatial box, but
still the type of terms ought to coincide with ours. They found, except
for $G_{1}, \ G_{2},\ G_{3}$ the terms:
\begin{displaymath}
\frac{m^{4}}{F^{4}} \int G_{m}^{4} dx \ , \quad
\frac{1}{F^{4}} \int (\partial_{\mu }G_{m} \partial_{\mu } G_{m})^{2}dx \ ,
\quad \frac{1}{F^{4}} (\partial_{\mu \nu} G_{m}(x)\vert_{x=0})^{2}
\end{displaymath}
We want to show that this corresponds to the type of terms
we found up to the $3^{rd}$ order.
\begin{eqnarray*}
i) && m^{4} \int \left( \frac{1}{Vm^{2}} + G - m^{2}\dot G
+ \frac{m^{4}}{2} \ddot G \dots \right) ^{4} dx \\
 & = & \frac{1}{V^{3}m^{4}} + \frac{6}{V} G_{2} + \frac{4m^{2}}{V}
(J_{3} - \frac{3}{2V}G_{3}) + O(m^{4})
\end{eqnarray*}
If we proceed one order by $- m^{2} \frac{d}{dm^{2}}$ and put $m=0$,
we can identify $J_{3}$ as a $3^{rd}$ order term,
which is relevant
for us since $m^{2}$ introduces an $H$-dependence. (The singularity
at $m=0$ \ corresponds to the 0-mode.)
\begin{displaymath}
ii) \quad \int \partial_{\mu}G_{m} \partial_{\mu}G_{m} \partial_{\nu}G_{m}
\partial_{\nu}G_{m} dx = J_{2} + 4m^{2}\Gamma_{3} + 0(m^{4})
\end{displaymath}
Here we can confirm in exactly the same way the term $\Gamma_{3}$.\\

$iii)$ For $d=4$ : $\frac{1}{F^{4}} (G_{\mu \nu})^{2} \propto L^{-8}$, so
$ \frac{1}{F^{4}} G_{\mu \nu}
\dot G_{\mu \nu}$ can enter the $3^{rd}$ order, but not with a factor
$m^{2}$. As we know from section 6, it does occur, with factors
$\frac{\gamma}{F^{2}}k_{j}$ instead of $m^{2}$. For $d=3$ this term
is only of $4^{th}$ order, in contrast to the results of $i)$ and $ii)$
that are classified equally for $d=3$ and $d=4$. \\

To look at $iii)$ one might wonder if one couldn't include terms of
even lower orders that one gets by further dot-derivatives.

But introducing such terms -- or also the corresponding terms
that could be produced in $i)$ and $ii)$ -- would contradict
our low energy expansion based on two leading coupling constants
as well as the perturbative renormalizability.

\section{Transformation to the modified Bessel function}

The differential equation (\ref{diffY}) and its first derivative state:
\begin{eqnarray}
\int d\Omega e^{z\Omega ^{00}}\Big[ z^{2}\Omega^{00~2} + (N-1)z\Omega^{00}
-z^{2} \Big] & = & 0 \label{a} \\
\int d\Omega e^{z\Omega^{00}} \Big[ z^{3}\Omega^{00~3} + Nz^{2}\Omega^{00~2}
-z^{3}\Omega^{00} -z^{2} \Big] & = & 0 \label{b}
\end{eqnarray}
We apply on
\footnote{As usual the index coincides with the order of magnitude.}
\begin{displaymath}
\begin{array}{ll}
\lefteqn{\int d\Omega e^{\gamma \Omega^{00}(1+\alpha_{1} +\alpha_{2}
 +\alpha_{3})
+(\gamma \Omega^{00})^{2}(\beta_{2}+\beta_{3})+(\gamma \Omega^{00})^{3}
\gamma_{3} } } \\
\cong \int d\Omega e^{\gamma \Omega^{00}} & \Big[ 1+ \gamma \Omega^{00}
(\alpha_{1} + \alpha_{2} + \alpha_{3} ) + (\gamma \Omega^{00})^{2}
(\frac{\alpha_{1}^{2}}{2} \alpha_{1} \alpha_{2} + \beta_{2} +\beta_{3}) \\
 & + (\gamma \Omega^{00})^{3} (\frac{1}{6} \alpha_{1}^{3} + \alpha_{1}
\beta_{2} + \gamma_{3}) \Big]
\end{array}
\end{displaymath}
the transformations enabled by (\ref{a}) and (\ref{b}) with the factors
$-a,\ -b$ respectively:
\begin{displaymath}
\begin{array}{ll}
= \int d\Omega e^{\gamma \Omega^{00}} & \Big[ 1+\gamma^{2}(a+b) +
\gamma \Omega^{00}(\alpha_{1}+\alpha_{2}+\alpha_{3}-(N-1)a +\gamma^{2}b) \\
& + (\gamma \Omega^{00})^{2}(\frac{\alpha_{1}^{2}}{2} + \alpha_{1}\alpha_{2}
+ \beta_{2} +\beta_{3} -a-Nb) \\
& +(\gamma \Omega^{00})^{3}(\frac{\alpha_{1}^{3}}{6}
+\alpha_{1}\beta_{2}+\gamma_{3}-b) \Big]
\end{array}
\end{displaymath}
This shall take the form: $e^{\delta_{1}+\delta_{2}+\delta_{3}}
\int d\Omega e^{\gamma \Omega^{00}(1+\varepsilon_{1}+\varepsilon_{2}
+\varepsilon_{3}) } $. Expanding this and comparing the coefficients
to the  orders of $(\gamma \Omega^{00})$ we get four equations for
the unknown $a,b,\delta ,\varepsilon $. On the 3 levels these are
12 variables to be determined by the 3 levels of the 4 equations,
which impose 12 constraints. $a$ and $b$ we don't need to know
explicitly, so we first eliminate them. From the remaining 2
equations we can determine on the level $\ell : \quad \delta_{\ell}$ and
$\varepsilon_{\ell }$.
\begin{eqnarray*}
\delta_{1} =&~ 0 \qquad \qquad \qquad \qquad \quad
& \varepsilon_{1}= \alpha_{1}  \\
\delta_{2} =&~ \gamma^{2}\beta_{2} \qquad \qquad \qquad \qquad
& \varepsilon_{2}=\alpha_{2}-(N-1)\beta_{2}\\
\delta_{3} =& \gamma^{2}(\beta_{3}-(N-1)\gamma_{3}) \quad &
\varepsilon_{3} = \alpha_{3}+(N-1)(\alpha_{1}\beta_{2}-\beta_{3})
+\gamma_{3}(\gamma^{2}+N(N-1))
\end{eqnarray*}
This has been inserted in sections 5 and 6 in order to transform
the row result for the partition function into the desired form.\\

Corresponding transformations into this form
are possible for results to any order.\\
For the order $\ell $ we derivate (\ref{diffY}) \ $ \ell -2$ times,
get a differential equation of degree $\ell $ that allows to
eliminate the term with $(\gamma \Omega^{00})^{\ell }$, etc.
Thus we arrive inductively at the desired from.

\section{Conclusions about the non-linear $\sigma $-model
in lower dimensions}
For the three and the four dimensional case that we
discussed in the main part of this paper, all the $O(N)$-invariant
terms had to be included in order to enable a perturbative
renormalization. This is different for the lower
dimensions; there it is sufficient to include the leading
coupling constants $\Sigma $ and $F$. Thus in one dimension
all singularities disappear (in the source dependent part we consider);
in two dimensions all the singularities become logarithmic and
can be renormalized solely by these two couplings.\\

\subsection{The one dimensional case}

Since we expand generally in powers of $L^{d-2}/F^{2}$, here
the quantity that has to be small to permit our expansion
is $L/F^{2}$, i.e. in contrast to the higher dimensions $L$
must be small (one could imagine it to be a time slice)
or the energy must be high when we translate our considerations
to a scattering process. With respect to the partition function
we deal with a high temperature expansion. Thus physics is turned
upside down (note also that there are no GB's any more) but the mathematical
methods remain applicable. In particular the zero mode is not weighted
strongly any more, but its contribution still diverges and its
treatment with collective variables is still a solution of this problem.

The explicit, finite terms that remain if we include only
$\Sigma $ and $F$ are easily obtained:
\begin{eqnarray}
G_{1} &=& \frac{L}{2^{2}3} \ , \qquad \frac{1}{V}G_{2} = \frac{L^{2}}
{2^{4}3^{2}5} \qquad  \label{d1id}
\frac{1}{V^{2}} G_{3} = \frac{L^{3}}{2^{4}3^{3}5 \cdot 7 } \\
\frac{1}{V}J_{3} &=& \frac{1}{2V^{2}}G_{3} \ , \qquad
\Gamma_{3} = \frac{9}{4V^{2}}G_{3} \nonumber
\end{eqnarray}
Clearly, including non-leading derivative couplings -- or higher powers of
the magnetic field -- does not make sense if we expand
in positive powers of $L$. (This concerns also non-leading
couplings in the measure.)

Hence we let $k_{i} \equiv 0 $ and find the partition function to be of
the form (\ref{Z3}) with
\begin{eqnarray}
\varepsilon_{1} &=& -\frac{N-1}{2^{3}3} L \\
\varepsilon_{2} &=& -\frac{(N-1)(N-3)}{2^{7}3 \cdot 5} L^{2} \\
\varepsilon_{3} &=& \frac{(N-1)(5N^{2}-440N+843-16 \gamma^{2})}{2^{10}
3^{4}5 \cdot 7} L^{3} \\
\delta_{2} &=& \frac{N-1}{2^{6}3^{2}5} L^{2} \\
\delta_{3} &=& \frac{(N-1)(13N-43)}{2^{8}3^{4}5 \cdot 7} L^{3}
\end{eqnarray}
This model describes a free quantum mechanical spin 0 particle moving on a
$N$ dimensional unit sphere where $L$ is the inverse temperature
(or the time for a short time transition amplitude).
Turning on the external magnetic field means geometrically a shift
of the center of the sphere away from the origin.
(Higher powers of $\vec H $ in ${\cal L}^{(sb)}$ would additionally
deform the sphere).\\

\subsection{The two dimensional case}

This case has attracted very much attention in the literature because
it is renormalizable and it represents an interesting toy model for
QCD, see below.\\

We consider again the simplification $k_{i} \equiv 0 $.
Then the three loop partition
function can be taken from section 5 or 6. The only terms that remain
singular in two dimensions are $G_{1}$ and $\Gamma_{3}$.

Here dimensional regularization looses its special meaning since all
the divergences are logarithmic. In particular:
\begin{displaymath}
\gle = c \cdot ln (\Lambda / \mu )
\end{displaymath}
where $\Lambda $ refers to the regularization parameter introduced in
section 7, $c$ is a positive constant and $\mu $ determines the mass scale.
$G_{1}$ is independent of this scale, so
\begin{equation} \label{gmug}
\mu \frac{\partial }{\partial \mu } \gle =
- \mu \frac{\partial }{\partial \mu } g_{1} \quad .
\end{equation}
$\Gamma_{3} $ contains a term quadratic and a term linear in $ln \Lambda $,
so it can be written in the form:
\begin{displaymath}
\Gamma_{3} = c_{A} \gle {^{2}} + c_{B} g_{1} \gle + \gamma_{3}
\end{displaymath}
where $c_{A}, \ c_{B}$ are constants and $\gamma_{3}$ is regular.
Analogously to (\ref{gmug}) we find
\begin{eqnarray}
\mu \frac{\partial }{\partial \mu } \gamma_{3} &=& - c \cdot c_{B} g_{1} \\
\qquad 2c_{A} &=& c_{B} \quad .
\end{eqnarray}

Now we come to the renormalized coupling constants $\Sigma_{r} $ and
$\Phi_{r}$, where we rename $\Phi \doteq 1/F^{2}$.
They can be taken from appendix A with slight modifications:
\begin{eqnarray*}
\Sigma_{r} &=& \Sigma \Big[ 1 - \frac{N-1}{2} \Phi \gle -
\frac{(N-1)(N-3)}{8} \Phi^{2} \gle {^{2}} \\
&& -\frac{(N-1)(N-3)(3N-7)}{48} \Phi^{3} \gle {^{3}} - (N-1)(N-2)
c_{A} \Phi^{3} \gle {^{2}} +O(\Phi^{4}) \Big] \\
\Phi_{r} &=& \Phi \Big[ 1 + (N-2) \Phi \gle + (N-2)^{2} \Phi^{2} \gle{^{2}}
+2(N-2) c_{B} \Phi^{2} \gle + O(\Phi^{3}) \Big]
\end{eqnarray*}

Hence their $\beta $-functions are :
\begin{eqnarray}
\beta_{\Sigma} & \doteq & \mu \frac{\partial \Sigma_{r}}{\partial \mu }
= \frac{N-1}{2} \cdot c  \cdot \Sigma_{r} \Phi_{r} + O(\Phi^{4}_{r}) \\
\beta_{\Phi} & \doteq & \mu \frac{\partial \Phi_{r}}{\partial \mu} =
-(N-2) \cdot c \cdot \Phi_{r}^{2} - 2(N-2) \cdot c \cdot c_{B} \Phi_{r}^{3}
+ O(\Phi_{r}^{4})
\end{eqnarray}
The cancellation of the second and third order in the $\beta $-function
of $\Sigma $ is a consequence of
our choice of the renormalization prescription. \\

If we include only the leading order of $\beta_{\Phi} $ , we find the
solution:
\begin{equation}
\Phi_{r}(\mu ) =
\frac{\Phi_{r,s}}{1+(N-2)\cdot c \cdot \Phi_{r,s} ln(\mu / \mu_{s})}
\end{equation}
where $\mu_{s}$ is a particular scale (i.e. the choice of one among the
trajectories that solve the differential equation) and $\Phi_{r,s} \doteq
\Phi_{r} (\mu_{s})$ .

For $N>2$ this solution has a ``Landau pole'' at
\begin{displaymath}
\mu_{L} = \mu_{s} exp(-[(N-2)\cdot c \cdot \Phi_{r,s}]^{-1})
\end{displaymath}
but of course in this regime the perturbative expansion is not
applicable.

For large $\mu $, however, $\Phi_{r}$ goes asymptotically to zero
and perturbation theory becomes reliable.
We clearly recognize asymptotic freedom for high energies
(large $\mu $), {\em just as in QCD}.

For higher orders in $\Phi $,
the fact that the $\beta $-functions must be independent of $\Lambda $
yields in $\Sigma_{r},\ \Phi_{r}$ immediately the 2, 3 counter terms
(respectively), which are leading in the power of $ln (\Lambda / \mu)$.

The renormalization group equation also provides us with some
knowledge about the partition function itself. Z has to be
independent of $\mu $ and a variation of the quantity $\gamma =
\Sigma HV$ shows that this independence holds separately for the
argument of the modified Bessel function
and the exponent of the $\Omega $-independent prefactor.
Also there it determines the coefficients
of the leading divergences to higher orders of $\Phi $. \\

But in this way we do not get any information about the regular
contributions of the multiloop terms, which are actually of physical
interest. Work about the way to deduce such information
and its limits is in progress.

\end{document}